\documentclass[aps,superscriptaddress,showpacs,amsmath,amssymb,amsfonts,altaffillsymbol,notitlepage,pre]{revtex4-2}
\usepackage[dvips]{graphicx}
\usepackage{color}
\usepackage{soul}
\usepackage{epsfig}
\usepackage{bm}
\usepackage[normalem]{ulem}
\usepackage[utf8]{inputenc}
\usepackage[T1]{fontenc}
\usepackage{mathptmx}
\usepackage{etoolbox}
\usepackage{placeins}
\usepackage[caption=false,font=normalsize,labelfont=bf]{subfig}

\newcommand{\marktext}[1]{\textcolor{black}{ #1}}

\begin{document} 

\title{Yielding behaviour of active particles in bulk and in confinement}

% \preprint{APS/123-QED}

\author{Yagyik Goswami}
\affiliation{Laboratory of Nanoscale Biology, Paul Scherrer Institut, Villigen, Aargau, Switzerland}
\affiliation{Theoretical Sciences Unit and School of Advanced Materials, Jawaharlal Nehru Centre for Advanced Scientific Research, Rachenahalli Lake Road, Bengaluru-560064, India}

\author{G. V. Shivashankar}
\affiliation{Laboratory of Nanoscale Biology, Paul Scherrer Institut, Villigen, Aargau, Switzerland}
\affiliation{Department of Health Sciences and Technology, ETH Zürich, Zürich, Switzerland}

\author{Srikanth Sastry}
\email{sastry@jncasr.ac.in}
\affiliation{Theoretical Sciences Unit and School of Advanced Materials, Jawaharlal Nehru Centre for Advanced Scientific Research, Rachenahalli Lake Road, Bengaluru-560064, India}

% \author{Ann Author}
%  \altaffiliation[Also at ]{Physics Department, XYZ University.}%Lines break automatically or can be forced with \\
% \author{Second Author}%
%  \email{Second.Author@institution.edu}
% \affiliation{%
%  Authors' institution and/or address\\
%  This line break forced with \textbackslash\textbackslash
% }%

% \collaboration{MUSO Collaboration}%\noaffiliation

% \author{Charlie Author}
%  \homepage{http://www.Second.institution.edu/~Charlie.Author}
% \affiliation{
%  Second institution and/or address\\
%  This line break forced% with \\
% }%
% \affiliation{
%  Third institution, the second for Charlie Author
% }%
% \author{Delta Author}
% \affiliation{%
%  Authors' institution and/or address\\
%  This line break forced with \textbackslash\textbackslash
% }%

% \collaboration{CLEO Collaboration}%\noaffiliation

\date{\today}% It is always \today, today,
             %  but any date may be explicitly specified
%%%% Springer-Nature title and author
% \title[Active yielding in bulk and confinement]{Yielding behaviour of active particles in bulk and confinement} 
% \author[1,3]{{\small \fnm{Yagyik} \sur{Goswami}}}
% % \email{iauthor@gmail.com}
% {\author[1,2]{{\small \fnm{G. V.} \sur{Shivashankar}}} 
% \author*[1]{{\small \fnm{Srikanth} \sur{Sastry}}}\email{sastry@jncasr.ac.in}

% \affil*[1]{\orgdiv{{\small Laboratory of Nanoscale Biology}}, \orgname{{\small Paul Scherrer Institut}},\orgaddress{\street{{\small Forschungsstrasse 111}}, \city{{\small Villigen}}, \postcode{{\small 5232}}, \state{{\small Aargau}}, \country{{\small Switzerland}}}}
% \affil*[2]{{\small \orgdiv{Department of Health Sciences and Technology}, \orgname{ETH Z\:urich, Z\:urich}, \orgaddress{\street{Universit\:tstrasse 2}, \city{Z\:urich}, \postcode{8092}, \country{Swizterland}}}}
% % \normalsize{$^{2}$, , Switzerland,}\\}

% \affil*[3]{{\small \orgdiv{Theoretical Sciences Unit and School of Advanced Materials}, \orgname{Jawaharlal Nehru Centre for Advanced Scientific Research}, \orgaddress{\street{Jakkur}, \city{Bengaluru}, \postcode{560064}, \state{Karnataka}, \country{India}}}}
%%%%% EN Spring-Nture format

\begin{abstract}
The investigation of collective behaviour in dense assemblies of self-propelled active particles has been motivated by a wide range of biological phenomena. Of particular interest are dynamical transitions of cellular and sub-cellular biological assemblies, including the cytoskeleton and the cell nucleus. Motivated by observations of mechanically induced changes in the dynamics of such systems, and the apparent role of confinement geometry, we study the transition between jammed and fluidized states of active particles assemblies, as a function of the strength and temporal persistence of active forces, and in different confinement geometries. Our results show that the fluidization transition broadly resembles yielding in amorphous solids, consistently with recent suggestions. More specifically, however, we find that a detailed analogy holds with the yielding transition under cyclic shear deformation, for finite persistence times. The fluidization transition is accompanied by driving induced annealing, strong dependence on the initial state of the system, a divergence of time scales to reach steady states, and a discontinuous onset of diffusive motion. We also observe a striking dependence of the transition on persistence times and on the nature of the confinement. Collectively, our results have implications in epigenetic cell state transitions induced by alterations in confinement geometry.
% } %250 word limit. 249 now. 
\end{abstract}
\maketitle

% \section{Significance statement}
% Dense assemblies of self-driven or active particles have been studied recently to investigate the transitions they exhibit between jammed and unjammed states. Such assemblies and transitions are of interest as models to describe biological phenomena in diverse settings 
% \marktext{such as cell migration in tissues and changes in the spatial organisation of chromatin that accompany state changes in cells. We}
% % Motivated by the spatial organization of chromatin in differentiated cell nuclei and observations of mechanically induced changes in dynamics and in differentiation state, we
% study the transition between jammed to fluidized states of assemblies of active particles, including in different confinement geometries. We find detailed analogies between the fluidization transition in such assemblies with yielding under cyclic shear deformation, and striking dependence on confinement, and discuss the implications of these findings to understanding mechanically driven changes in differentiation states in cell nuclei.

\section{Introduction}
Active matter, composed of interacting self-propelled particles or entities, has been investigated intensely in recent years \cite{SRARCMP2010,ActiveMatterRMP2013}, and displays novel aspects of dynamics and organization not realised in conventional condensed matter. Although avenues for synthesizing active materials have been well explored, many obvious examples of active matter arise in biological systems across length scales, from bird flocks and animal herds\cite{ACIGARCMP2014}, bacterial suspensions\cite{Yhat2004}, tissues \cite{BiNPhys2015,BiPRX2016,DasPRX2021}, to the cytoskeleton\cite{JULICHER20073}, which have been investigated theoretically and computationally \cite{CompActiveNatRevPhys2020} to comprehend experimental findings. In several contexts, the systems of interest may be characterised as dense, disordered assemblies, displaying features characteristic of  glassy systems or phenomenology associated with packings that undergo jamming. 

In particular, novel behaviour arising in models of {\it active glasses} \cite{Janssen_2019,Henkes2011,MandalSM2016,mandal2020extreme,Mo2020,BiNPhys2015,BiPRX2016,DasPRX2021,Xu2018,MorsePNAS21,Agoritsas_2021,During2021,mandal2021study,mandal2022random,amiri2022random} have been explored, seeking similarities with and departures from glass and jamming phenomenology in the absence of active forces or self-propulsion. In addition to investigations of the nature of relaxation dynamics, and approach to structurally arrested states with a decrease in thermal fluctuations and active forces and/or increase in density \cite{mandal2020extreme,BiNPhys2015,BiPRX2016,DasPRX2021}, some recent works \cite{Xu2018,Mo2020,Agoritsas_2021,During2021,amiri2022random} have addressed the converse problem of loss of rigidity or fluidization, with an increase in the magnitude of active forces. Such a fluidization transition has an appealing analogy with the phenomenon of yielding in amorphous solids, upon an increase of external driving through the application of, {\it e.~g.}, shear stress or deformation. The active forces on individual particles, in the simple case where their orientations remain fixed, may be viewed as a spatially randomised generalisation of shear induced forces or displacements that arise when a global stress or strain is applied. Such an analogy has indeed been fruitfully pursued for the case when the orientations of the active forces is fixed. 

However, if one considers such active glass models in specific experimental contexts, including biological contexts where the active forces are generated by energy-consumption, one must consider a situation where the orientation of the active forces is not fixed but may evolve in time, in a manner characterised by a persistence time. For active glasses with finite persistence times, we find that the fluidization transition bears striking similarities to the yielding transition under cyclic shear deformation, wherein the shear deformation is  applied cyclically (typically) around the zero strain state. The response of amorphous solids to cyclic shear is characterised \cite{leishangthem2017,BhaumikPNAS21} by (i) annealing effects at small deformations, (ii) strong  dependence on the initial annealing state of the amorphous solids, (iii) a discontinuous transition from an {\it absorbing} state with zero diffusivity of particles to  a {\it diffusive} state with finite diffusivities, and (iv) diverging times to reach steady states as the yielding transition is approached from either side. As we show, active glasses with finite (but large) persistence times show a transition from arrested to fluidized states display these features remarkably. 

\marktext{
In addition to the persistence time, we also study the role of confinement on the transition to the fluidized state. 
The presence of a confining boundary has been found to induce non-trivial behaviour in active systems\cite{yang2014aggregation,fily2014dynamics,vishen2018soft}. Such effects can be expected to play an important role in determining the properties of many biological assemblies that experience strong confinement conditions. A particular example of this kind is the organization of chromatin within the nucleus, where an interplay of active forces and the effect of the confinement boundary are of potential interest to understand in mechanically induced cell-state changes\cite{talwar2013correlated,makhija2016nuclear,UhlerGVSReview2017}. While we do not attempt to model the specific details of such biological assemblies in this work, the results we obtain should be of relevance in several such contexts.
}

Instead, we aim to consider an idealised system, studying which we aim to comprehend the the role of perturbing factors which may be biologically relevant. 
We thus consider the arrested-to-fluidised, or yielding, transition in a dense assembly of particles at very low temperatures, subjected to active forces which have a finite persistence time. We investigate the yielding behaviour of local energy minimum configurations, referred to as glasses or amorphous solids, which are prepared with different degrees of annealing (see {\it Materials and Methods} and {\it SI Appendix}, Fig. S1 for details), followed by local energy minimization, so that their initial energies vary over a wide range.  We simulate a two dimensional $65:35$ (Kob-Andersen) binary mixture of particles interacting with the Lennard-Jones potential, whose dynamics is described by the Langevin equation, with forces from interparticle interactions, and active forces whose orientation changes diffusively with a specified persistence time. Further details of the model and simulations methods are given under {\it Materials and Methods}. 

We first investigate the yielding behaviour as a function of the strength of the activity and persistence time, and demonstrate that the yielding behaviour indeed strikingly resembles the yielding transition under cyclic shear. Since we also expect the effects of confinement to be significant, we consider different confinement geometries, where the interacting particles experience short range attractive interactions and strong repulsion at smaller distances to the confining boundary.  

%------------

\section{Results}

We first consider a system of $N = 1000$ particles in two dimensions with a number density of $\rho = 1.2$, with a small reduced temperature of $T = 10^{-3}$, with periodic boundary conditions. Each of these cases is simulated using Langevin dynamics with forces arising from interparticle interactions and an active force, with a time step of $dt=0.01$, over a range of values of the active force magnitudes $f$, which reorients on a time scale given by the persistence time $\tau_p$ (see {\it Materials and Methods}). We thus present our results for a fixed value of $\tau_p$  as a function of $f$, considering amorphous solids with a range of initial energies as described earlier (see {\it SI Appendix}, Fig. S1). In all cases, the systems evolve to reach steady states and we report the properties of the system in the steady state (unless otherwise stated). 

\subsection{Yielding transition}

\begin{figure}[htb!]
\centering
\subfloat{\includegraphics[height=3.5cm,width=4cm]{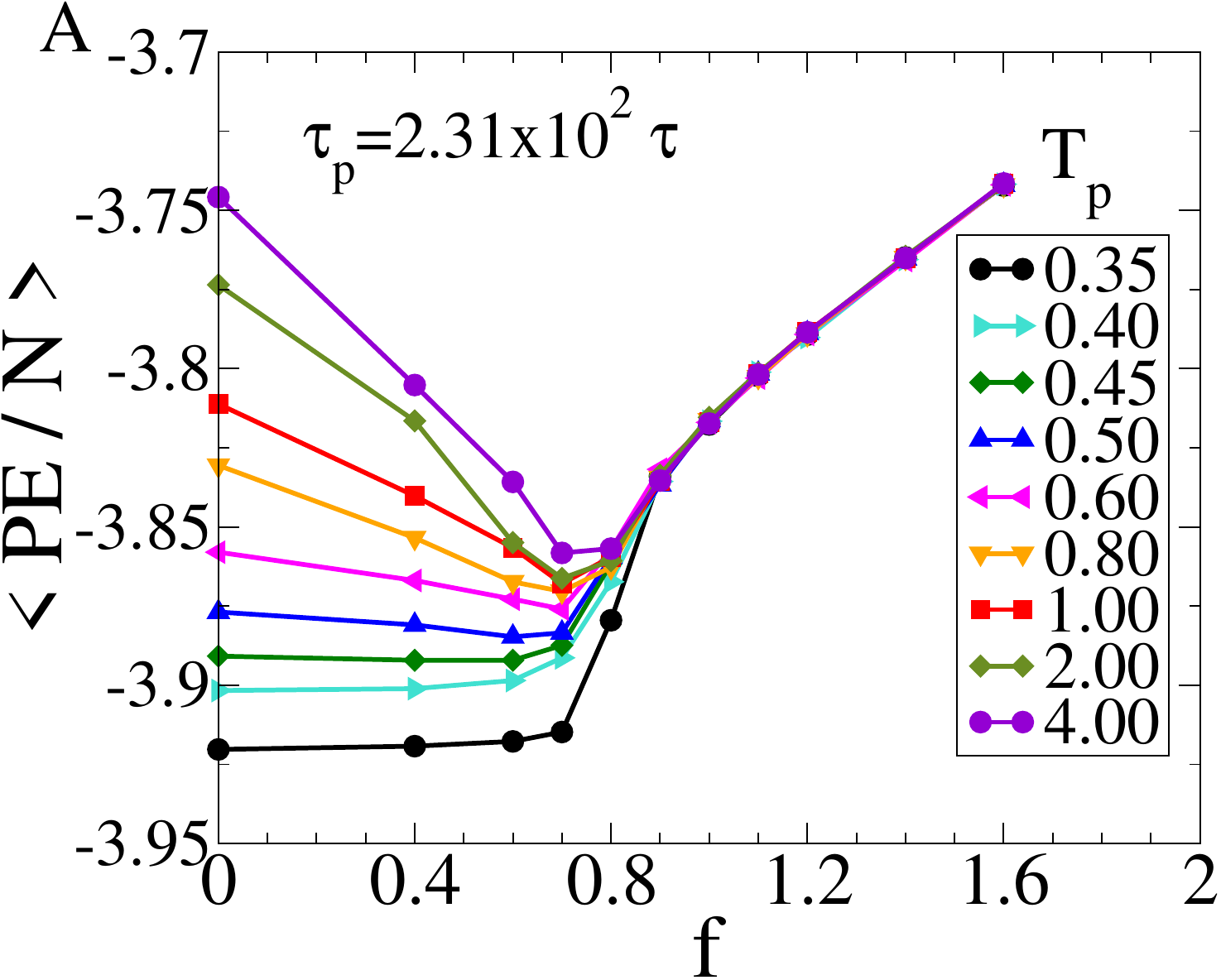}}
\subfloat{\includegraphics[height=3.5cm,width=4cm]{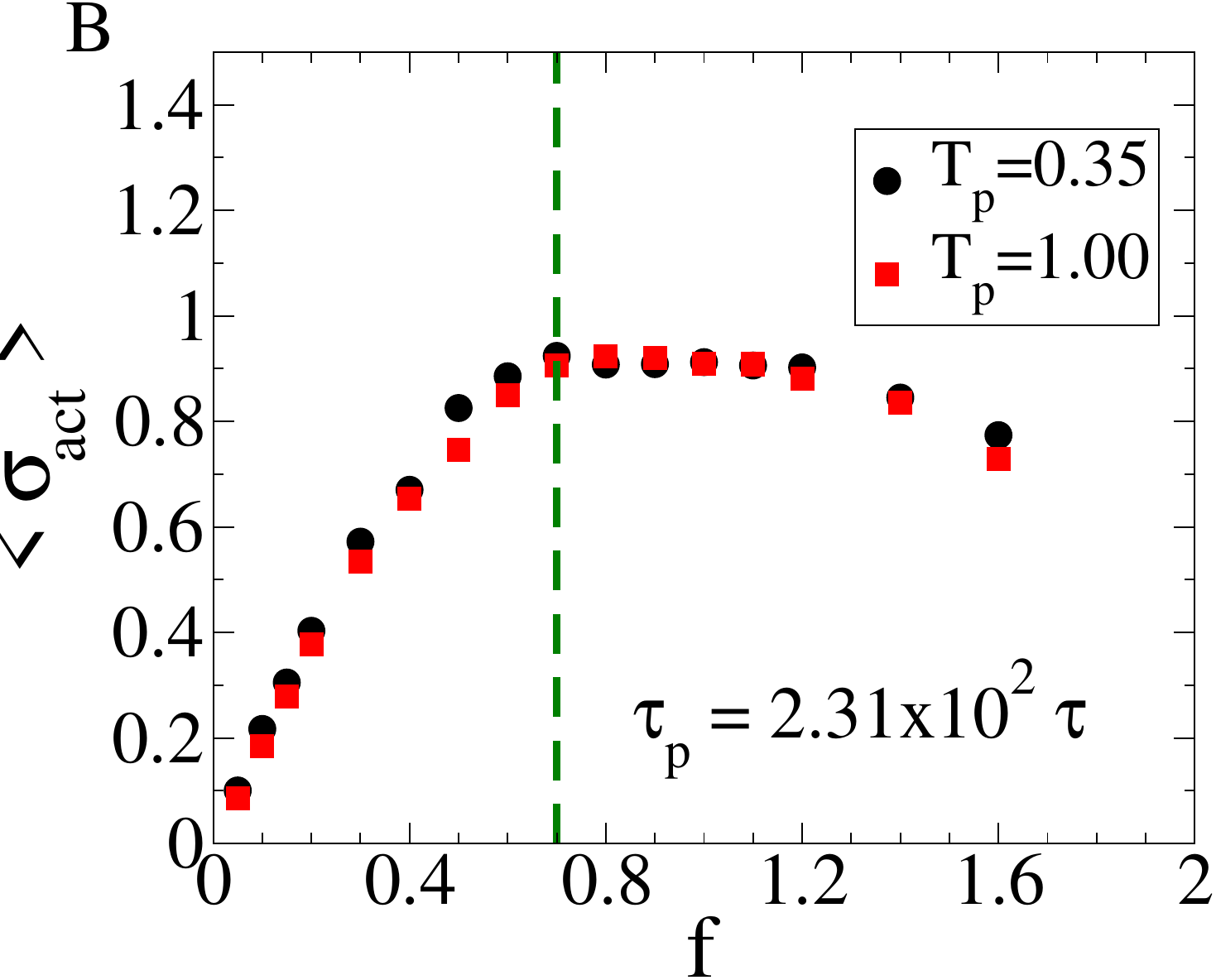}}

\subfloat{\includegraphics[height=3.5cm,width=4cm]{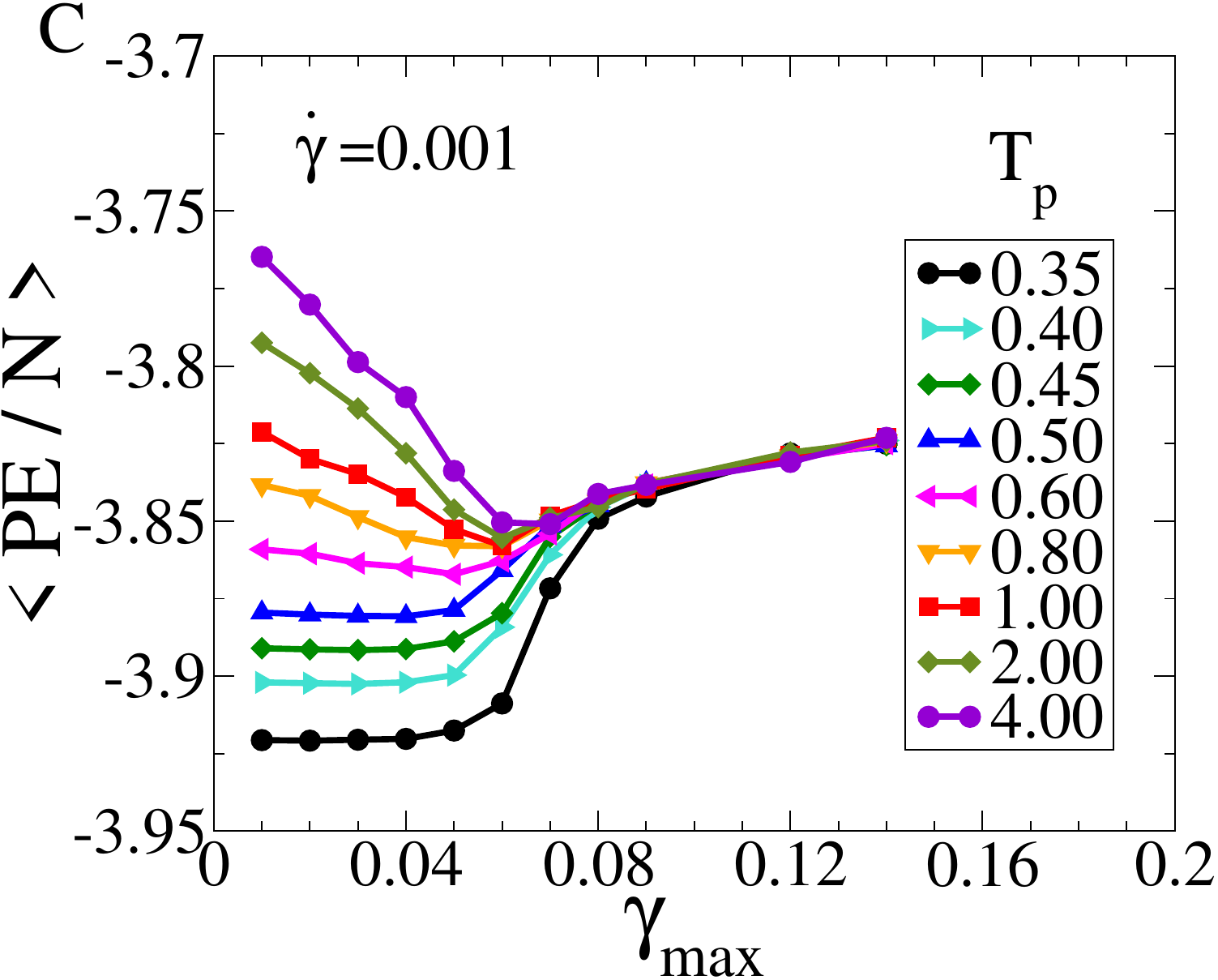}}
\subfloat{\includegraphics[height=3.5cm,width=4cm]{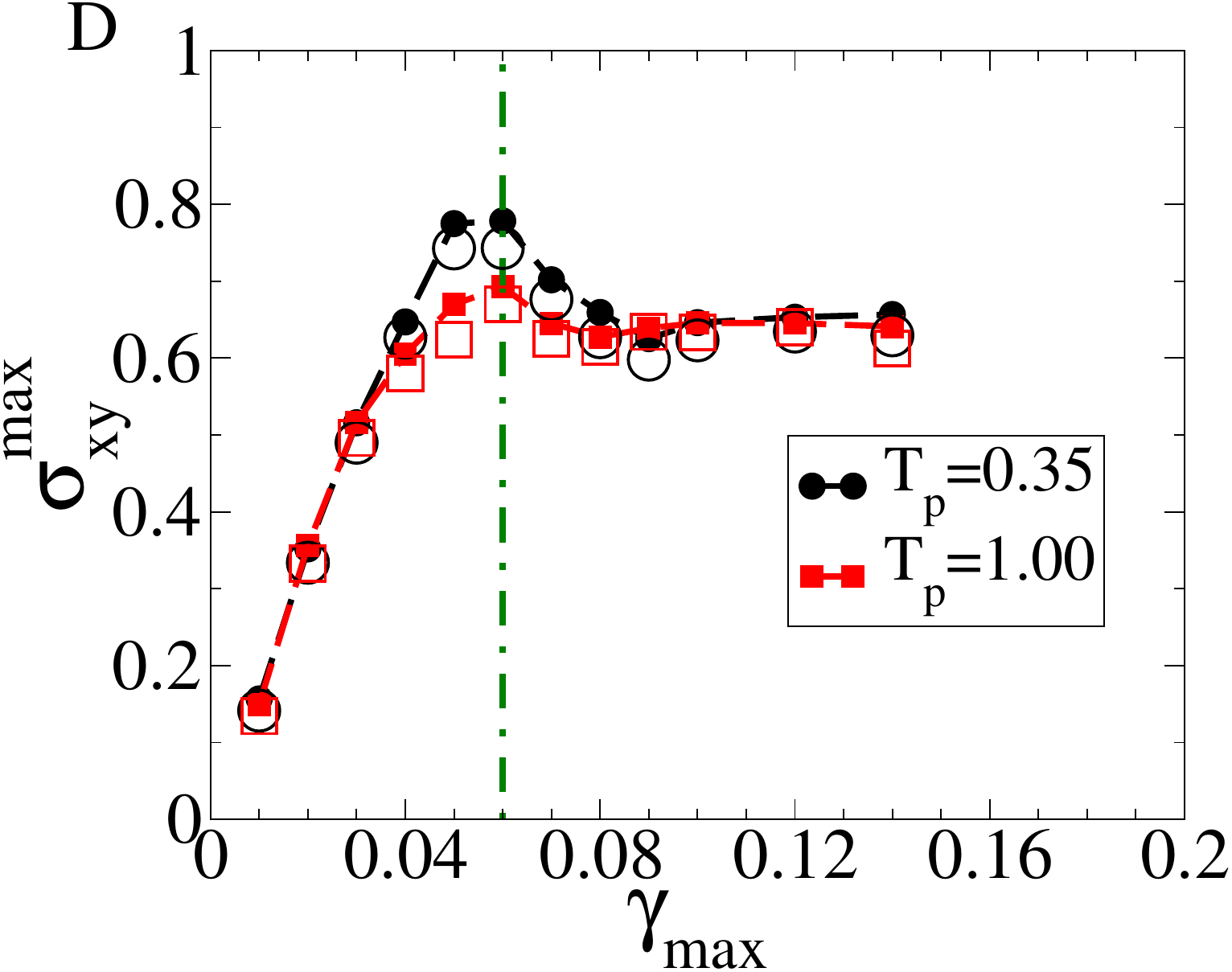}}

\caption{Yielding diagram under active driving. (A) Amorphous solids subject to active forces display annealing to lower energies for high energy initial configurations for small active forces, with the degree of annealing diminished or absent for lower energy initial states. The steady state energies are shown. A common, {\it ergodic}, fluidized state is observed above a critical value of active forces. These observations are analogous to those observed for yielding under cyclic shear deformation. (B) The yielding transition is accompanied by a saturation of the active stress, computed as explained in Materials and Methods. Corresponding results for (C) energy at zero strain, and (D) stress at strain amplitude $\gamma_{max}$, for cyclic shear are shown for comparison. Also shown in (D) is the comparison of stresses obtained from the procedure outlined in Materials and Methods (open symbols) and the conventional virial stress (filled symbols). 
}
\label{fig:fig1}
\end{figure}

In cyclic shear simulations, it is found that for high initial energies (or, poorly annealed amorphous solids)  the final (non-diffusive or absorbing) steady states exhibit progressively larger degrees of mechanically induced annealing (decrease in the internal energy) with increasing amplitude of strain (the control parameter, like the strength of active force $f$ in the present case) till the yielding or fluidization transition, beyond which there is a drop in stress and a discontinuous onset of diffusive behavior. Well annealed (low initial energy) amorphous solids exhibit little annealing before, and strongly discontinuous change at, the yield point. We thus consider the corresponding behaviour of the steady state energies and stresses (Fig.~\ref{fig:fig1}) and diffusion (Fig.~\ref{fig:fig2}) for the active glasses we now investigate. Fig.~\ref{fig:fig1} (A) shows, as a function of the active force strength $f$, the final (steady state) energies reached, for a number of initial energies, for $\tau_p = 2.31\times10^2$.  The resulting curves bear striking similarities with the cyclic shear yielding diagram \cite{BhaumikPNAS21} wherein, as a function of the strain amplitude (i) poorly annealed (high energy) glasses show progressive mechanical annealing before yielding, and (ii) well annealed glasses show strong discontinuous yielding. In Fig. S2 in the {\it SI Appendix}, we show corresponding steady state energies at two other, higher, values of $\tau_p$.

The stress can be computed from the first derivative of the energy with respect to strain, and corresponding procedures for computing the active stress (discussed, {\it e.g} in  \cite{MorsePNAS21,During2021}) requires identifying a strain step  $\Delta \gamma$ associated with displacements in the presence of active forces.  Following \cite{During2021}, we first consider that the alignment of particle velocities with the direction of active force produces a net strain rate (see Eq.~\ref{eq:active_strain_rate} in {\it Materials and Methods} and {\it SI Appendix}), which can be used to obtain the strain step $\Delta \gamma$ over a time interval (see Materials and Methods).
We analyse the parametric dependence of $\Delta E$ on $\Delta \gamma$ to compute the active stress. The resultant average stress, $\langle \sigma_{act}\rangle $, defined in Eq.~\ref{eq:active_stress} is found to increase linearly with $f$ until the yield point where it attains a maximal value, as shown in Fig.~\ref{fig:fig1} (B). The net velocity alignment, from which we determine an effective strain rate, remains negligible in magnitude below the yield point, with a subsequent increase for active forces larger than the yield value (see {\it SI Appendix}, Fig. S3) as also noted in similar contexts~\cite{xu2021yielding,hopkins2022local}. 

For comparison, we show the corresponding data for cyclic shear, in Fig.~\ref{fig:fig1} (C) and (D), which indeed confirms the similarity of the results in these two different driving protocols. Similar results have also been reported recently for active `run and tumble' driving \cite{Sharma2023} at considerably smaller persistence times, in the {\it thermal} limit, as discussed, {\it e. g.}, in \cite{mandal2020extreme} and below. In Fig.~\ref{fig:fig1} (D), we compare the shear stress at maximal strain, $\gamma_{max}$, in the cyclically sheared system, obtained from the virial expression with 
that obtained from the procedure outlined above (see also {\it SI Appendix}) and employed for the active case (Fig.~\ref{fig:fig1} (B)). The close agreement validates the latter procedure, although improvements to the procedure merit further investigation. Comparing results in Fig.~\ref{fig:fig1} (B) and Fig.~\ref{fig:fig1} (D) we note that while the computed stresses clearly mark the fluidization transition for the active case, the stress overshoot, the presence of a constant flow stress, as well as the difference between the peak stresses of the well-annealed samples and the poorly annealed samples, is less pronounced. 

\subsection{Fluidisation transition under active forcing and relaxation to the steady state}

\begin{figure}[htp!]
\centering
\subfloat{\includegraphics[height=3.5cm,width=4cm]{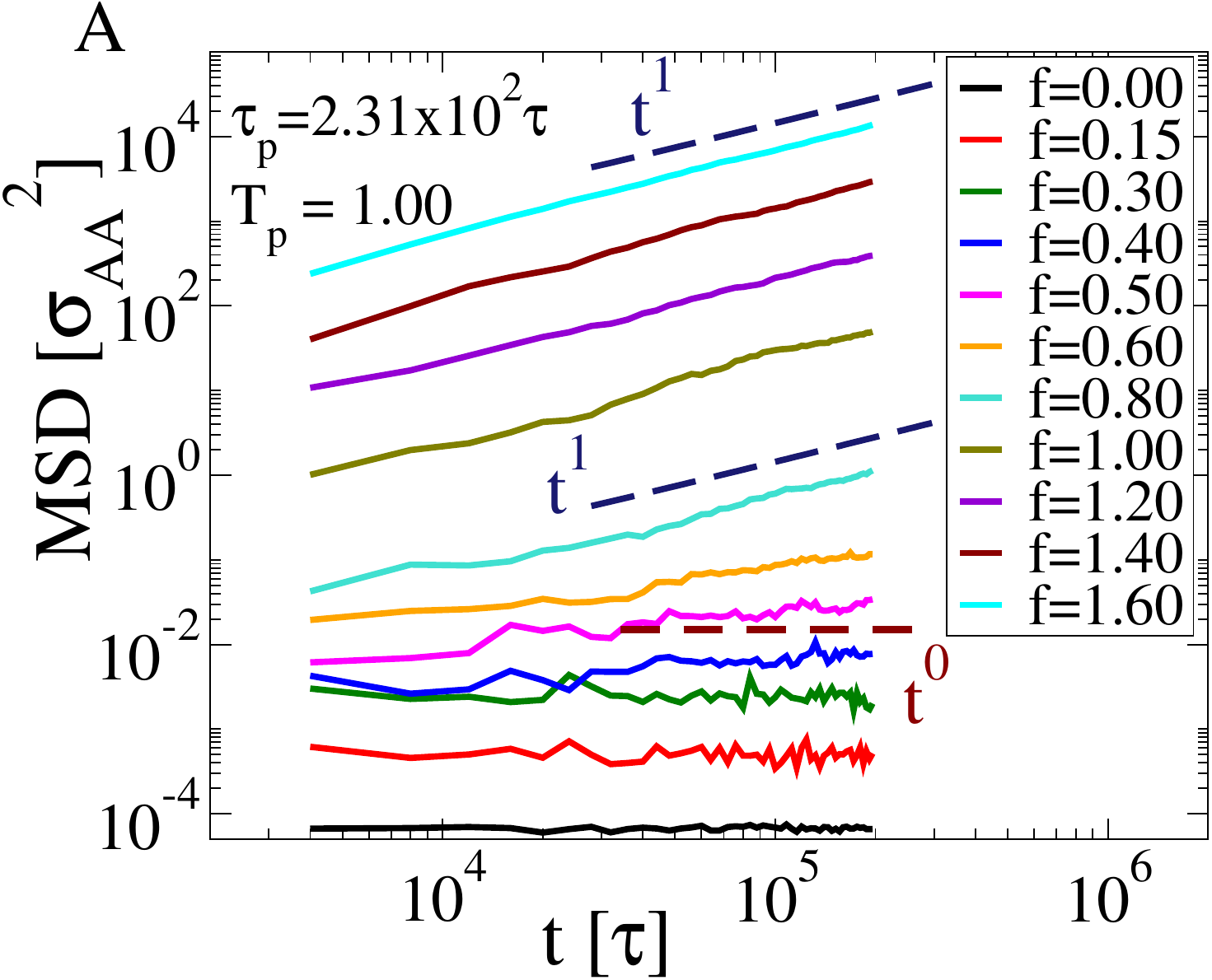}}
\subfloat{\includegraphics[height=3.5cm,width=4cm]{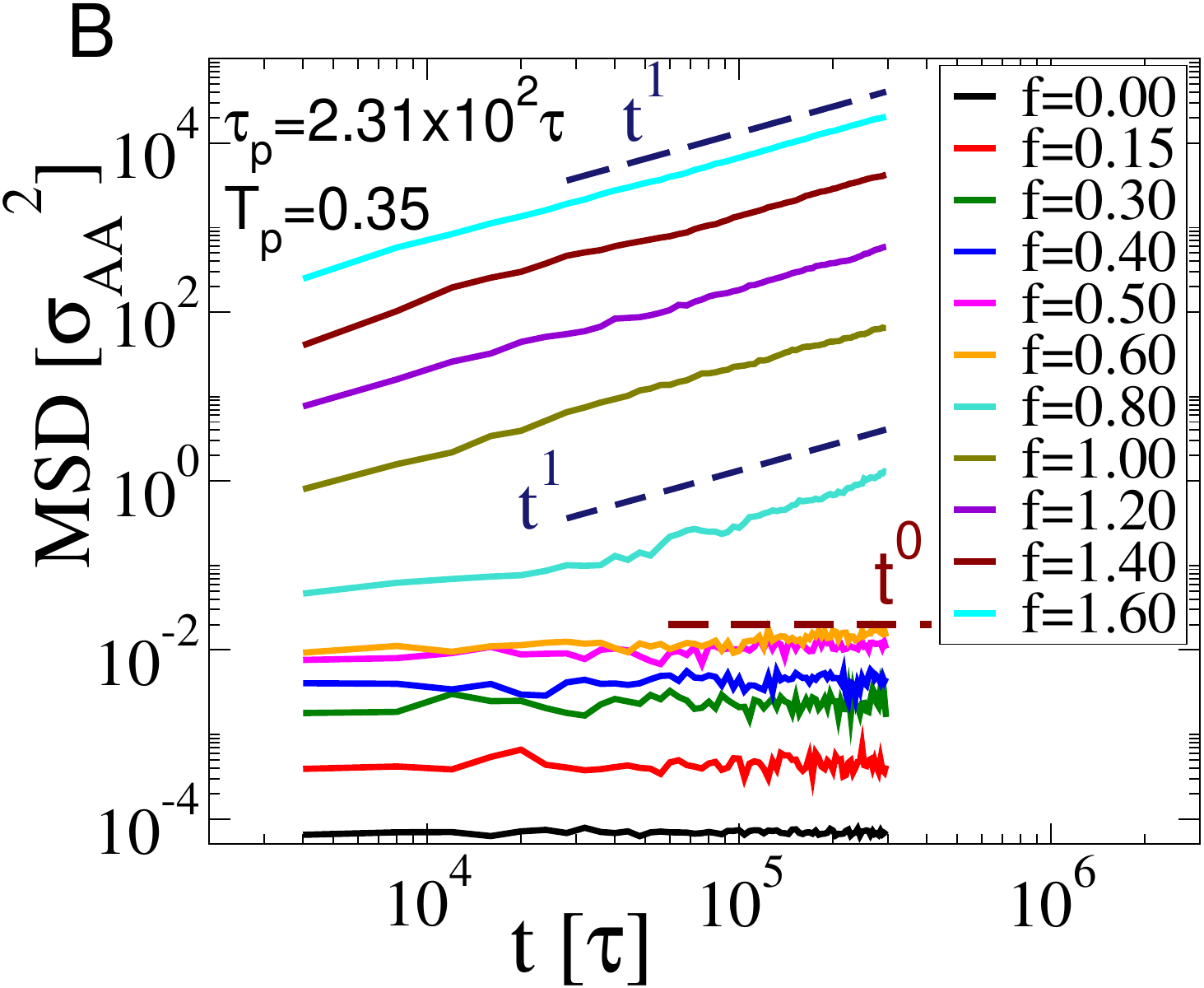}}

\subfloat{\includegraphics[height=3.5cm,width=4cm]{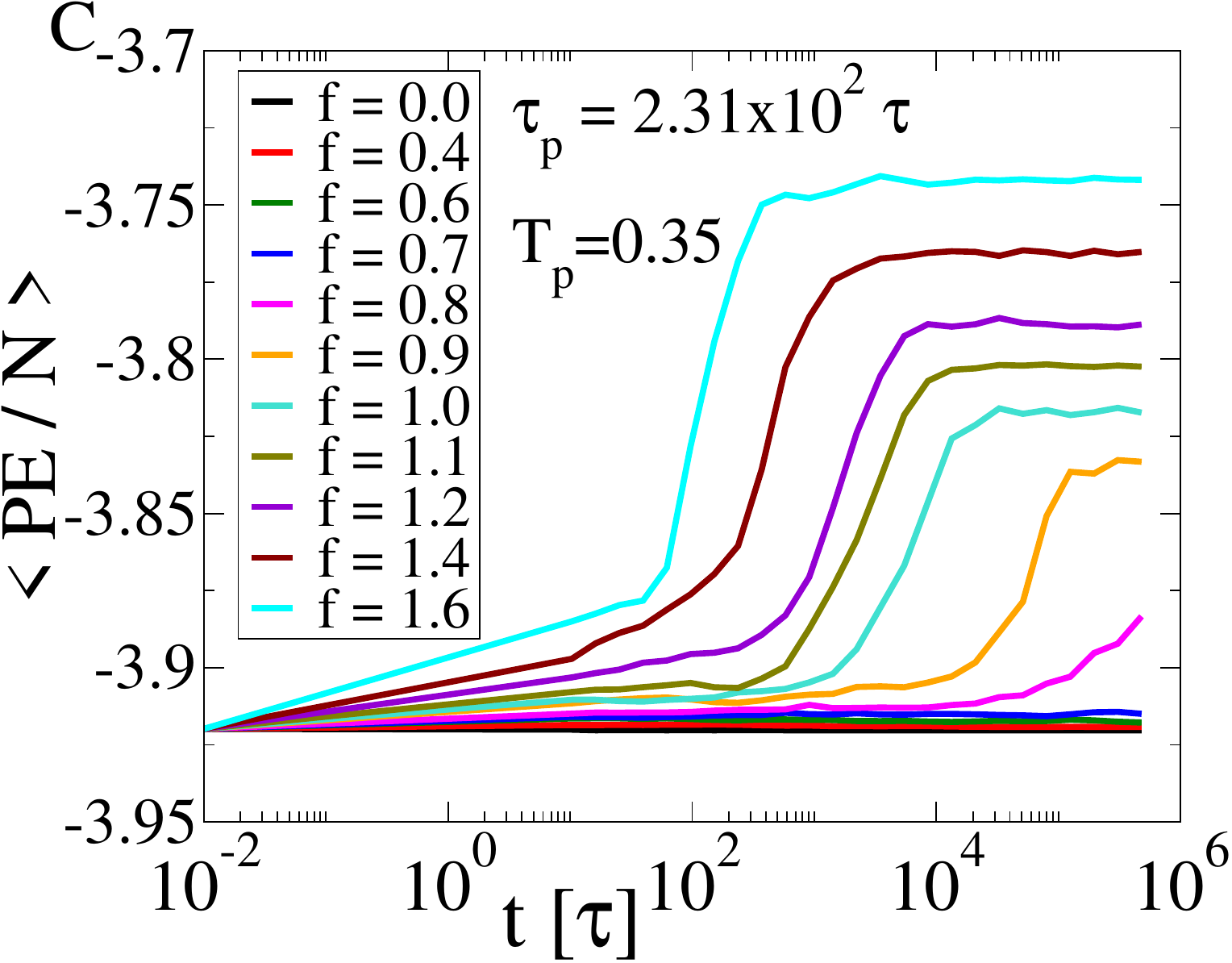}}
\subfloat{\includegraphics[height=3.5cm,width=4cm]{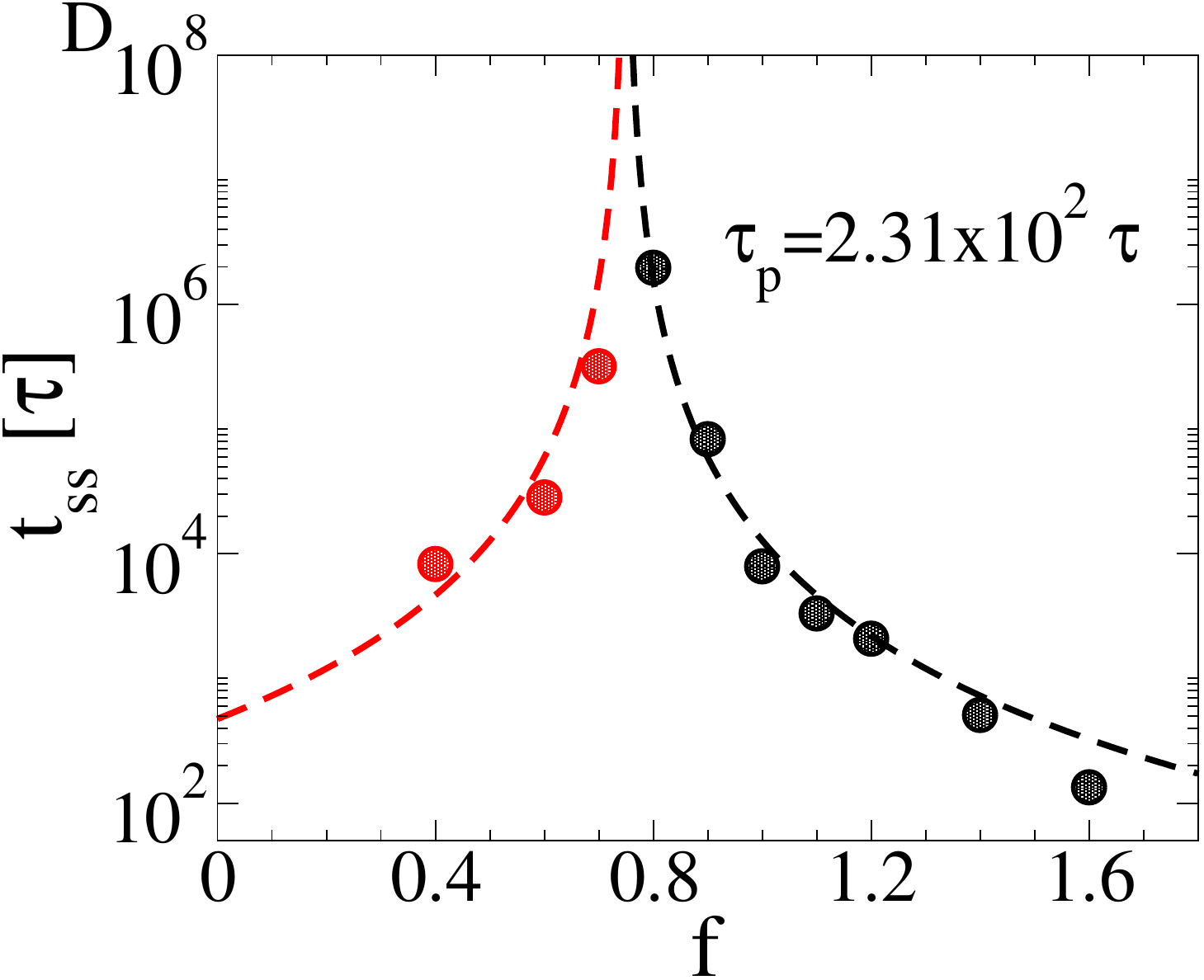}}
\caption{Diffusivity and time to reach steady state. (A) Mean squared displacement for different magnitudes of the active force showing a transition from an absorbing (non-diffusive) to a diffusive state for the poorly annealed case, and (B) the well-annealed case. Lines are shown with linear (navy blue) and no time dependence (dashed maroon) for reference.  
(C) Relaxation curves for energy {\it vs.} active force, obtained by averaging over $8$ independent trajectories in logarithmically spaced time intervals (D) Relaxation times exhibiting divergence at yielding transition. Dashed orange and turquoise lines are guides to the eye, indicating the diverging time taken to reach the steady state. }
\label{fig:fig2}
\end{figure}

We next consider the mean squared displacements (MSD) of particles as a function of elapsed time, in the steady state, for different values of $f$, as shown in Fig.~\ref{fig:fig2}, for two different initial conditions (Fig.~\ref{fig:fig2}(A) and 2(B)). It can be seen that the MSD curves exhibit a saturation below a critical value of $f$ (which can be deduced from data in Fig.~\ref{fig:fig1}) and a linear dependence on time for higher $f$ values. The corresponding diffusion coefficients jump discontinuously at the yield point, analogously with cyclic shear yielding (see {\it SI Appendix}, Fig. S4 for supporting data).

Another characteristic of the cyclic shear yielding transition is the time taken to reach the steady state for different strain amplitudes. To obtain the corresponding data, we consider the relaxation of the energy (shown in Fig.~\ref{fig:fig2} (C) for $T_p  = 0.35$) for different values of $f$.  
A stretched exponential fit to the average energy over time yields a relaxation timescale, $\langle t_{ss} \rangle$ (see {\it SI Appendix}, Fig. S5, S6 for the fits) which is shown in Fig.~\ref{fig:fig2} (D), indicating that the time to approach the steady state becomes progressively longer as the yielding transition is approached from either side, with an apparent divergence at the transition. Data for force magnitudes smaller than the critical force are shown for a poorly annealed sample, prepared at $T_p=1.00$, while those for larger forces are shown for $T_p=0.35$.
The estimated times to steady state obtained from the procedure described above compare well with corresponding estimates from the average first passage time to reach the steady state energy, as shown in the {\it SI Appendix} in Fig. S7.

The above results clearly demonstrate that yielding in a system with active forces with a finite persistence time has the characteristics of yielding under cyclic shear. Intuitively, a way to understand this is to note that in addition to having no macroscopically  well-defined direction of driving as in, say, simple shear deformation, with finite persistence time, the direction of driving also changes over time, thus bearing a broad similarity to the `back and forth' driving under cyclic shear. 

In the {\it thermal} limit of low persistence times $\tau_p$, the active driving we consider is equivalent to a higher effective temperature, with the contribution from active driving increasing quadratically with $f$ (see {\it SI Appendix}, Fig. S8-S9 for comparison with kinetic energies in the low persistence limit and passive dynamics). Extensions of such a relationship to finite and large persistence times have also been discussed \cite{nandi2018random,mandal2020extreme,mandal2022random}. The persistence times we consider are larger than those in the {\it thermal} regime \cite{mandal2020extreme}. Indeed, they span a range that includes {\it intermediate} (up to $\sim 10^3$ and  {\it large} persistence times ($10^4  - 10^7$) for which one observes clear deviations from the expectations in \cite{nandi2018random} and the yielding active force value (see below) increases with active force $f$, rather than saturate to a constant value. 

We next focus on the dependence of this yielding transition on parameters that may be of relevance not only in the biological context of our interest, but more generically. We thus consider the effect of the change in persistence times, and the nature of the confinement, on the yielding transition. 

%SS till here

\subsection{Role of persistence time in the yield point}
\begin{figure}[htb!]
    \centering
    \subfloat{\includegraphics[height=3.5cm,width=4cm]{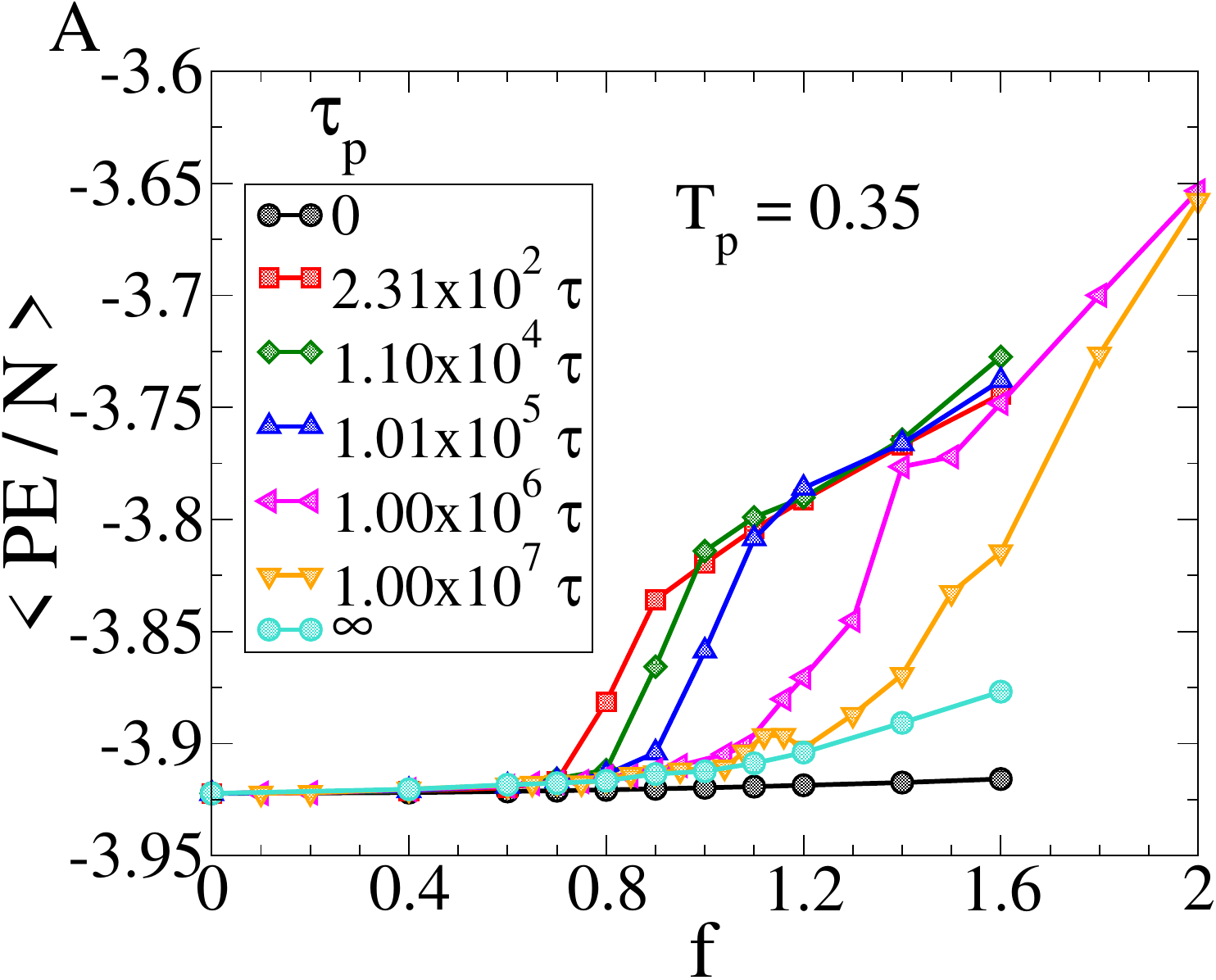}}
    \subfloat{\includegraphics[height=3.5cm,width=4cm]{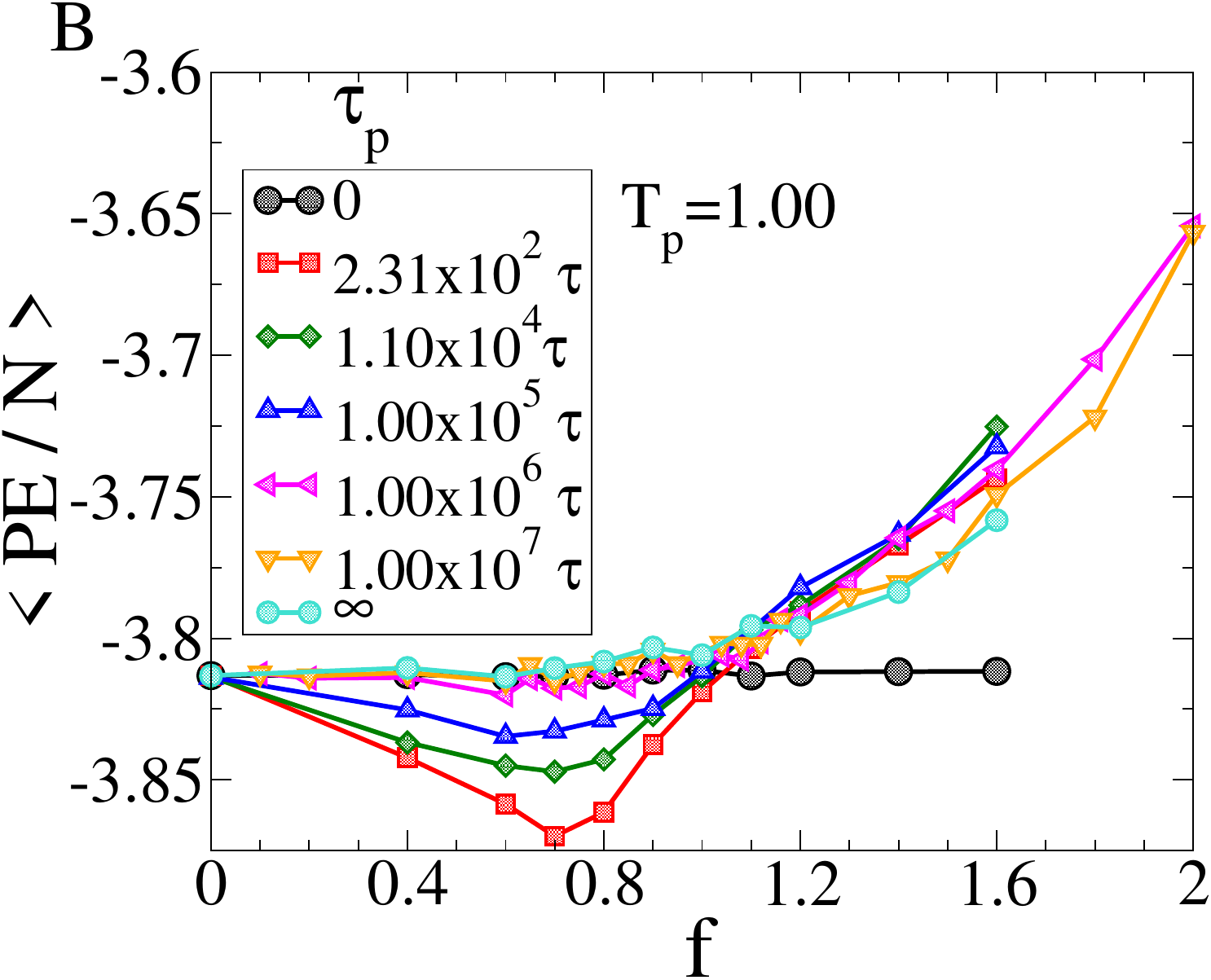}}    
    
    \subfloat{\includegraphics[height=3.45cm,width=4.2cm]{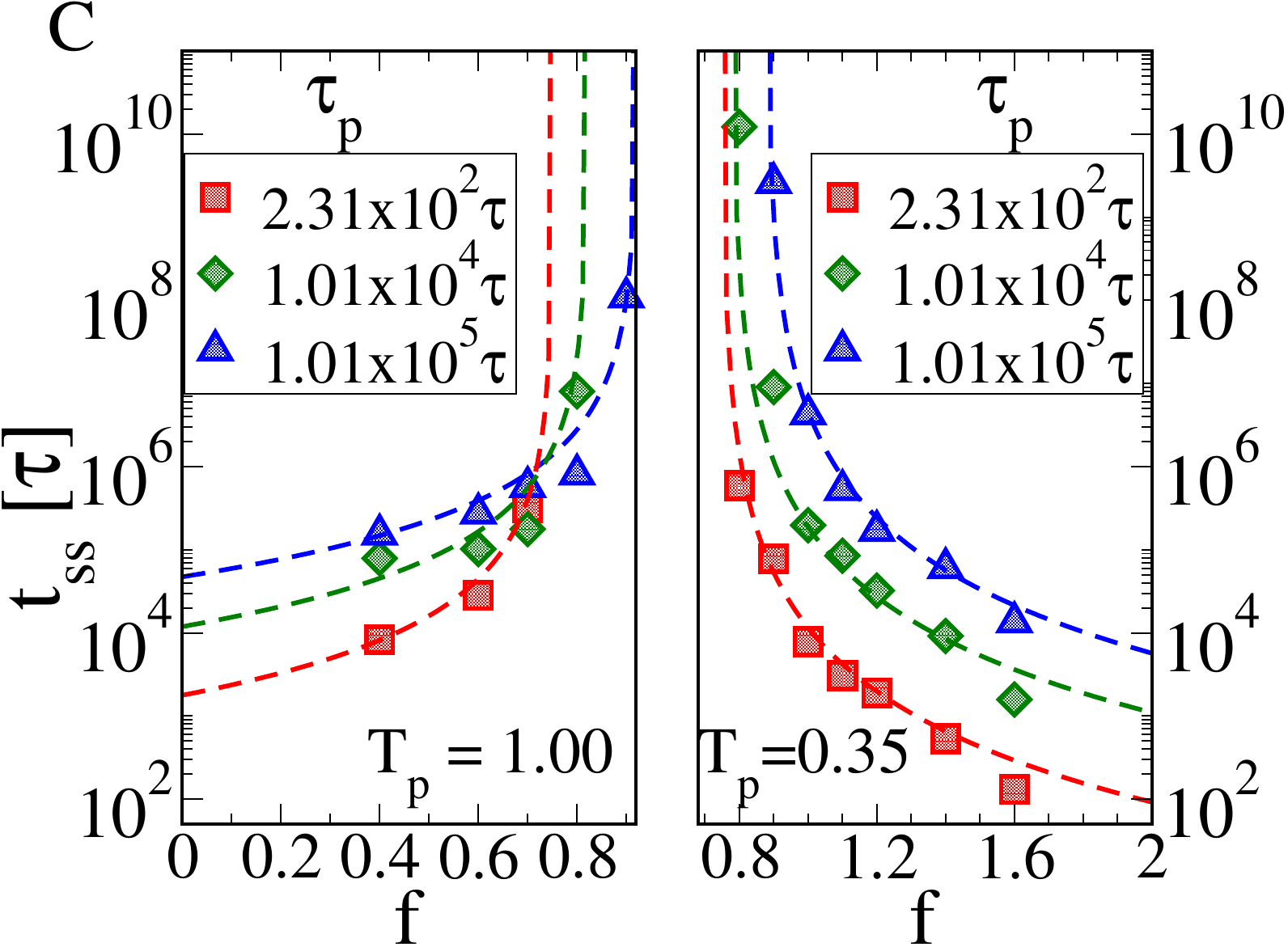}}      
    \subfloat{\includegraphics[height=3.5cm,width=3.8cm]{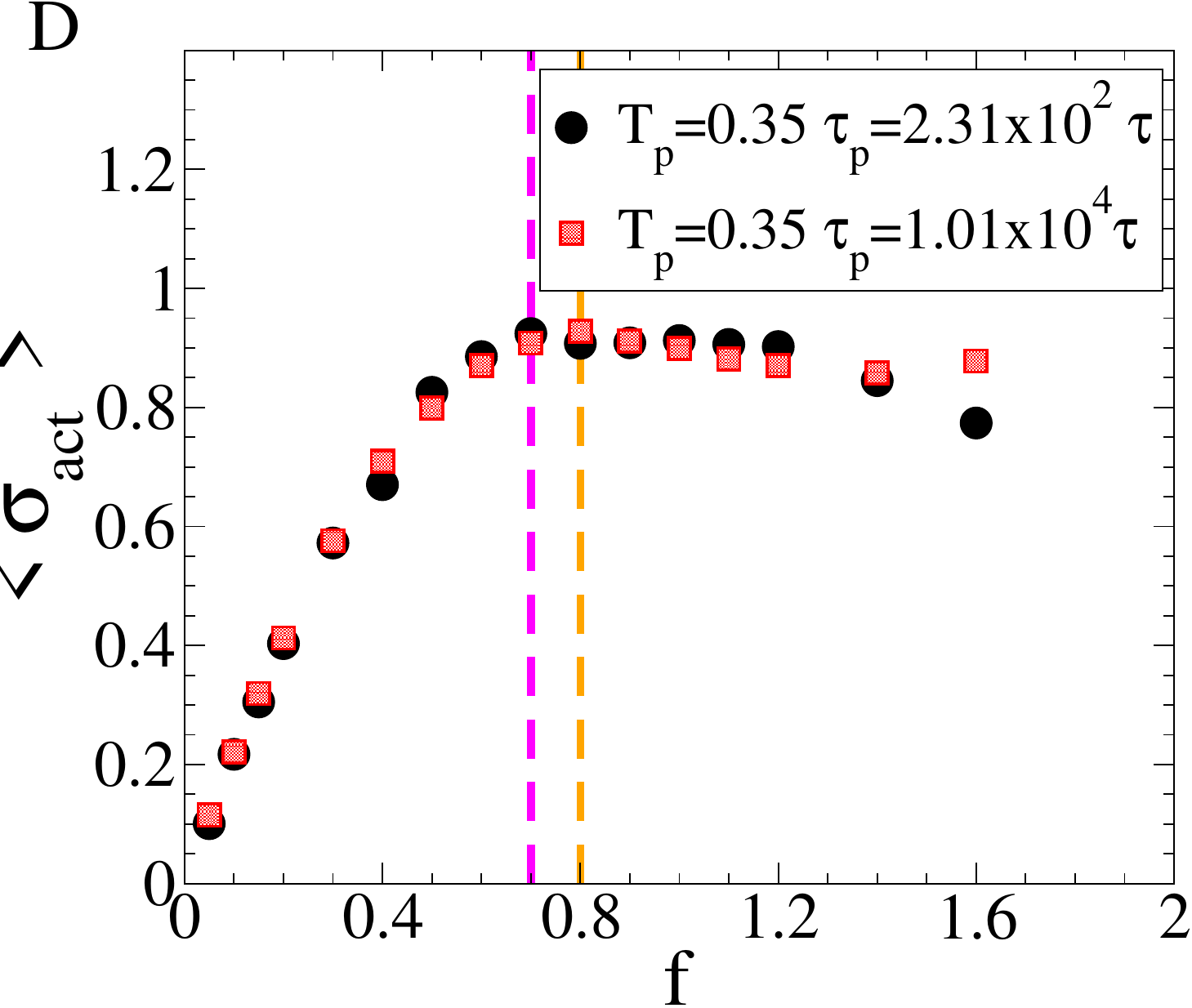}}
    
    \subfloat{\includegraphics[trim=0 10 0 00,clip,height=3cm,width=5.8cm]{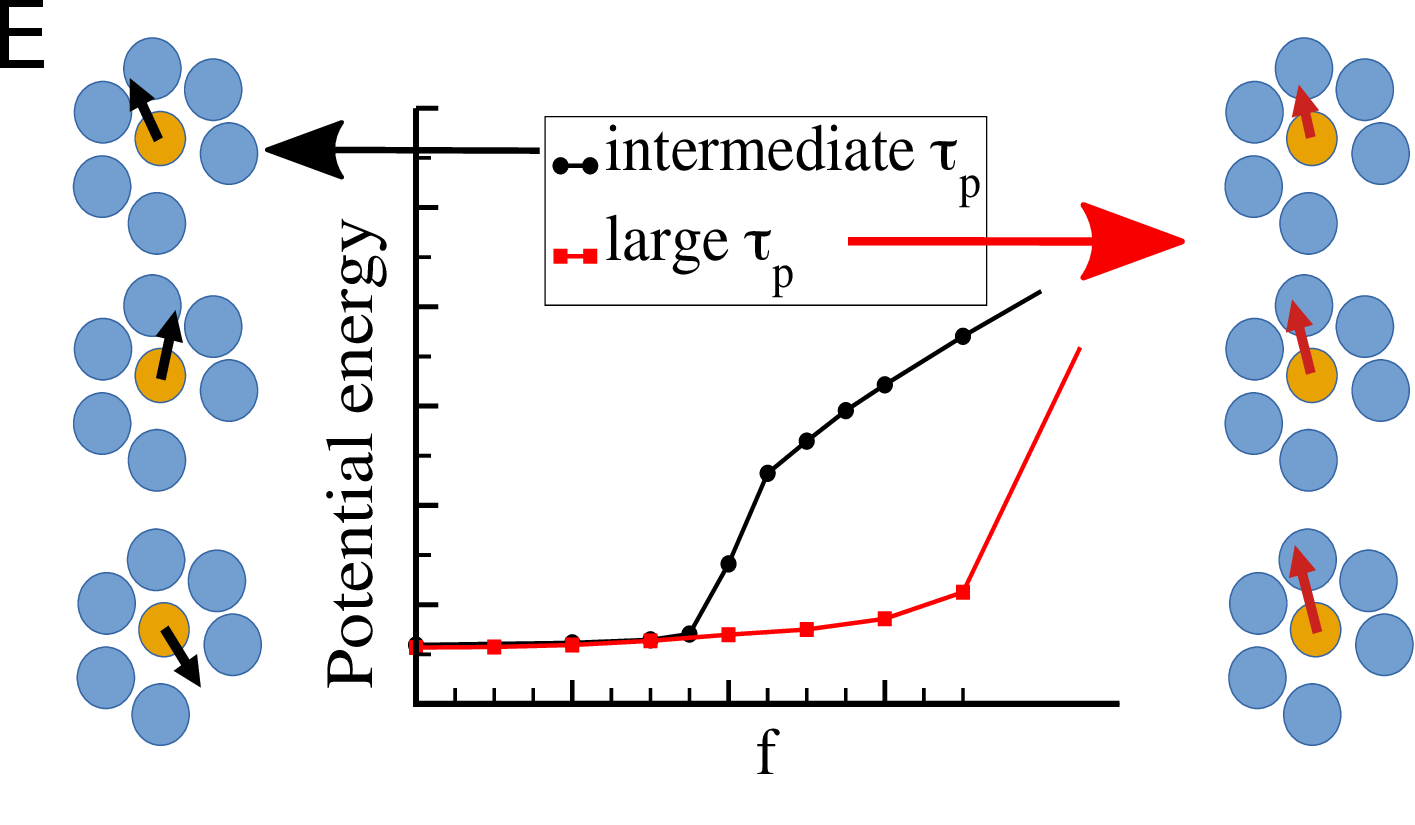}}
    \caption{Dependence of yielding behaviour on persistence time. The yielding transition shifts to larger active force magnitudes $f$ as the persistence time increases, for (A) a well annealed and (B) poorly annealed initial configurations. The mechanical annealing in the latter case diminishes with an increase in the persistence time. (C) The divergence of the time to reach the steady state at different persistence times. The dashed lines are guides to the eye. (D) Stress as a function of active force, with a shift in the value of force ($f$) where deviation from linearity first occurs. The value of $f$ corresponding to the peak stress is indicated by a vertical magenta line for $\tau_p=2.31\times 10^2~\tau$ and a vertical orange for $\tau_p=1.01\times10^4\tau$.    (E) schematic description of the role of persistence time. Small persistence times facilitate rapid exploration of routes to escape the cages constituted by their nearest neighbours, leading to yielding at small active driving magnitudes. As the persistence time increases, this capacity for exploration decreases and particles instead ``break through" their cages, which requires larger magnitude active forces.
    }
    \label{fig:fig3}
\end{figure}

A key control factor of relevance in various biological contexts is the persistence time, which determines the time scale over which the active force reorients. Changes to the persistence time have strong implications for the efficacy of exploration of phase space, producing different dynamical regimes\cite{hopkins2022local,mandal2020extreme,szamel2015glassy,keta2023intermittent,mandal2021study,mandal2022random}, motivating us to study changes in yielding behaviour arising from changes in the persistence time. We study the effect of changing the persistence time on mechanical annealing below, and on the location of, the yield point, for both well annealed (Fig.~\ref{fig:fig3} (A)) and poorly annealed samples (Fig.~\ref{fig:fig3} (B)). In both cases, the yield point shifts to larger values of active force $f$ as the persistence time increases. For the poorly annealed case, the mechanical annealing effects diminish as the persistence time increases.
In both cases, the apparently continuous increase of energies in the yielded branch have to do with the difficulty in reaching the steady-state close to the transition, as we explain further in the {\it SI Appendix}. 
Thus, with an increase in the persistence time, the particles are more and more readily `jammed' in kinetically arrested higher energy states, reminiscent of phenomenology explored in \cite{mandal2020extreme,mandal2021study,mandal2022random}. Unlike the small persistence time limit, yielding occurs at higher values of the active force with increasing persistence time, consistently with data and discussion in \cite{mandal2020extreme,mandal2022random}. 
We also show the results for the limit of zero persistence time ($\tau_p = dt$, mentioned above), which reveal negligible change in the range of $f$ values shown (which correspond to very small increments in the bath temperature, as elaborated in the {\it SI Appendix}, Fig. S8).

In Fig.~\ref{fig:fig3} (C), we show the change in $t_{ss}$ as the persistence time is increased, where $t_{ss}$ below the yield point are obtained for poorly annealed initial samples, and for well annealed samples above the yield point, as before. 
The point of divergence shifts to larger values of $f$ with larger $\tau_p$, consistently with results in Fig.~\ref{fig:fig3} (A) and (B).
Fig.~\ref{fig:fig3} (D) shows the active stress, $\sigma_{act}$, which increases up to the yield point (as seen earlier (Fig.~\ref{fig:fig1} (B)), which are slightly different for the two cases shown. 
The schematic in Fig.~\ref{fig:fig3} (E) illustrates how the efficiency of exploration of phase space, and in turn the activity induced annealing and the yield point, may be altered by the persistence time. Such a picture, while intuitive for the range of $\tau_p$ that we have explored, is not expected to hold for small $\tau_p$ values, where the system begins to resemble a thermal system as the $\tau_p \rightarrow 0$ limit is approached. Preliminary data also indicate that the yield values increase non-monotically upon  lowering $\tau_p$, consistently with results in \cite{mandal2020extreme}.

% \FloatBarrier

\subsection{Simulations in confinement}
\vspace*{-2mm}

We next consider the influence on the yielding behaviour of confinement geometry, a factor that may be relevant in several, including biological, contexts. For example, experiments studying the dynamics and spatial organisation of chromatin within the nucleus have pointed to the significant role of confinement geometry, with possible implications on cell-state transitions\cite{RoyPNAS2017,RoyPNAS2020}. Active particles in confinement have previously been noted to display rich behaviour such as wall-aggregation and the formation of cavities\cite{yang2014aggregation,fily2014dynamics}. We consider here a case of high density and investigate yielding behaviour as the strength of active driving is increased.
We compare the yielding behaviour for two confinement geometries: (i) a nearly circular confinement with aspect ratio (ratio of major axis to minor axis) $e=1.2$, and (ii) a strongly elliptical confinement with aspect ratio $e=1.8$. Particles experience attractive interactions of the Lennard-Jones form with the wall, with parameters as given in {\it Materials and Methods}. 
Initial configurations with different annealing are prepared by a procedure analogous to the bulk case as described in {\it Materials and Methods} and the {\it SI Appendix, Fig. S12}. These configurations are then subjected to active forces, at a persistence time of $\tau_p=200\tau$ and the potential energy is tracked to identify a transition from an absorbing to a diffusive state (see {\it SI Appendix}, Fig. S13).

\begin{figure}[htb!]
    \centering
    \subfloat{\includegraphics[trim = -10 0 -40 0, clip, width=3.8cm,height=3.6cm]{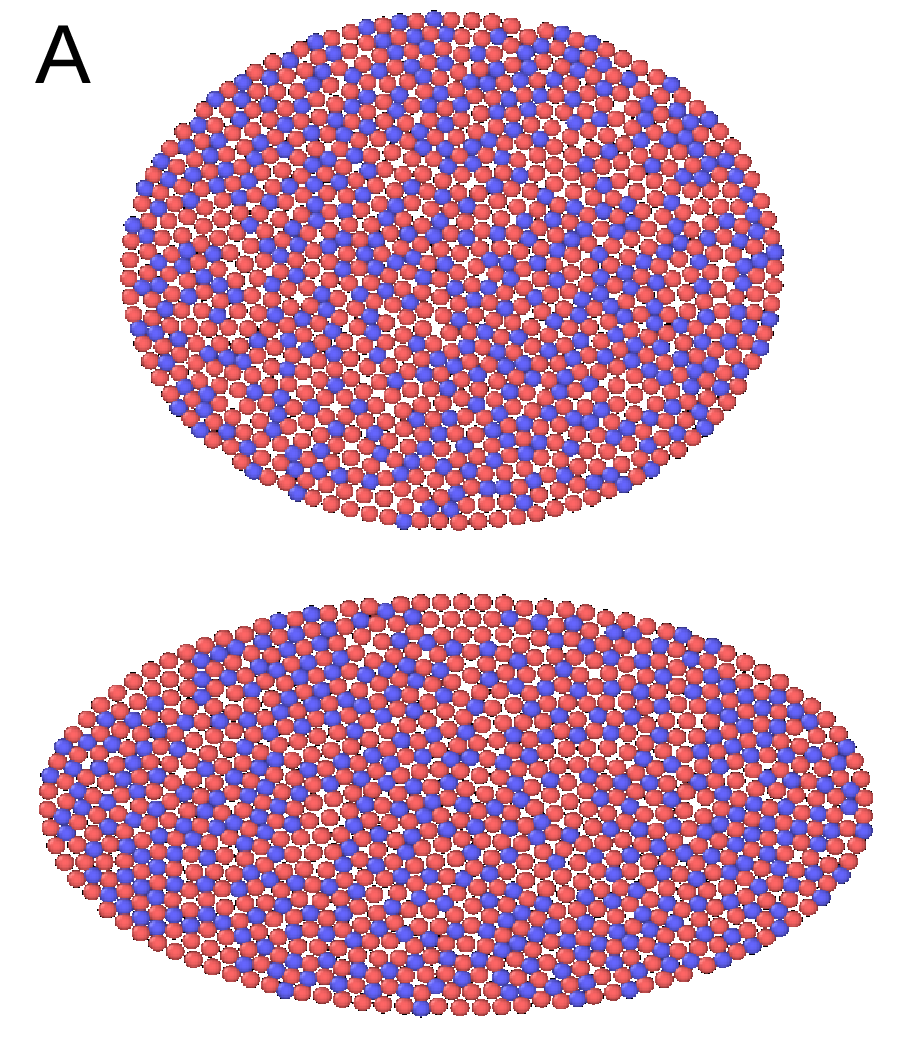}}
    \subfloat{\includegraphics[width=4cm,height=3.6cm]{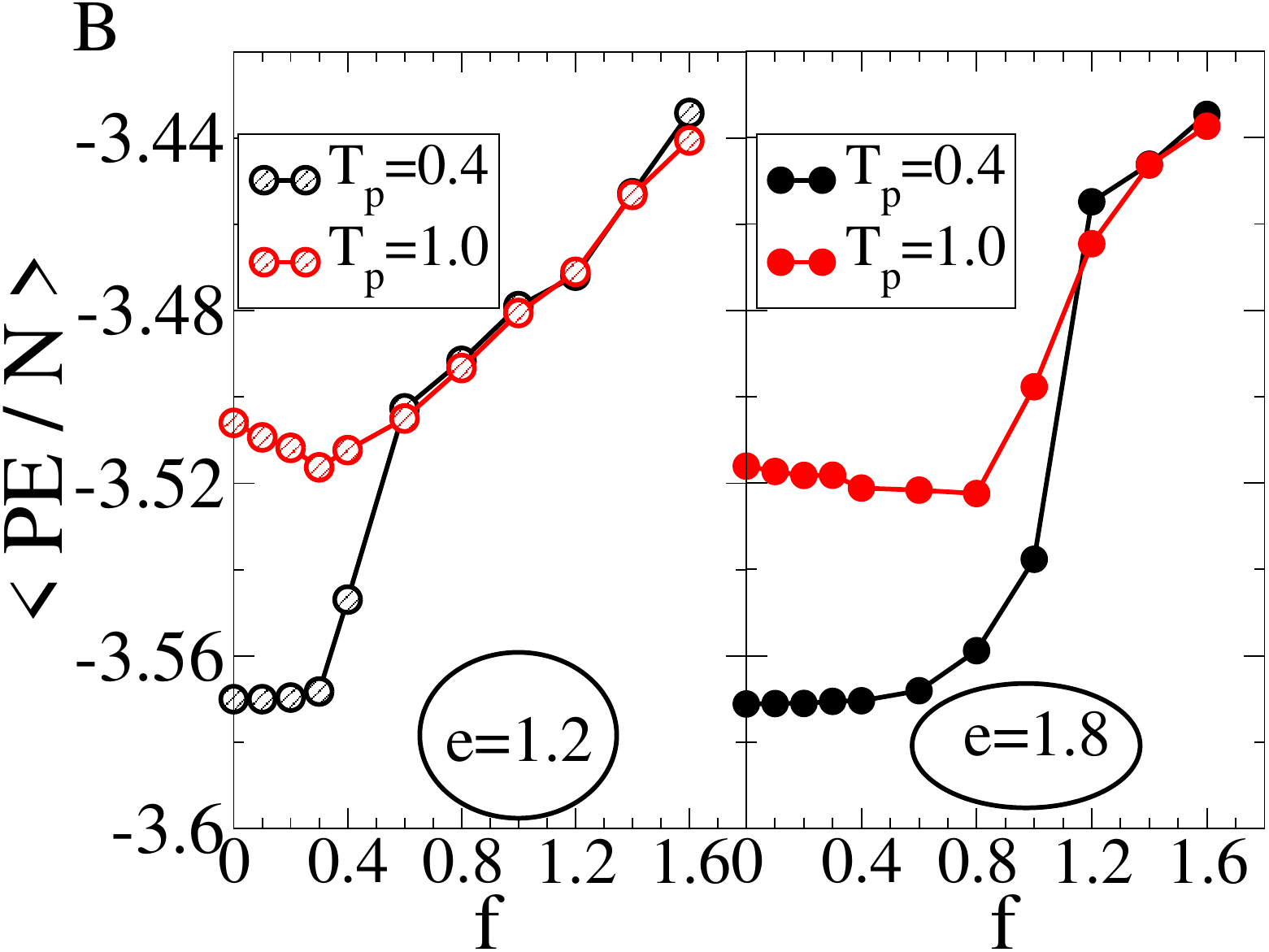}}

    \subfloat{\includegraphics[width=4cm,height=3.6cm]{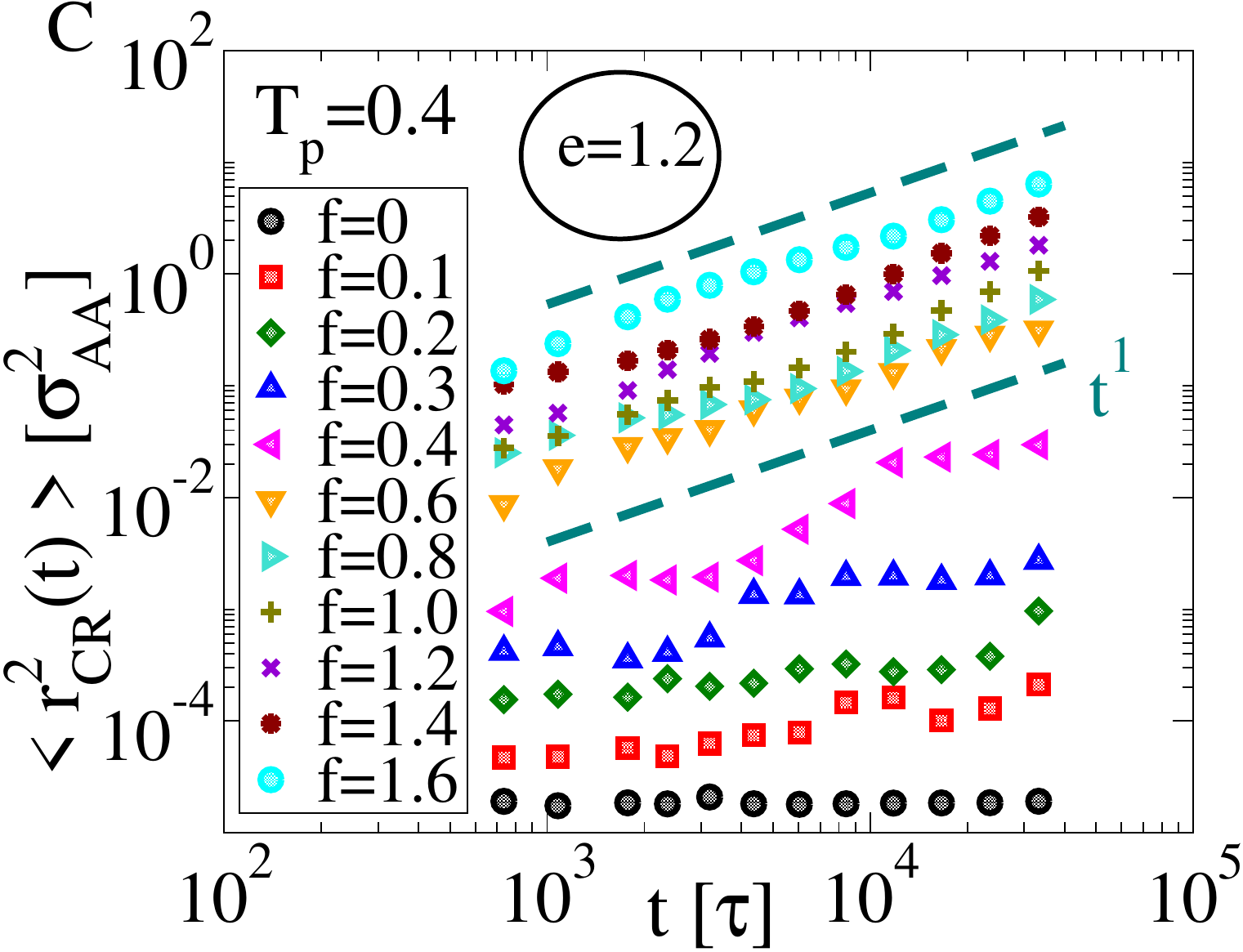}}
    \subfloat{\includegraphics[width=4cm,height=3.6cm]{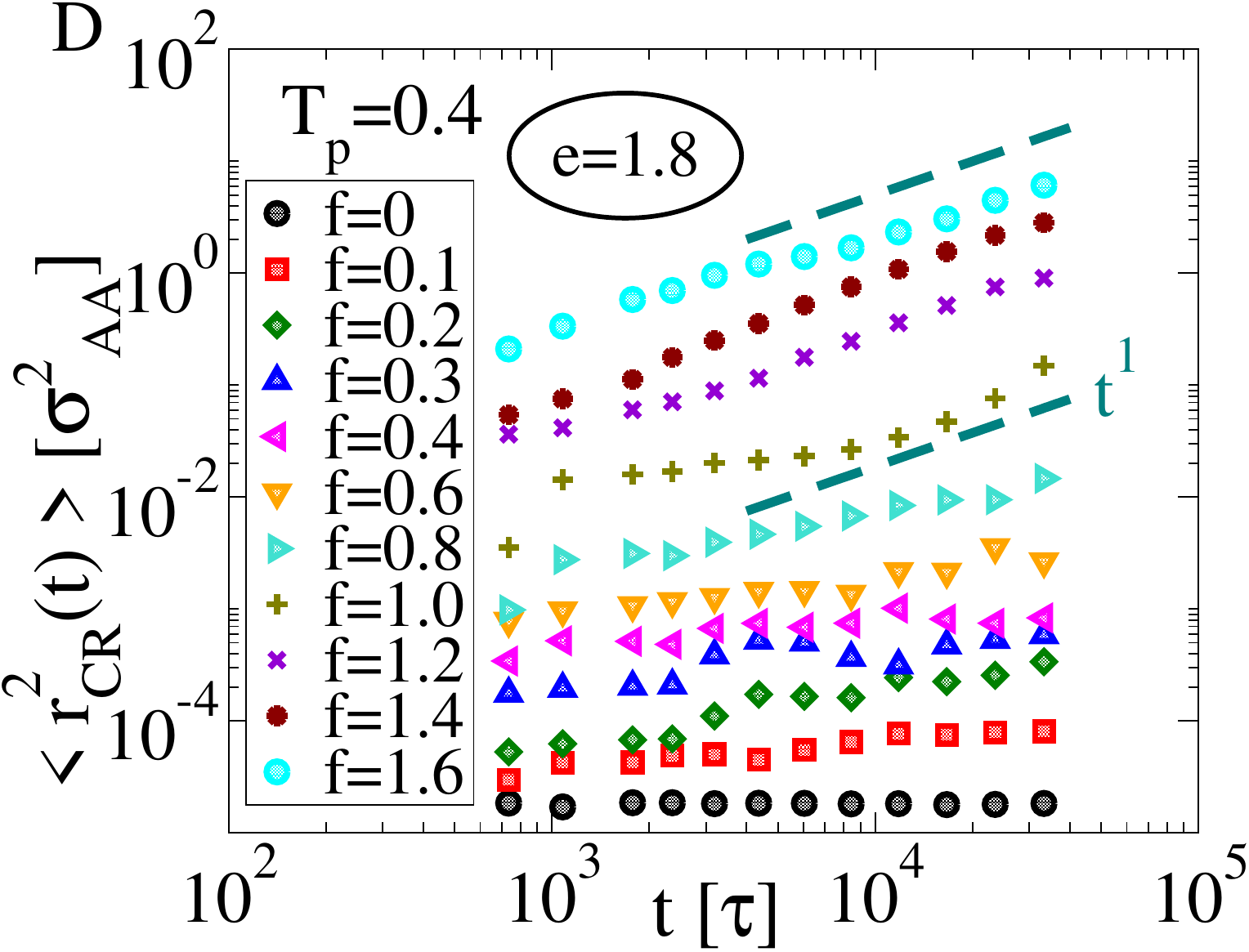}}
    \caption{Dependence of yielding behaviour on confinement. Simulations are performed in the two confinement geometries shown in (A), with eccentricities of $e=1.2$ (more isotropic) and $e=1.8$, more anisotropic, respectively. Red (blue) discs represent A(B) type particles. (B) The yielding diagram with the steady state potential energy per particle at different values of the active force in the two geometries, comparing the behaviour for a well-annealed case ($T_p=0.4$) with that of a poorly annealed case ($T_p=1.0$). The cage-relative mean squared displacements, $r_{CR}^2(t)$, are shown for (C) ($e=1.2$ and (D) $e=1.8$, for $T_p = 0.4$, the well annealed case. Note that in the absorbing regime of low force, $r_{CR}^2(t)$, remains below $0.01$, indicating an average displacement of $\approx~10 \%\sigma_{AA}$ for each particle with respect to its caging neighbours. The dashed teal lines in panels (C) and (D) indicate a lineaar dependence characteristic of the diffusive regime.}
    \label{fig:fig4}
\end{figure}

The geometries considered, and the resulting yielding behaviour is shown in Fig.~\ref{fig:fig4}. The results indicate that in going from a more circular to a more elliptical geometry, the yielding point is pushed to larger values of the active force. In both geometries, the post-yield states are ergodic, as observed from the dependence of the potential energy per particle on $f$ in Fig.~\ref{fig:fig4} (B). It has been observed that confined active systems exhibit global rotations, which should be subtracted in computing diffusive displacements of particles. This is accomplished, following  \cite{shiba2016unveiling}, by computing the cage-relative mean-squared displacements, $\langle |r_{CR}|^2(t) \rangle$(see {\it Materials and Methods}), which are shown in Fig.~\ref{fig:fig4} (C) for the more circular case, and in Fig.~\ref{fig:fig4} (D) for the more elliptical case starting from well-annealed samples. The superposition of global rotations and local rearrangements results in a less reliable measure of diffusive motion despite the correction employed. Nevertheless, one observes that the system remains in a non-diffusive absorbing state in the pre-yield regime,  whereas diffusive motion sets in beyond the yielding $f$ value in each of the two geometries, $f_{yield}\approx 0.3$ for $e=1.2$ and $f_{yield}\approx 0.6$ for $e=1.8$. Such a large change in the yield value of the active force is not readily explained by the perimeter length, of about 6\%, and point to a sensitive dependence of the fluidization transition on the confinement geometry. Such dependence, including the role of wall interactions, needs to be explored in greater detail.

These results show that at a given magnitude of the active forces, a change to a more symmetric geometry can induce yielding and fluidization. They highlight the role of confinement geometry in state transitions in assemblies subject to active forces, including the biological assemblies mentioned in the introduction, as discussed briefly below.  

% \vspace*{-2mm}
% \FloatBarrier
\section{Discussion}

Our results demonstrate the existence of a strong analogy between cyclically sheared amorphous assemblies and actively driven assemblies. Both modes of drive result in a yielding transition from an absorbing state to a fluidised state beyond a threshold. We find a strong correspondence in the yielding phase diagram, with regimes of annealing and yielding to an ``ergodic'' state, similar dependence on the initial state of the amorphous solids, the onset of a diffusive state for large perturbations and the divergence in the time to reach the steady state. 

\marktext{
Active driving with a finite persistence time, and in confinement, as implemented in this work, is motivated by our interest in biological systems that display a transition from rigid to flowing states when subjected to analogous mechanical conditions. Our results demonstrate that the persistence time is indeed a key parameter, with an increase in the persistence time leading to an increase in the critical perturbation strength (active force) at which the system fluidizes.
We also find that changing the confinement geometry leads to a change in the yielding behaviour with the transition to fluidised states occurring at much lower values of the active force for the more isotropic confinement geometry. That confinement geometry alone affects yielding behaviour in such a manner is, in our view, an interesting effect with numerous possible implications in various contexts. While the model system we investigate does not serve as a model for specific biological assemblies, we believe that the results are of interest to contexts such as the organization and dynamics of chromatin in the nucleus. In such contexts, the interplay and complex causal relationships of the nature of active driving, confinement and the fluidization state are important to elucidate in order to develop faithful theoretical models. The analysis presented here, we believe, forms an important first step in that direction. 
}

\section{Methods}

\subsection*{Model - 2D binary mixture} \label{subsec:potential}
We simulate a bidisperse Lennard-Jones mixture at constant $N, V$, with periodic boundary conditions (wraparound), at high density, $\rho=1.2$ in 2D. The interaction is truncated at a cut-off of $r_c^{\alpha \beta} = \sigma_{\alpha \beta}$, where $\alpha$ and $\beta$ denote the type of the interacting particles $i$ and $j$, either $A$ or $B$, and is given by:
\begin{align}
\phi_{LJ}(r_{ij}) &=
\begin{cases}
 4\epsilon_{\alpha \beta}\left [\left ({\sigma_{\alpha \beta}\over r_{ij}}\right )^{12} - \left ({\sigma_{\alpha\beta} \over r_{ij}} \right )^{6}\right ] \quad \text{if} \quad r_{ij} \leq r_c^{\alpha \beta} \nonumber \\
 = 0 \quad \text{otherwise}
 \end{cases}
\end{align}
We choose $N = 1000$ particles in a $65:35$ mixture with $N_A=650$. Reduced units are used throughout, in terms of $\sigma_{AA}$, $\epsilon_{AA}$ and the particle mass $M$, with the unit of time $\tau =  \sqrt{M \sigma_{AA}^2/\epsilon_{AA}}$, ${\sigma_{AB} \over \sigma_{AA}} = 0.8$, ${\sigma_{BB} \over \sigma_{AA}} = 0.88$, ${\epsilon_{AB} \over \epsilon{AA}} = 1.5$ and ${\epsilon_{BB} \over \epsilon{AA}} = 0.5$.

\subsection*{Initial sample preparation}
We perform constant temperature ($NVT$) molecular dynamics simulations using the Nos\'e-Hoover thermostat with a time step of $dt=0.01$ in the LAMMPS software suite\cite{plimpton1995fast} at different preparation temperatures, $T_p$, to generate equilibrated liquid configurations. Energy minimum configurations (inherent structures) are produced by performing conjugate gradient minimisation using LAMMPS. The inherent structure energies show a $T_{p}$ dependence as shown in the {\it SI Appendix}.

\subsection*{Shear simulations}
Cyclic shear simulations are performed at a finite shear rate and temperature, propagating the SLLOD equations of motion using the LAMMPS software suite\cite{plimpton1995fast}. Simulations are performed with a timestep of $dt=0.01$, with the strain being varied sinusoidally, for a $100$ cycles for each strain amplitude $\gamma_{max}$. The strain rate, computed at zero strain, is $\dot\gamma=0.001$. 

%%%%%%%%%%%%%%%%%%%%%%%%%%%%%%%%%%%%%%%%%%%%%%%%%%%%%%%%%%%%
\subsection*{Equations of motion for active dynamics}
Active dynamics are performed starting from the energy minimised configurations by integrating the following stochastic equations of motion.
\begin{align}
\mathbf{\dot p}_i &= -\gamma \mathbf{p}_i + \sum\limits_{j \ne i = 1}^{N} \mathbf{f_{ij}} + f \mathbf{\hat n}_i + \mathbf{\xi_t}^i \nonumber \\
\mathbf{\dot x}_i &= \mathbf{v}_i \equiv M^{-1} \mathbf{p}_i \nonumber \\
\mathbf{\dot \theta}_i &= \mathbf{\xi_r}^i
\end{align} 
The translational noise $\xi_t$ has zero mean and variance given by  $<~\xi_{t}^{i\alpha}(t) ~ \xi_{t}^{j\beta}(t^{'})~> = 2 k_B T M \gamma \delta_{ij} \delta_{\alpha \beta} \delta(t - t^{'}) $. The rotational diffusion of the direction of active forcing likewise has zero mean and variance given by $<\xi_{r}^{i}(t)  ~ \xi_{r}^{j}(t^{'})> = 2 \tau_{p}^{-1} \delta_{ij}  \delta(t - t^{'}) $.
%$\mathbf{\xi_t,\xi_r}$ $\left (\sqrt{2\gamma k_BT/\Delta t},\sqrt{2/\tau_p \Delta t}\right )$\\
The persistence time, $\tau_p$, determines the timescale of change in $\mathbf{\hat{n}_i}$ ($(\cos(\theta_i),\sin(\theta_i))$ in 2D), the direction of the active force with strength $f$. $\gamma$ is the dissipative friction term, and $\mathbf{f_{ij}} = -\nabla \phi(\mathbf{\{r_{ij}\}})$ are the interparticle forces. We integrate the equations of motion with a time step of $dt=0.01$ following a {\bf BAOAB} update scheme described in~\cite{leimkuhler2013rational,allen2017computer}. Details are included in the {\it SI Appendix}.
%%%%%%%%%%%%%%%%%%%%%%%%%%%%%%%%%%%%%%%%%%%%%%%%%%%%%%%%%%
\subsection*{Strain, stress and strain rate resulting from active forcing}
% We measure the alignment of particle velocities with the direction of active force to identify a strain rate, $\dot \gamma_{act}$.
The alignment of velocities with the active force direction produces a deformation field with which we identify a strain rate, $\gamma_{act}$\cite{Xu2018,During2021}.
% \begin{align}
%     v_{par}^{act} &= \frac{1}{N} \left  \langle \sum\limits_{i=1}^{N}\mathbf{v_i(t)}\cdot\mathbf{n_i(t)} \right \rangle_t \nonumber \\
%     \dot \gamma_{act}  &= \frac{\sqrt{12N}}{L} v_{par}^{act}
%     \label{eq:active_strain_rate}
% \end{align}
\begin{equation}
     v_{par} = \left  \langle  \sum\limits_{i=1}^{N}\mathbf{v_i(t)}\cdot\mathbf{n_i(t)} \right \rangle
\end{equation}
Following the approach in~\cite{MorsePNAS21}, we consider the case of cyclic shear at finite strain rate, where the affine component of the velocity of particle $i$ is of the form $\dot \gamma \mathbf{r}_i\cdot \hat y$, and the magnitude of the resultant field is given by $\dot \gamma [ \sum (y_i)^2 ]^{1/2}$. Discounting centre of mass motion and assuming a uniform distribution of particles along the $y$ direction one obtains a relationship of $v_{par}^{cs} = \dot \gamma L_y\sqrt{N/12}$. Retaining the pre-factor as an approximation for active dynamics, one can write the effective strain rate in active systems in terms of the deformation field as:
\begin{equation}
    \dot \gamma_{act}  = \frac{\sqrt{12}}{\sqrt{N}L} v_{par}^{act}
    \label{eq:active_strain_rate}
\end{equation}
One can then define an instantaneous strain step, $\Delta \gamma = \dot \gamma_{act}\Delta t$.
The derivative of the potential energy with respect to the strain, $dPE / d\gamma$, yields the stress\cite{MorsePNAS21} $\sigma_{act}$, which we estimate from the parametric dependence of $\Delta PE$ with respect to $\Delta \gamma$ over a given timestep, $t \rightarrow t +\Delta t$(see {\it SI Appendix}): 

\begin{equation}
\sigma_{act} = \frac{1}{L^2} \frac{dE}{d \gamma},
\label{eq:active_stress}
\end{equation}
while $L$ is the box length. We employ a similar procedure in cyclically sheared simulations in order to compare the stress measured using this procedure, with that obtained directly from the virial stress tensor at the maximal strain value, $\gamma_{max}$.

\subsection*{Boundary interaction in confinement}
We perform simulations of the same $65:35$ Lennard-Jones mixture of $N=1000$ particles in elliptical confinement of constant shape and volume. Each particle interacts with the boundary wall with a truncated Lennard-Jones interaction where the wall-specific parameters are given by $\epsilon_w = 0.5\epsilon_{\alpha \alpha}$, $\sigma_w = 0.5\sigma_{\alpha\alpha}$ and $r_{cw} = 3 \sigma_{AA}$ ($\alpha$ denotes the particle type, either $A$ or $B$). 

The instantaneous distance of each particle from the confinement boundary is obtained by implementing an iterative procedure is described in~\cite{Chatfield2017} (see {\it SI Appendix} for details). Heatmaps showing the distance from an ellipse boundary calculated using this scheme are shown in the {\it SI Appendix}.
Visualisations of the particle assembly in confinement are prepared using the OVITO software suite\cite{stukowski2009visualization}.

\subsection*{Initial sample preparation in confinement}
The initial configurations are prepared by first equilibrating a high temperature liquid at $T=2.5$. Uncorrelated configurations are then thermally annealed at different preparation temperatures $T_p$ to obtain samples in confinement (see {\it SI Appendix} for time evolution during thermal annealing). The annealed samples are then subjected to an instantaneous thermal quench to a very low temperature, $T=10^{-4}$, to approximate an energy minimisation procedure. 

\subsection*{The cage-relative mean squared displacement}
The cage-relative mean squared displacement\cite{shiba2016unveiling} is given by:
\begin{align}
    \Delta \mathbf{r}_{i,CR}(t) = \Delta \mathbf{r}_i - \frac{1}{N_{nn}} \sum\limits_{j=1}^{N_{nn}} \Delta \mathbf{r}_j(t) \nonumber \\
    \langle \| \Delta r_{CR}(t)\|^2 \rangle = \left \langle \frac{1}{N} \sum\limits_{i=1}^{N} \| \Delta \mathbf{r}_{i,CR}(t)\|^2 \right \rangle
    \label{eq:CR_MSD}
\end{align}
where $\Delta \mathbf{r}_i(t) = \mathbf{r}_i(t) - \mathbf{r}_i(0)$, $N_{nn}$ is the number of nearest neighbours within a shell of $1.5\sigma_{\alpha_i \alpha_j}$ from particle $i$ at time $t=0$. 
Further, in order to correct for motion due to global rotations without rearrangement,
we identify the optimally rotated configuration, $\{\mathbf{R}(\theta_{min})\mathbf{r_i}(t)\}$, where $\mathbf{R(\theta)}$ is a matrix that performs a rotation by $\theta$ in 2D. The angle, $\theta_{min}$, is determined by an optimisation procedure that rotates each particle along a scaled ellipse of identical aspect ratio as the confinement geometry (but with scaled axes).
\begin{equation}
    \theta_{min} = \text{argmin}_{\theta} \sqrt{\langle \left [\mathbf{r_i(t+\Delta t)} - \mathbf{R}(\theta)\mathbf{r_i(t)}\right ]^2\rangle}.
\end{equation}

The accumulated displacement vector between $\{\mathbf{r}(t+\Delta t)\}$ and the optimally rotated $\{\mathbf{R}(\theta_{min})\mathbf{r_i}(t)\}$, is tracked and it's magnitude gives the instantaneous cage relative mean squared displacement.
\begin{align}
\Delta \mathbf{r}_{i,CR}(t) &= \Delta \mathbf{r}_{i,CR}(t-\Delta t) + \Delta \mathbf{r}_i^{\theta_{min}}(t) - \frac{1}{N_{nn}} \sum\limits_{j=1}^{N_{nn}} \Delta \mathbf{r}_j(t)^{\theta_{min}} \nonumber \\
\Delta \mathbf{r}_i^{\theta_{min}}(t) &= \mathbf{r_i}(t) - \mathbf{R}(\theta_{min})\mathbf{r_i}(t-\Delta t) 
\end{align}
Note that while the initial centre of mass force and angular momentum is set to zero by appropriately adjusting the active force orientation vectors, the subsequent free diffusion of the orientation vectors does not prevent global motion.

% \bibliography{ActiveYieldingConfinementBib}

\begin{thebibliography}{44}%
\makeatletter
\providecommand \@ifxundefined [1]{%
 \@ifx{#1\undefined}
}%
\providecommand \@ifnum [1]{%
 \ifnum #1\expandafter \@firstoftwo
 \else \expandafter \@secondoftwo
 \fi
}%
\providecommand \@ifx [1]{%
 \ifx #1\expandafter \@firstoftwo
 \else \expandafter \@secondoftwo
 \fi
}%
\providecommand \natexlab [1]{#1}%
\providecommand \enquote  [1]{``#1''}%
\providecommand \bibnamefont  [1]{#1}%
\providecommand \bibfnamefont [1]{#1}%
\providecommand \citenamefont [1]{#1}%
\providecommand \href@noop [0]{\@secondoftwo}%
\providecommand \href [0]{\begingroup \@sanitize@url \@href}%
\providecommand \@href[1]{\@@startlink{#1}\@@href}%
\providecommand \@@href[1]{\endgroup#1\@@endlink}%
\providecommand \@sanitize@url [0]{\catcode `\\12\catcode `\$12\catcode
  `\&12\catcode `\#12\catcode `\^12\catcode `\_12\catcode `\%12\relax}%
\providecommand \@@startlink[1]{}%
\providecommand \@@endlink[0]{}%
\providecommand \url  [0]{\begingroup\@sanitize@url \@url }%
\providecommand \@url [1]{\endgroup\@href {#1}{\urlprefix }}%
\providecommand \urlprefix  [0]{URL }%
\providecommand \Eprint [0]{\href }%
\providecommand \doibase [0]{https://doi.org/}%
\providecommand \selectlanguage [0]{\@gobble}%
\providecommand \bibinfo  [0]{\@secondoftwo}%
\providecommand \bibfield  [0]{\@secondoftwo}%
\providecommand \translation [1]{[#1]}%
\providecommand \BibitemOpen [0]{}%
\providecommand \bibitemStop [0]{}%
\providecommand \bibitemNoStop [0]{.\EOS\space}%
\providecommand \EOS [0]{\spacefactor3000\relax}%
\providecommand \BibitemShut  [1]{\csname bibitem#1\endcsname}%
\let\auto@bib@innerbib\@empty
%</preamble>
\bibitem [{\citenamefont {Ramaswamy}(2010)}]{SRARCMP2010}%
  \BibitemOpen
  \bibfield  {author} {\bibinfo {author} {\bibfnamefont {S.}~\bibnamefont
  {Ramaswamy}},\ }\bibfield  {title} {\bibinfo {title} {The mechanics and
  statistics of active matter},\ }\href
  {https://doi.org/10.1146/annurev-conmatphys-070909-104101} {\bibfield
  {journal} {\bibinfo  {journal} {Annual Review of Condensed Matter Physics}\
  }\textbf {\bibinfo {volume} {1}},\ \bibinfo {pages} {323} (\bibinfo {year}
  {2010})},\ \Eprint
  {https://arxiv.org/abs/https://doi.org/10.1146/annurev-conmatphys-070909-104101}
  {https://doi.org/10.1146/annurev-conmatphys-070909-104101} \BibitemShut
  {NoStop}%
\bibitem [{\citenamefont {Marchetti}\ \emph {et~al.}(2013)\citenamefont
  {Marchetti}, \citenamefont {Joanny}, \citenamefont {Ramaswamy}, \citenamefont
  {Liverpool}, \citenamefont {Prost}, \citenamefont {Rao},\ and\ \citenamefont
  {Simha}}]{ActiveMatterRMP2013}%
  \BibitemOpen
  \bibfield  {author} {\bibinfo {author} {\bibfnamefont {M.~C.}\ \bibnamefont
  {Marchetti}}, \bibinfo {author} {\bibfnamefont {J.~F.}\ \bibnamefont
  {Joanny}}, \bibinfo {author} {\bibfnamefont {S.}~\bibnamefont {Ramaswamy}},
  \bibinfo {author} {\bibfnamefont {T.~B.}\ \bibnamefont {Liverpool}}, \bibinfo
  {author} {\bibfnamefont {J.}~\bibnamefont {Prost}}, \bibinfo {author}
  {\bibfnamefont {M.}~\bibnamefont {Rao}},\ and\ \bibinfo {author}
  {\bibfnamefont {R.~A.}\ \bibnamefont {Simha}},\ }\bibfield  {title} {\bibinfo
  {title} {Hydrodynamics of soft active matter},\ }\href
  {https://doi.org/10.1103/RevModPhys.85.1143} {\bibfield  {journal} {\bibinfo
  {journal} {Rev. Mod. Phys.}\ }\textbf {\bibinfo {volume} {85}},\ \bibinfo
  {pages} {1143} (\bibinfo {year} {2013})}\BibitemShut {NoStop}%
\bibitem [{\citenamefont {Cavagna}\ and\ \citenamefont
  {Giardina}(2014)}]{ACIGARCMP2014}%
  \BibitemOpen
  \bibfield  {author} {\bibinfo {author} {\bibfnamefont {A.}~\bibnamefont
  {Cavagna}}\ and\ \bibinfo {author} {\bibfnamefont {I.}~\bibnamefont
  {Giardina}},\ }\bibfield  {title} {\bibinfo {title} {Bird flocks as condensed
  matter},\ }\href {https://doi.org/10.1146/annurev-conmatphys-031113-133834}
  {\bibfield  {journal} {\bibinfo  {journal} {Annual Review of Condensed Matter
  Physics}\ }\textbf {\bibinfo {volume} {5}},\ \bibinfo {pages} {183} (\bibinfo
  {year} {2014})},\ \Eprint
  {https://arxiv.org/abs/https://doi.org/10.1146/annurev-conmatphys-031113-133834}
  {https://doi.org/10.1146/annurev-conmatphys-031113-133834} \BibitemShut
  {NoStop}%
\bibitem [{\citenamefont {Hatwalne}\ \emph {et~al.}(2004)\citenamefont
  {Hatwalne}, \citenamefont {Ramaswamy}, \citenamefont {Rao},\ and\
  \citenamefont {Simha}}]{Yhat2004}%
  \BibitemOpen
  \bibfield  {author} {\bibinfo {author} {\bibfnamefont {Y.}~\bibnamefont
  {Hatwalne}}, \bibinfo {author} {\bibfnamefont {S.}~\bibnamefont {Ramaswamy}},
  \bibinfo {author} {\bibfnamefont {M.}~\bibnamefont {Rao}},\ and\ \bibinfo
  {author} {\bibfnamefont {R.~A.}\ \bibnamefont {Simha}},\ }\bibfield  {title}
  {\bibinfo {title} {Rheology of active-particle suspensions},\ }\href
  {https://doi.org/10.1103/PhysRevLett.92.118101} {\bibfield  {journal}
  {\bibinfo  {journal} {Phys. Rev. Lett.}\ }\textbf {\bibinfo {volume} {92}},\
  \bibinfo {pages} {118101} (\bibinfo {year} {2004})}\BibitemShut {NoStop}%
\bibitem [{\citenamefont {Bi}\ \emph {et~al.}(2015)\citenamefont {Bi},
  \citenamefont {Lopez}, \citenamefont {Schwarz},\ and\ \citenamefont
  {Manning}}]{BiNPhys2015}%
  \BibitemOpen
  \bibfield  {author} {\bibinfo {author} {\bibfnamefont {D.}~\bibnamefont
  {Bi}}, \bibinfo {author} {\bibfnamefont {J.~H.}\ \bibnamefont {Lopez}},
  \bibinfo {author} {\bibfnamefont {J.~M.}\ \bibnamefont {Schwarz}},\ and\
  \bibinfo {author} {\bibfnamefont {L.~M.}\ \bibnamefont {Manning}},\
  }\bibfield  {title} {\bibinfo {title} {A density-independent rigidity
  transition in biological tissues},\ }\href
  {https://doi.org/10.1038/nphys3471} {\bibfield  {journal} {\bibinfo
  {journal} {Nature Physics}\ }\textbf {\bibinfo {volume} {11}},\ \bibinfo
  {pages} {1074–1079} (\bibinfo {year} {2015})}\BibitemShut {NoStop}%
\bibitem [{\citenamefont {Bi}\ \emph {et~al.}(2016)\citenamefont {Bi},
  \citenamefont {Yang}, \citenamefont {Marchetti},\ and\ \citenamefont
  {Manning}}]{BiPRX2016}%
  \BibitemOpen
  \bibfield  {author} {\bibinfo {author} {\bibfnamefont {D.}~\bibnamefont
  {Bi}}, \bibinfo {author} {\bibfnamefont {X.}~\bibnamefont {Yang}}, \bibinfo
  {author} {\bibfnamefont {M.~C.}\ \bibnamefont {Marchetti}},\ and\ \bibinfo
  {author} {\bibfnamefont {M.~L.}\ \bibnamefont {Manning}},\ }\bibfield
  {title} {\bibinfo {title} {Motility-driven glass and jamming transitions in
  biological tissues},\ }\href {https://doi.org/10.1103/PhysRevX.6.021011}
  {\bibfield  {journal} {\bibinfo  {journal} {Phys. Rev. X}\ }\textbf {\bibinfo
  {volume} {6}},\ \bibinfo {pages} {021011} (\bibinfo {year}
  {2016})}\BibitemShut {NoStop}%
\bibitem [{\citenamefont {Das}\ \emph {et~al.}(2021)\citenamefont {Das},
  \citenamefont {Sastry},\ and\ \citenamefont {Bi}}]{DasPRX2021}%
  \BibitemOpen
  \bibfield  {author} {\bibinfo {author} {\bibfnamefont {A.}~\bibnamefont
  {Das}}, \bibinfo {author} {\bibfnamefont {S.}~\bibnamefont {Sastry}},\ and\
  \bibinfo {author} {\bibfnamefont {D.}~\bibnamefont {Bi}},\ }\bibfield
  {title} {\bibinfo {title} {Controlled neighbor exchanges drive glassy
  behavior, intermittency, and cell streaming in epithelial tissues},\ }\href
  {https://doi.org/10.1103/PhysRevX.11.041037} {\bibfield  {journal} {\bibinfo
  {journal} {Phys. Rev. X}\ }\textbf {\bibinfo {volume} {11}},\ \bibinfo
  {pages} {041037} (\bibinfo {year} {2021})}\BibitemShut {NoStop}%
\bibitem [{\citenamefont {Jülicher}\ \emph {et~al.}(2007)\citenamefont
  {Jülicher}, \citenamefont {Kruse}, \citenamefont {Prost},\ and\
  \citenamefont {Joanny}}]{JULICHER20073}%
  \BibitemOpen
  \bibfield  {author} {\bibinfo {author} {\bibfnamefont {F.}~\bibnamefont
  {Jülicher}}, \bibinfo {author} {\bibfnamefont {K.}~\bibnamefont {Kruse}},
  \bibinfo {author} {\bibfnamefont {J.}~\bibnamefont {Prost}},\ and\ \bibinfo
  {author} {\bibfnamefont {J.-F.}\ \bibnamefont {Joanny}},\ }\bibfield  {title}
  {\bibinfo {title} {Active behavior of the cytoskeleton},\ }\href
  {https://doi.org/https://doi.org/10.1016/j.physrep.2007.02.018} {\bibfield
  {journal} {\bibinfo  {journal} {Physics Reports}\ }\textbf {\bibinfo {volume}
  {449}},\ \bibinfo {pages} {3} (\bibinfo {year} {2007})},\ \bibinfo {note}
  {nonequilibrium physics: From complex fluids to biological systems III.
  Living systems}\BibitemShut {NoStop}%
\bibitem [{\citenamefont {Shaebani}\ \emph {et~al.}(2020)\citenamefont
  {Shaebani}, \citenamefont {Wysocki}, \citenamefont {Winkler}, \citenamefont
  {Gompper},\ and\ \citenamefont {Rieger}}]{CompActiveNatRevPhys2020}%
  \BibitemOpen
  \bibfield  {author} {\bibinfo {author} {\bibfnamefont {M.}~\bibnamefont
  {Shaebani}}, \bibinfo {author} {\bibfnamefont {A.}~\bibnamefont {Wysocki}},
  \bibinfo {author} {\bibfnamefont {R.}~\bibnamefont {Winkler}}, \bibinfo
  {author} {\bibfnamefont {G.}~\bibnamefont {Gompper}},\ and\ \bibinfo {author}
  {\bibfnamefont {H.}~\bibnamefont {Rieger}},\ }\bibfield  {title} {\bibinfo
  {title} {Computational models for active matter},\ }\href
  {https://doi.org/10.1038/s42254-020-0152-1} {\bibfield  {journal} {\bibinfo
  {journal} {Nature Reviews Physics}\ }\textbf {\bibinfo {volume} {2}},\
  \bibinfo {pages} {181–199} (\bibinfo {year} {2020})}\BibitemShut {NoStop}%
\bibitem [{\citenamefont {Janssen}(2019)}]{Janssen_2019}%
  \BibitemOpen
  \bibfield  {author} {\bibinfo {author} {\bibfnamefont {L.~M.~C.}\
  \bibnamefont {Janssen}},\ }\bibfield  {title} {\bibinfo {title} {Active
  glasses},\ }\href {https://doi.org/10.1088/1361-648X/ab3e90} {\bibfield
  {journal} {\bibinfo  {journal} {Journal of Physics: Condensed Matter}\
  }\textbf {\bibinfo {volume} {31}},\ \bibinfo {pages} {503002} (\bibinfo
  {year} {2019})}\BibitemShut {NoStop}%
\bibitem [{\citenamefont {Henkes}\ \emph {et~al.}(2011)\citenamefont {Henkes},
  \citenamefont {Fily},\ and\ \citenamefont {Marchetti}}]{Henkes2011}%
  \BibitemOpen
  \bibfield  {author} {\bibinfo {author} {\bibfnamefont {S.}~\bibnamefont
  {Henkes}}, \bibinfo {author} {\bibfnamefont {Y.}~\bibnamefont {Fily}},\ and\
  \bibinfo {author} {\bibfnamefont {M.~C.}\ \bibnamefont {Marchetti}},\
  }\bibfield  {title} {\bibinfo {title} {Active jamming: Self-propelled soft
  particles at high density},\ }\href
  {https://doi.org/10.1103/PhysRevE.84.040301} {\bibfield  {journal} {\bibinfo
  {journal} {Phys. Rev. E}\ }\textbf {\bibinfo {volume} {84}},\ \bibinfo
  {pages} {040301} (\bibinfo {year} {2011})}\BibitemShut {NoStop}%
\bibitem [{\citenamefont {Mandal}\ \emph {et~al.}(2016)\citenamefont {Mandal},
  \citenamefont {Bhuyan}, \citenamefont {Rao},\ and\ \citenamefont
  {Dasgupta}}]{MandalSM2016}%
  \BibitemOpen
  \bibfield  {author} {\bibinfo {author} {\bibfnamefont {R.}~\bibnamefont
  {Mandal}}, \bibinfo {author} {\bibfnamefont {P.~J.}\ \bibnamefont {Bhuyan}},
  \bibinfo {author} {\bibfnamefont {M.}~\bibnamefont {Rao}},\ and\ \bibinfo
  {author} {\bibfnamefont {C.}~\bibnamefont {Dasgupta}},\ }\bibfield  {title}
  {\bibinfo {title} {Active fluidization in dense glassy systems},\ }\href
  {https://doi.org/10.1039/C5SM02950C} {\bibfield  {journal} {\bibinfo
  {journal} {Soft Matter}\ }\textbf {\bibinfo {volume} {12}},\ \bibinfo {pages}
  {6268} (\bibinfo {year} {2016})}\BibitemShut {NoStop}%
\bibitem [{\citenamefont {Mandal}\ \emph {et~al.}(2020)\citenamefont {Mandal},
  \citenamefont {Bhuyan}, \citenamefont {Chaudhuri}, \citenamefont {Dasgupta},\
  and\ \citenamefont {Rao}}]{mandal2020extreme}%
  \BibitemOpen
  \bibfield  {author} {\bibinfo {author} {\bibfnamefont {R.}~\bibnamefont
  {Mandal}}, \bibinfo {author} {\bibfnamefont {P.~J.}\ \bibnamefont {Bhuyan}},
  \bibinfo {author} {\bibfnamefont {P.}~\bibnamefont {Chaudhuri}}, \bibinfo
  {author} {\bibfnamefont {C.}~\bibnamefont {Dasgupta}},\ and\ \bibinfo
  {author} {\bibfnamefont {M.}~\bibnamefont {Rao}},\ }\bibfield  {title}
  {\bibinfo {title} {Extreme active matter at high densities},\ }\href@noop {}
  {\bibfield  {journal} {\bibinfo  {journal} {Nature communications}\ }\textbf
  {\bibinfo {volume} {11}},\ \bibinfo {pages} {2581} (\bibinfo {year}
  {2020})}\BibitemShut {NoStop}%
\bibitem [{\citenamefont {Mo}\ \emph {et~al.}(2020)\citenamefont {Mo},
  \citenamefont {Liao},\ and\ \citenamefont {Xu}}]{Mo2020}%
  \BibitemOpen
  \bibfield  {author} {\bibinfo {author} {\bibfnamefont {R.}~\bibnamefont
  {Mo}}, \bibinfo {author} {\bibfnamefont {Q.}~\bibnamefont {Liao}},\ and\
  \bibinfo {author} {\bibfnamefont {N.}~\bibnamefont {Xu}},\ }\bibfield
  {title} {\bibinfo {title} {Rheological similarities between dense
  self-propelled and sheared particulate systems},\ }\href
  {https://doi.org/10.1039/D0SM00101E} {\bibfield  {journal} {\bibinfo
  {journal} {Soft Matter}\ }\textbf {\bibinfo {volume} {16}},\ \bibinfo {pages}
  {3642} (\bibinfo {year} {2020})}\BibitemShut {NoStop}%
\bibitem [{\citenamefont {Liao}\ and\ \citenamefont {Xu}(2018)}]{Xu2018}%
  \BibitemOpen
  \bibfield  {author} {\bibinfo {author} {\bibfnamefont {Q.}~\bibnamefont
  {Liao}}\ and\ \bibinfo {author} {\bibfnamefont {N.}~\bibnamefont {Xu}},\
  }\bibfield  {title} {\bibinfo {title} {Criticality of the zero-temperature
  jamming transition probed by self-propelled particles},\ }\href
  {https://doi.org/10.1039/C7SM01909B} {\bibfield  {journal} {\bibinfo
  {journal} {Soft Matter}\ }\textbf {\bibinfo {volume} {14}},\ \bibinfo {pages}
  {853} (\bibinfo {year} {2018})}\BibitemShut {NoStop}%
\bibitem [{\citenamefont {Morse}\ \emph {et~al.}(2021)\citenamefont {Morse},
  \citenamefont {Roy}, \citenamefont {Agoritsas},\ and\ \citenamefont
  {Manning}}]{MorsePNAS21}%
  \BibitemOpen
  \bibfield  {author} {\bibinfo {author} {\bibfnamefont {P.~K.}\ \bibnamefont
  {Morse}}, \bibinfo {author} {\bibfnamefont {S.}~\bibnamefont {Roy}}, \bibinfo
  {author} {\bibfnamefont {E.}~\bibnamefont {Agoritsas}},\ and\ \bibinfo
  {author} {\bibfnamefont {M.~L.}\ \bibnamefont {Manning}},\ }\bibfield
  {title} {\bibinfo {title} {A direct link between active matter and sheared
  granular systems},\ }\bibfield  {journal} {\bibinfo  {journal} {Proceedings
  of the National Academy of Sciences}\ }\textbf {\bibinfo {volume} {118}},\
  \href {https://doi.org/10.1073/pnas.2019909118} {10.1073/pnas.2019909118}
  (\bibinfo {year} {2021})\BibitemShut {NoStop}%
\bibitem [{\citenamefont {Agoritsas}(2021)}]{Agoritsas_2021}%
  \BibitemOpen
  \bibfield  {author} {\bibinfo {author} {\bibfnamefont {E.}~\bibnamefont
  {Agoritsas}},\ }\bibfield  {title} {\bibinfo {title} {Mean-field dynamics of
  infinite-dimensional particle systems: global shear versus random local
  forcing},\ }\href {https://doi.org/10.1088/1742-5468/abdd18} {\bibfield
  {journal} {\bibinfo  {journal} {Journal of Statistical Mechanics: Theory and
  Experiment}\ }\textbf {\bibinfo {volume} {2021}},\ \bibinfo {pages} {033501}
  (\bibinfo {year} {2021})}\BibitemShut {NoStop}%
\bibitem [{\citenamefont {Villarroel}\ and\ \citenamefont
  {Düring}(2021)}]{During2021}%
  \BibitemOpen
  \bibfield  {author} {\bibinfo {author} {\bibfnamefont {C.}~\bibnamefont
  {Villarroel}}\ and\ \bibinfo {author} {\bibfnamefont {G.}~\bibnamefont
  {Düring}},\ }\bibfield  {title} {\bibinfo {title} {Critical yielding
  rheology: from externally deformed glasses to active systems},\ }\href
  {https://doi.org/10.1039/D1SM00948F} {\bibfield  {journal} {\bibinfo
  {journal} {Soft Matter}\ }\textbf {\bibinfo {volume} {17}},\ \bibinfo {pages}
  {9944} (\bibinfo {year} {2021})}\BibitemShut {NoStop}%
\bibitem [{\citenamefont {Amiri}\ \emph {et~al.}(2022)\citenamefont {Amiri},
  \citenamefont {Duclut}, \citenamefont {Jülicher},\ and\ \citenamefont
  {Popović}}]{amiri2022random}%
  \BibitemOpen
  \bibfield  {author} {\bibinfo {author} {\bibfnamefont {A.}~\bibnamefont
  {Amiri}}, \bibinfo {author} {\bibfnamefont {C.}~\bibnamefont {Duclut}},
  \bibinfo {author} {\bibfnamefont {F.}~\bibnamefont {Jülicher}},\ and\
  \bibinfo {author} {\bibfnamefont {M.}~\bibnamefont {Popović}},\ }\href@noop
  {} {\bibinfo {title} {Random traction yielding transition in epithelial
  tissues}} (\bibinfo {year} {2022}),\ \Eprint
  {https://arxiv.org/abs/2211.02159} {arXiv:2211.02159 [cond-mat.soft]}
  \BibitemShut {NoStop}%
\bibitem [{\citenamefont {Leishangthem}\ \emph {et~al.}(2017)\citenamefont
  {Leishangthem}, \citenamefont {Parmar},\ and\ \citenamefont
  {Sastry}}]{leishangthem2017}%
  \BibitemOpen
  \bibfield  {author} {\bibinfo {author} {\bibfnamefont {P.}~\bibnamefont
  {Leishangthem}}, \bibinfo {author} {\bibfnamefont {A.~D.~S.}\ \bibnamefont
  {Parmar}},\ and\ \bibinfo {author} {\bibfnamefont {S.}~\bibnamefont
  {Sastry}},\ }\bibfield  {title} {\bibinfo {title} {The yielding transition in
  amorphous solids under oscillatory shear deformation},\ }\href
  {http://dx.doi.org/10.1038/ncomms14653} {\bibfield  {journal} {\bibinfo
  {journal} {Nature Communications}\ }\textbf {\bibinfo {volume} {8}},\
  \bibinfo {pages} {14653} (\bibinfo {year} {2017})}\BibitemShut {NoStop}%
\bibitem [{\citenamefont {Bhaumik}\ \emph {et~al.}(2021)\citenamefont
  {Bhaumik}, \citenamefont {Foffi},\ and\ \citenamefont
  {Sastry}}]{BhaumikPNAS21}%
  \BibitemOpen
  \bibfield  {author} {\bibinfo {author} {\bibfnamefont {H.}~\bibnamefont
  {Bhaumik}}, \bibinfo {author} {\bibfnamefont {G.}~\bibnamefont {Foffi}},\
  and\ \bibinfo {author} {\bibfnamefont {S.}~\bibnamefont {Sastry}},\
  }\bibfield  {title} {\bibinfo {title} {The role of annealing in determining
  the yielding behavior of glasses under cyclic shear deformation},\ }\bibfield
   {journal} {\bibinfo  {journal} {Proceedings of the National Academy of
  Sciences}\ }\textbf {\bibinfo {volume} {118}},\ \href
  {https://doi.org/10.1073/pnas.2100227118} {10.1073/pnas.2100227118} (\bibinfo
  {year} {2021})\BibitemShut {NoStop}%
\bibitem [{\citenamefont {Yang}\ \emph {et~al.}(2014)\citenamefont {Yang},
  \citenamefont {Manning},\ and\ \citenamefont
  {Marchetti}}]{yang2014aggregation}%
  \BibitemOpen
  \bibfield  {author} {\bibinfo {author} {\bibfnamefont {X.}~\bibnamefont
  {Yang}}, \bibinfo {author} {\bibfnamefont {M.~L.}\ \bibnamefont {Manning}},\
  and\ \bibinfo {author} {\bibfnamefont {M.~C.}\ \bibnamefont {Marchetti}},\
  }\bibfield  {title} {\bibinfo {title} {Aggregation and segregation of
  confined active particles},\ }\href@noop {} {\bibfield  {journal} {\bibinfo
  {journal} {Soft matter}\ }\textbf {\bibinfo {volume} {10}},\ \bibinfo {pages}
  {6477} (\bibinfo {year} {2014})}\BibitemShut {NoStop}%
\bibitem [{\citenamefont {Fily}\ \emph {et~al.}(2014)\citenamefont {Fily},
  \citenamefont {Baskaran},\ and\ \citenamefont {Hagan}}]{fily2014dynamics}%
  \BibitemOpen
  \bibfield  {author} {\bibinfo {author} {\bibfnamefont {Y.}~\bibnamefont
  {Fily}}, \bibinfo {author} {\bibfnamefont {A.}~\bibnamefont {Baskaran}},\
  and\ \bibinfo {author} {\bibfnamefont {M.~F.}\ \bibnamefont {Hagan}},\
  }\bibfield  {title} {\bibinfo {title} {Dynamics of self-propelled particles
  under strong confinement},\ }\href@noop {} {\bibfield  {journal} {\bibinfo
  {journal} {Soft matter}\ }\textbf {\bibinfo {volume} {10}},\ \bibinfo {pages}
  {5609} (\bibinfo {year} {2014})}\BibitemShut {NoStop}%
\bibitem [{\citenamefont {Vishen}\ \emph {et~al.}(2018)\citenamefont {Vishen},
  \citenamefont {Rupprecht}, \citenamefont {Shivashankar}, \citenamefont
  {Prost},\ and\ \citenamefont {Rao}}]{vishen2018soft}%
  \BibitemOpen
  \bibfield  {author} {\bibinfo {author} {\bibfnamefont {A.~S.}\ \bibnamefont
  {Vishen}}, \bibinfo {author} {\bibfnamefont {J.-F.}\ \bibnamefont
  {Rupprecht}}, \bibinfo {author} {\bibfnamefont {G.}~\bibnamefont
  {Shivashankar}}, \bibinfo {author} {\bibfnamefont {J.}~\bibnamefont
  {Prost}},\ and\ \bibinfo {author} {\bibfnamefont {M.}~\bibnamefont {Rao}},\
  }\bibfield  {title} {\bibinfo {title} {Soft inclusion in a confined
  fluctuating active gel},\ }\href@noop {} {\bibfield  {journal} {\bibinfo
  {journal} {Physical Review E}\ }\textbf {\bibinfo {volume} {97}},\ \bibinfo
  {pages} {032602} (\bibinfo {year} {2018})}\BibitemShut {NoStop}%
\bibitem [{\citenamefont {Talwar}\ \emph {et~al.}(2013)\citenamefont {Talwar},
  \citenamefont {Kumar}, \citenamefont {Rao}, \citenamefont {Menon},\ and\
  \citenamefont {Shivashankar}}]{talwar2013correlated}%
  \BibitemOpen
  \bibfield  {author} {\bibinfo {author} {\bibfnamefont {S.}~\bibnamefont
  {Talwar}}, \bibinfo {author} {\bibfnamefont {A.}~\bibnamefont {Kumar}},
  \bibinfo {author} {\bibfnamefont {M.}~\bibnamefont {Rao}}, \bibinfo {author}
  {\bibfnamefont {G.~I.}\ \bibnamefont {Menon}},\ and\ \bibinfo {author}
  {\bibfnamefont {G.}~\bibnamefont {Shivashankar}},\ }\bibfield  {title}
  {\bibinfo {title} {Correlated spatio-temporal fluctuations in chromatin
  compaction states characterize stem cells},\ }\href@noop {} {\bibfield
  {journal} {\bibinfo  {journal} {Biophysical journal}\ }\textbf {\bibinfo
  {volume} {104}},\ \bibinfo {pages} {553} (\bibinfo {year}
  {2013})}\BibitemShut {NoStop}%
\bibitem [{\citenamefont {Makhija}\ \emph {et~al.}(2016)\citenamefont
  {Makhija}, \citenamefont {Jokhun},\ and\ \citenamefont
  {Shivashankar}}]{makhija2016nuclear}%
  \BibitemOpen
  \bibfield  {author} {\bibinfo {author} {\bibfnamefont {E.}~\bibnamefont
  {Makhija}}, \bibinfo {author} {\bibfnamefont {D.}~\bibnamefont {Jokhun}},\
  and\ \bibinfo {author} {\bibfnamefont {G.}~\bibnamefont {Shivashankar}},\
  }\bibfield  {title} {\bibinfo {title} {Nuclear deformability and telomere
  dynamics are regulated by cell geometric constraints},\ }\href@noop {}
  {\bibfield  {journal} {\bibinfo  {journal} {Proceedings of the National
  Academy of Sciences}\ }\textbf {\bibinfo {volume} {113}},\ \bibinfo {pages}
  {E32} (\bibinfo {year} {2016})}\BibitemShut {NoStop}%
\bibitem [{\citenamefont {Uhler}\ and\ \citenamefont
  {Shivashankar}(2017)}]{UhlerGVSReview2017}%
  \BibitemOpen
  \bibfield  {author} {\bibinfo {author} {\bibfnamefont {C.}~\bibnamefont
  {Uhler}}\ and\ \bibinfo {author} {\bibfnamefont {G.~V.}\ \bibnamefont
  {Shivashankar}},\ }\bibfield  {title} {\bibinfo {title} {Regulation of genome
  organization and gene expression by nuclear mechanotransduction},\ }\href
  {https://doi.org/10.1038/nrm.2017.101} {\bibfield  {journal} {\bibinfo
  {journal} {Nature Reviews Molecular Cell Biology}\ }\textbf {\bibinfo
  {volume} {18}},\ \bibinfo {pages} {717–727} (\bibinfo {year}
  {2017})}\BibitemShut {NoStop}%
\bibitem [{\citenamefont {Xu}\ \emph {et~al.}(2021)\citenamefont {Xu},
  \citenamefont {Andresen},\ and\ \citenamefont {Regev}}]{xu2021yielding}%
  \BibitemOpen
  \bibfield  {author} {\bibinfo {author} {\bibfnamefont {H.}~\bibnamefont
  {Xu}}, \bibinfo {author} {\bibfnamefont {J.~C.}\ \bibnamefont {Andresen}},\
  and\ \bibinfo {author} {\bibfnamefont {I.}~\bibnamefont {Regev}},\ }\bibfield
   {title} {\bibinfo {title} {Yielding in an amorphous solid subject to
  constant stress at finite temperatures},\ }\href@noop {} {\bibfield
  {journal} {\bibinfo  {journal} {Physical Review E}\ }\textbf {\bibinfo
  {volume} {103}},\ \bibinfo {pages} {052604} (\bibinfo {year}
  {2021})}\BibitemShut {NoStop}%
\bibitem [{\citenamefont {Hopkins}\ \emph {et~al.}(2022)\citenamefont
  {Hopkins}, \citenamefont {Chiang}, \citenamefont {Loewe}, \citenamefont
  {Marenduzzo},\ and\ \citenamefont {Marchetti}}]{hopkins2022local}%
  \BibitemOpen
  \bibfield  {author} {\bibinfo {author} {\bibfnamefont {A.}~\bibnamefont
  {Hopkins}}, \bibinfo {author} {\bibfnamefont {M.}~\bibnamefont {Chiang}},
  \bibinfo {author} {\bibfnamefont {B.}~\bibnamefont {Loewe}}, \bibinfo
  {author} {\bibfnamefont {D.}~\bibnamefont {Marenduzzo}},\ and\ \bibinfo
  {author} {\bibfnamefont {M.~C.}\ \bibnamefont {Marchetti}},\ }\bibfield
  {title} {\bibinfo {title} {Local yield and compliance in active cell
  monolayers},\ }\href@noop {} {\bibfield  {journal} {\bibinfo  {journal}
  {Physical Review Letters}\ }\textbf {\bibinfo {volume} {129}},\ \bibinfo
  {pages} {148101} (\bibinfo {year} {2022})}\BibitemShut {NoStop}%
\bibitem [{\citenamefont {Sharma}\ and\ \citenamefont
  {Karmakar}(2023)}]{Sharma2023}%
  \BibitemOpen
  \bibfield  {author} {\bibinfo {author} {\bibfnamefont {R.}~\bibnamefont
  {Sharma}}\ and\ \bibinfo {author} {\bibfnamefont {S.}~\bibnamefont
  {Karmakar}},\ }\bibfield  {title} {\bibinfo {title} {Activity-induced
  annealing leads to ductile-to-brittle transition in amorphous solids},\
  }\href@noop {} {\bibfield  {journal} {\bibinfo  {journal} {arXiv preprint
  arXiv:2305.17545}\ } (\bibinfo {year} {2023})},\ \Eprint
  {https://arxiv.org/abs/2305.17545} {arXiv:2305.17545 [cond-mat.soft]}
  \BibitemShut {NoStop}%
\bibitem [{\citenamefont {Nandi}\ \emph {et~al.}(2018)\citenamefont {Nandi},
  \citenamefont {Mandal}, \citenamefont {Bhuyan}, \citenamefont {Dasgupta},
  \citenamefont {Rao},\ and\ \citenamefont {Gov}}]{nandi2018random}%
  \BibitemOpen
  \bibfield  {author} {\bibinfo {author} {\bibfnamefont {S.~K.}\ \bibnamefont
  {Nandi}}, \bibinfo {author} {\bibfnamefont {R.}~\bibnamefont {Mandal}},
  \bibinfo {author} {\bibfnamefont {P.~J.}\ \bibnamefont {Bhuyan}}, \bibinfo
  {author} {\bibfnamefont {C.}~\bibnamefont {Dasgupta}}, \bibinfo {author}
  {\bibfnamefont {M.}~\bibnamefont {Rao}},\ and\ \bibinfo {author}
  {\bibfnamefont {N.~S.}\ \bibnamefont {Gov}},\ }\bibfield  {title} {\bibinfo
  {title} {A random first-order transition theory for an active glass},\
  }\href@noop {} {\bibfield  {journal} {\bibinfo  {journal} {Proceedings of the
  National Academy of Sciences}\ }\textbf {\bibinfo {volume} {115}},\ \bibinfo
  {pages} {7688} (\bibinfo {year} {2018})}\BibitemShut {NoStop}%
\bibitem [{\citenamefont {Mandal}\ \emph {et~al.}(2022)\citenamefont {Mandal},
  \citenamefont {Nandi}, \citenamefont {Dasgupta}, \citenamefont {Sollich},\
  and\ \citenamefont {Gov}}]{mandal2022random}%
  \BibitemOpen
  \bibfield  {author} {\bibinfo {author} {\bibfnamefont {R.}~\bibnamefont
  {Mandal}}, \bibinfo {author} {\bibfnamefont {S.~K.}\ \bibnamefont {Nandi}},
  \bibinfo {author} {\bibfnamefont {C.}~\bibnamefont {Dasgupta}}, \bibinfo
  {author} {\bibfnamefont {P.}~\bibnamefont {Sollich}},\ and\ \bibinfo {author}
  {\bibfnamefont {N.~S.}\ \bibnamefont {Gov}},\ }\bibfield  {title} {\bibinfo
  {title} {The random first-order transition theory of active glass in the
  high-activity regime},\ }\href@noop {} {\bibfield  {journal} {\bibinfo
  {journal} {Journal of Physics Communications}\ }\textbf {\bibinfo {volume}
  {6}},\ \bibinfo {pages} {115001} (\bibinfo {year} {2022})}\BibitemShut
  {NoStop}%
\bibitem [{\citenamefont {Szamel}\ \emph {et~al.}(2015)\citenamefont {Szamel},
  \citenamefont {Flenner},\ and\ \citenamefont {Berthier}}]{szamel2015glassy}%
  \BibitemOpen
  \bibfield  {author} {\bibinfo {author} {\bibfnamefont {G.}~\bibnamefont
  {Szamel}}, \bibinfo {author} {\bibfnamefont {E.}~\bibnamefont {Flenner}},\
  and\ \bibinfo {author} {\bibfnamefont {L.}~\bibnamefont {Berthier}},\
  }\bibfield  {title} {\bibinfo {title} {Glassy dynamics of athermal
  self-propelled particles: Computer simulations and a nonequilibrium
  microscopic theory},\ }\href@noop {} {\bibfield  {journal} {\bibinfo
  {journal} {Physical Review E}\ }\textbf {\bibinfo {volume} {91}},\ \bibinfo
  {pages} {062304} (\bibinfo {year} {2015})}\BibitemShut {NoStop}%
\bibitem [{\citenamefont {Keta}\ \emph {et~al.}(2023)\citenamefont {Keta},
  \citenamefont {Mandal}, \citenamefont {Sollich}, \citenamefont {Jack},\ and\
  \citenamefont {Berthier}}]{keta2023intermittent}%
  \BibitemOpen
  \bibfield  {author} {\bibinfo {author} {\bibfnamefont {Y.-E.}\ \bibnamefont
  {Keta}}, \bibinfo {author} {\bibfnamefont {R.}~\bibnamefont {Mandal}},
  \bibinfo {author} {\bibfnamefont {P.}~\bibnamefont {Sollich}}, \bibinfo
  {author} {\bibfnamefont {R.~L.}\ \bibnamefont {Jack}},\ and\ \bibinfo
  {author} {\bibfnamefont {L.}~\bibnamefont {Berthier}},\ }\bibfield  {title}
  {\bibinfo {title} {Intermittent relaxation and avalanches in extremely
  persistent active matter},\ }\href@noop {} {\bibfield  {journal} {\bibinfo
  {journal} {Soft Matter}\ }\textbf {\bibinfo {volume} {19}},\ \bibinfo {pages}
  {3871} (\bibinfo {year} {2023})}\BibitemShut {NoStop}%
\bibitem [{\citenamefont {Mandal}\ and\ \citenamefont
  {Sollich}(2021)}]{mandal2021study}%
  \BibitemOpen
  \bibfield  {author} {\bibinfo {author} {\bibfnamefont {R.}~\bibnamefont
  {Mandal}}\ and\ \bibinfo {author} {\bibfnamefont {P.}~\bibnamefont
  {Sollich}},\ }\bibfield  {title} {\bibinfo {title} {How to study a persistent
  active glassy system},\ }\href@noop {} {\bibfield  {journal} {\bibinfo
  {journal} {Journal of Physics: Condensed Matter}\ }\textbf {\bibinfo {volume}
  {33}},\ \bibinfo {pages} {184001} (\bibinfo {year} {2021})}\BibitemShut
  {NoStop}%
\bibitem [{\citenamefont {Roy}\ \emph {et~al.}(2017)\citenamefont {Roy},
  \citenamefont {Venkatachalapathy}, \citenamefont {Ratna}, \citenamefont
  {Wang}, \citenamefont {Jokhun}, \citenamefont {Nagarajana},\ and\
  \citenamefont {Shivashankar}}]{RoyPNAS2017}%
  \BibitemOpen
  \bibfield  {author} {\bibinfo {author} {\bibfnamefont {B.}~\bibnamefont
  {Roy}}, \bibinfo {author} {\bibfnamefont {S.}~\bibnamefont
  {Venkatachalapathy}}, \bibinfo {author} {\bibfnamefont {P.}~\bibnamefont
  {Ratna}}, \bibinfo {author} {\bibfnamefont {Y.}~\bibnamefont {Wang}},
  \bibinfo {author} {\bibfnamefont {D.~S.}\ \bibnamefont {Jokhun}}, \bibinfo
  {author} {\bibfnamefont {M.}~\bibnamefont {Nagarajana}},\ and\ \bibinfo
  {author} {\bibfnamefont {G.~V.}\ \bibnamefont {Shivashankar}},\ }\bibfield
  {title} {\bibinfo {title} {Laterally confined growth of cells induces nuclear
  reprogramming in the absence of exogenous biochemical factors},\ }\href
  {https://doi.org/10.1073/pnas.1714770115} {\bibfield  {journal} {\bibinfo
  {journal} {Proceedings of the National Academy of Sciences}\ }\textbf
  {\bibinfo {volume} {115}},\ \bibinfo {pages} {E4741–E4750} (\bibinfo {year}
  {2017})}\BibitemShut {NoStop}%
\bibitem [{\citenamefont {Roy}\ \emph {et~al.}(2020)\citenamefont {Roy},
  \citenamefont {Yuan}, \citenamefont {Lee},\ and\ \citenamefont
  {Shivashankr}}]{RoyPNAS2020}%
  \BibitemOpen
  \bibfield  {author} {\bibinfo {author} {\bibfnamefont {B.}~\bibnamefont
  {Roy}}, \bibinfo {author} {\bibfnamefont {L.}~\bibnamefont {Yuan}}, \bibinfo
  {author} {\bibfnamefont {Y.}~\bibnamefont {Lee}},\ and\ \bibinfo {author}
  {\bibfnamefont {G.~V.}\ \bibnamefont {Shivashankr}},\ }\bibfield  {title}
  {\bibinfo {title} {Fibroblast rejuvenation by mechanical reprogramming and
  redifferentiation},\ }\href {https://doi.org/10.1073/pnas.1911497117}
  {\bibfield  {journal} {\bibinfo  {journal} {Proceedings of the National
  Academy of Sciences}\ }\textbf {\bibinfo {volume} {117}},\ \bibinfo {pages}
  {10131} (\bibinfo {year} {2020})}\BibitemShut {NoStop}%
\bibitem [{\citenamefont {Shiba}\ \emph {et~al.}(2016)\citenamefont {Shiba},
  \citenamefont {Yamada}, \citenamefont {Kawasaki},\ and\ \citenamefont
  {Kim}}]{shiba2016unveiling}%
  \BibitemOpen
  \bibfield  {author} {\bibinfo {author} {\bibfnamefont {H.}~\bibnamefont
  {Shiba}}, \bibinfo {author} {\bibfnamefont {Y.}~\bibnamefont {Yamada}},
  \bibinfo {author} {\bibfnamefont {T.}~\bibnamefont {Kawasaki}},\ and\
  \bibinfo {author} {\bibfnamefont {K.}~\bibnamefont {Kim}},\ }\bibfield
  {title} {\bibinfo {title} {Unveiling dimensionality dependence of glassy
  dynamics: 2d infinite fluctuation eclipses inherent structural relaxation},\
  }\href@noop {} {\bibfield  {journal} {\bibinfo  {journal} {Physical review
  letters}\ }\textbf {\bibinfo {volume} {117}},\ \bibinfo {pages} {245701}
  (\bibinfo {year} {2016})}\BibitemShut {NoStop}%
\bibitem [{\citenamefont {Plimpton}(1995)}]{plimpton1995fast}%
  \BibitemOpen
  \bibfield  {author} {\bibinfo {author} {\bibfnamefont {S.}~\bibnamefont
  {Plimpton}},\ }\bibfield  {title} {\bibinfo {title} {Fast parallel algorithms
  for short-range molecular dynamics},\ }\href@noop {} {\bibfield  {journal}
  {\bibinfo  {journal} {Journal of computational physics}\ }\textbf {\bibinfo
  {volume} {117}},\ \bibinfo {pages} {1} (\bibinfo {year} {1995})}\BibitemShut
  {NoStop}%
\bibitem [{\citenamefont {Leimkuhler}\ and\ \citenamefont
  {Matthews}(2013)}]{leimkuhler2013rational}%
  \BibitemOpen
  \bibfield  {author} {\bibinfo {author} {\bibfnamefont {B.}~\bibnamefont
  {Leimkuhler}}\ and\ \bibinfo {author} {\bibfnamefont {C.}~\bibnamefont
  {Matthews}},\ }\bibfield  {title} {\bibinfo {title} {Rational construction of
  stochastic numerical methods for molecular sampling},\ }\href@noop {}
  {\bibfield  {journal} {\bibinfo  {journal} {Applied Mathematics Research
  eXpress}\ }\textbf {\bibinfo {volume} {2013}},\ \bibinfo {pages} {34}
  (\bibinfo {year} {2013})}\BibitemShut {NoStop}%
\bibitem [{\citenamefont {Allen}\ and\ \citenamefont
  {Tildesley}(2017)}]{allen2017computer}%
  \BibitemOpen
  \bibfield  {author} {\bibinfo {author} {\bibfnamefont {M.~P.}\ \bibnamefont
  {Allen}}\ and\ \bibinfo {author} {\bibfnamefont {D.~J.}\ \bibnamefont
  {Tildesley}},\ }\href@noop {} {\emph {\bibinfo {title} {Computer simulation
  of liquids}}}\ (\bibinfo  {publisher} {Oxford university press},\ \bibinfo
  {year} {2017})\BibitemShut {NoStop}%
\bibitem [{\citenamefont {Chatfield}(2017)}]{Chatfield2017}%
  \BibitemOpen
  \bibfield  {author} {\bibinfo {author} {\bibfnamefont {C.}~\bibnamefont
  {Chatfield}},\ }\href@noop {} {\bibinfo {title} {A simple method for distance
  to ellipse}},\ \bibinfo {howpublished}
  {\url{https://chatfield.io/simple-method-for-distance-to-ellipse/.}}
  (\bibinfo {year} {2017})\BibitemShut {NoStop}%
\bibitem [{\citenamefont {Stukowski}(2009)}]{stukowski2009visualization}%
  \BibitemOpen
  \bibfield  {author} {\bibinfo {author} {\bibfnamefont {A.}~\bibnamefont
  {Stukowski}},\ }\bibfield  {title} {\bibinfo {title} {Visualization and
  analysis of atomistic simulation data with ovito--the open visualization
  tool},\ }\href@noop {} {\bibfield  {journal} {\bibinfo  {journal} {Modelling
  and simulation in materials science and engineering}\ }\textbf {\bibinfo
  {volume} {18}},\ \bibinfo {pages} {015012} (\bibinfo {year}
  {2009})}\BibitemShut {NoStop}%
\bibitem [{\citenamefont {Maisonobe}(2006)}]{maisonobe2006quick}%
  \BibitemOpen
  \bibfield  {author} {\bibinfo {author} {\bibfnamefont {L.}~\bibnamefont
  {Maisonobe}},\ }\href@noop {} {\bibinfo {title} {Quick computation of the
  distance between a point and an ellipse}} (\bibinfo {year}
  {2006})\BibitemShut {NoStop}%
\end{thebibliography}

\begin{thebibliography}{10}

\bibitem{sileimkuhler2013rational}
Benedict Leimkuhler and Charles Matthews.
\newblock Rational construction of stochastic numerical methods for molecular
  sampling.
\newblock {\em Applied Mathematics Research eXpress}, 2013(1):34--56, 2013.

\bibitem{siallen2017computer}
Michael~P Allen and Dominic~J Tildesley.
\newblock {\em Computer simulation of liquids}.
\newblock Oxford university press, 2017.

\bibitem{siDuring2021}
Carlos Villarroel and Gustavo Düring.
\newblock Critical yielding rheology: from externally deformed glasses to
  active systems.
\newblock {\em Soft Matter}, 17:9944--9949, 2021.

\bibitem{sixu2021yielding}
Haiyan Xu, Juan~Carlos Andresen, and Ido Regev.
\newblock Yielding in an amorphous solid subject to constant stress at finite
  temperatures.
\newblock {\em Physical Review E}, 103(5):052604, 2021.

\bibitem{sinandi2018random}
Saroj~Kumar Nandi, Rituparno Mandal, Pranab~Jyoti Bhuyan, Chandan Dasgupta,
  Madan Rao, and Nir~S Gov.
\newblock A random first-order transition theory for an active glass.
\newblock {\em Proceedings of the National Academy of Sciences},
  115(30):7688--7693, 2018.

\bibitem{simandal2022random}
Rituparno Mandal, Saroj~Kumar Nandi, Chandan Dasgupta, Peter Sollich, and Nir~S
  Gov.
\newblock The random first-order transition theory of active glass in the
  high-activity regime.
\newblock {\em Journal of Physics Communications}, 6(11):115001, 2022.

\bibitem{simandal2020extreme}
Rituparno Mandal, Pranab~Jyoti Bhuyan, Pinaki Chaudhuri, Chandan Dasgupta, and
  Madan Rao.
\newblock Extreme active matter at high densities.
\newblock {\em Nature communications}, 11(1):2581, 2020.

\bibitem{fodorPRL2016}
\'Etienne Fodor, Cesare Nardini, Michael~E. Cates, Julien Tailleur, Paolo
  Visco, and Fr\'ed\'eric van Wijland.
\newblock How far from equilibrium is active matter?
\newblock {\em Phys. Rev. Lett.}, 117:038103, Jul 2016.

\bibitem{simaisonobe2006quick}
L~Maisonobe.
\newblock Quick computation of the distance between a point and an ellipse,
  2006.

\bibitem{siChatfield2017}
Carl Chatfield.
\newblock A simple method for distance to ellipse.
\newblock \url{https://chatfield.io/simple-method-for-distance-to-ellipse/.},
  2017.

\end{thebibliography}
%apsrev4-2.bst 2019-01-14 (MD) hand-edited version of apsrev4-1.bst
%Control: key (0)
%Control: author (8) initials jnrlst
%Control: editor formatted (1) identically to author
%Control: production of article title (0) allowed
%Control: page (0) single
%Control: year (1) truncated
%Control: production of eprint (0) enabled
%

\eject

\part*{}   %%%% SI begins

% \part*{}   %%%% SI begins
\setcounter{equation}{0}
\setcounter{figure}{0}
\setcounter{section}{0}
\renewcommand{\theequation}{S\arabic{equation}}%
\renewcommand{\thefigure}{S\arabic{figure}}%
\renewcommand{\thesection}{S\arabic{section}}%

% \appendix
\section*{Supplementary Information}

\subsection*{Integration of equations of motions for active dynamics}

The discretized time integration is done following the prescription in~\cite{leimkuhler2013rational,allen2017computer} for the {\bf BAOAB} operator splitting to include the active force along the orientation vector, $\hat n = (cos(\theta),sin(\theta))$.

Given the equations of motion:
\begin{align}
\mathbf{\dot p}_i &= -\gamma \mathbf{p}_i + \sum\limits_{j \ne i = 1}^{N}\left ( \mathbf{f_{ij}} \right )+ f \mathbf{\hat n}_i + \mathbf{\xi_t}^i \nonumber \\
\mathbf{\dot x}_i &= \mathbf{v}_i \equiv M^{-1} \mathbf{p}_i \nonumber \\
\mathbf{\dot \theta}_i &= \mathbf{\xi_r}^i
\end{align} 

we write 
\begin{equation}
{\rm d} \left [ \!\!\begin{array}{c} \mathbf{x}\\\mathbf{p}\\\theta\end{array}\!\! \right ]=
\underbrace{\left [ \!\!\begin{array}{c} M^{-1} \mathbf{p} \\0\\0\end{array}\!\! \right ]{\rm d}t}_{\rm A} +
\underbrace{\left [\!\! \begin{array}{c} 0 \\\mathbf{-\nabla U} + f \mathbf{{\hat n}}\\ 0\end{array}\!\! \right ]{\rm d}t}_{\rm B} +
\underbrace{\left [\!\! \begin{array}{c} 0\\ -\gamma \mathbf{p}{\rm d}t  +\sqrt{2\gamma M k_BT}{\rm d} \mathbf{w} \\\sqrt{2/\tau_p}{\rm d} \mathbf{w}\end{array} \!\!\right ],}_{\rm O}
\label{gla2}
\end{equation}

The stochastic component is modelled on the Weiner process $\mathbf{w}$, which follows:
\begin{equation}
{\rm d}\mathbf{w} = \mathbf{w}(t+ {\rm d}t) - w(t) = \sqrt{{\rm d}t} \mathbf{GR(0,1;t)}
\end{equation}
where $\mathbf{GR(0,1;t)}$ is a vector of independent, Gaussian distributed random variables with means $0$ and variance $1$. 
For the active particle system, one needs to include the active forces in the velocity update of the {\bf BAOAB} integrator, and as well perform the time integration for the orientation, $\theta(t)$. 

\noindent Defining constants,
\[
c_1 = e^{-\gamma \delta t}, \quad c_2 = \gamma^{-1} (1-c_1),
\]

\[
c_3 = \sqrt{k_BT(1-c_1^2)}, \quad c_4= \sqrt{2~dt/\tau_p},
\]
we have the update steps: 
\begin{subequations}
\begin{align}
\mathbf{p}_i\left (t+\frac{{\rm d}t}{2} \right) &=  \mathbf{p}_i(t) + \frac{{\rm d} t}{2} \left [\, \mathbf{-\nabla U(x}_i(t)) + f \mathbf{{\hat n_i(t)}}\right ];&  \\
% \mathbf{v}_{\theta}\left (t+\frac{{\rm d}t}{2} \right) &=   \mathbf{v}_{\theta}(t)  ;&  \\
\mathbf{x}_i\left (t+\frac{{\rm d}t}{2} \right) &=  \mathbf{x}_i(t) + \frac{{\rm d}t}{2} \, \mathbf{M^{-1} p}_i\left (t+\frac{{\rm d}t}{2} \right);&\\
% \theta\left (t+\frac{{\rm d}t}{2} \right) &=  \theta(t) + \frac{{\rm d}t}{2} \, \mathbf{v}_{\theta}\left (t+\frac{{\rm d}t}{2} \right);&\\
\mathbf{\hat{p}}_i\left (t+\frac{{\rm d}t}{2} \right) &=   c_1  \mathbf{p}_i\left (t+\frac{{\rm d}t}{2} \right) + c_3  \mathbf{M^{1/2} GR}_i(0,1;t+dt);&\\
% \hat{v_{\theta}}\left (t+\frac{{\rm d}t}{2} \right) &=  c_4   GR(0,1;t+dt);&\\
\theta_i(t+dt) &= \theta_i(t) + c_4\mathbf{GR}^{'}_{i}(0,1;t+dt);&\\
\mathbf{x}_i(t+ {\rm d}t) &=   \mathbf{x}_i\left (t+\frac{{\rm d}t}{2} \right) + \frac{{\rm d}t}{2}\, \mathbf{M^{-1} \hat{p}}_i\left (t+\frac{{\rm d}t}{2} \right);& \\
% \theta(t+{\rm d}t) &=  \theta(t + \frac{{\rm d}t}{2}) + \frac{{\rm d}t}{2} \, \hat v_{\theta}\left (t+\frac{{\rm d}t}{2} \right);&\\
\mathbf{p}_i(t + {\rm d}t) &=   \mathbf{\hat{p}}_i\left (t+\frac{{\rm d}t}{2} \right) + \frac{{\rm d}t}{2} \left [ \, \mathbf{-\nabla U(x}_i(t + {\rm d}t)) + f \mathbf{{\hat n}}_i(t+{\rm d}t) \right ].& \\
% v_{\theta}(t+{\rm d}t) &=   \hat v_{\theta}\left (t+\frac{{\rm d}t}{2} \right)  ;&  \\
\end{align}
\end{subequations}
\clearpage

%%% Each figure should be on its own page
\subsection*{Sample preparation}
\begin{figure}[hpb!]
\centering
\includegraphics[scale=0.35]{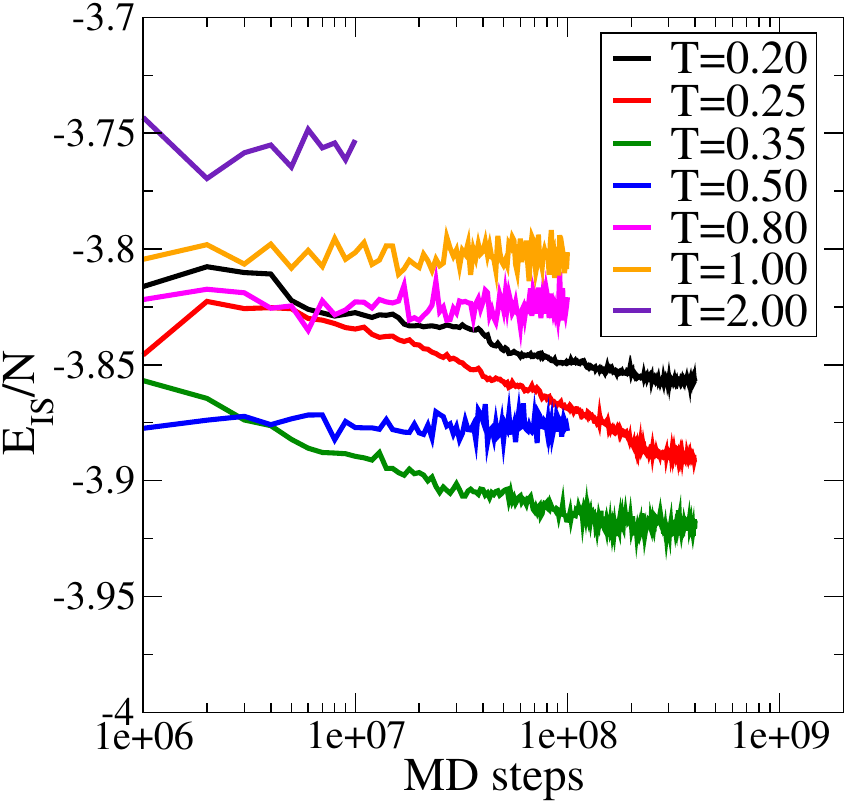}
\includegraphics[scale=0.35]{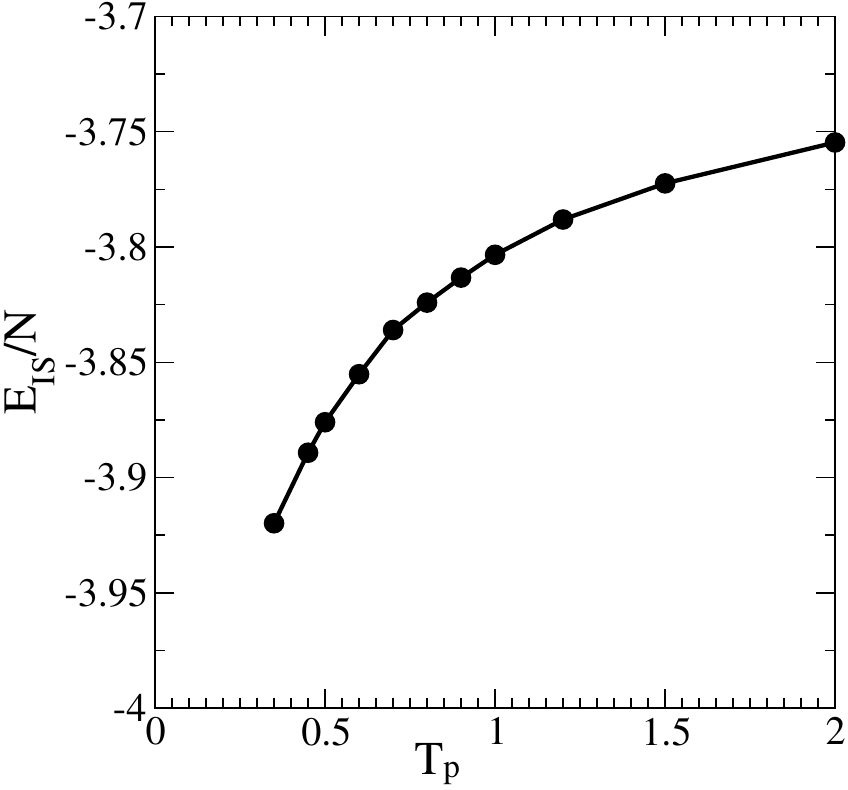}
\caption{Average inherent structure energy per particle as a function of time for $N=1000$ particles of the binary Lennard-Jones mixture in 2D, simulated in the NVT ensemble at different temperatures. The steady state average inherent structure energy per particle, $E_{IS}$, is shown as a function of simulation temperature for the cases at which a steady state is reached, i.e., $T~\geq~0.45$.}
\label{fig:EIS_v_T}
\end{figure}

We show the inherent structure potential energies, $E_{IS}$, as a function of time as well as the dependence of the steady state inherent structure energies, $E_{IS}$, as a function of the  temperature at which the simulation is performed, $T_p$, in Fig.~\ref{fig:EIS_v_T}.
% \clearpage
\subsection*{Yielding phase diagram at different $\mathbf{\tau_p}$}
\begin{figure}[hpb!]
\centering
\includegraphics[scale=0.25]{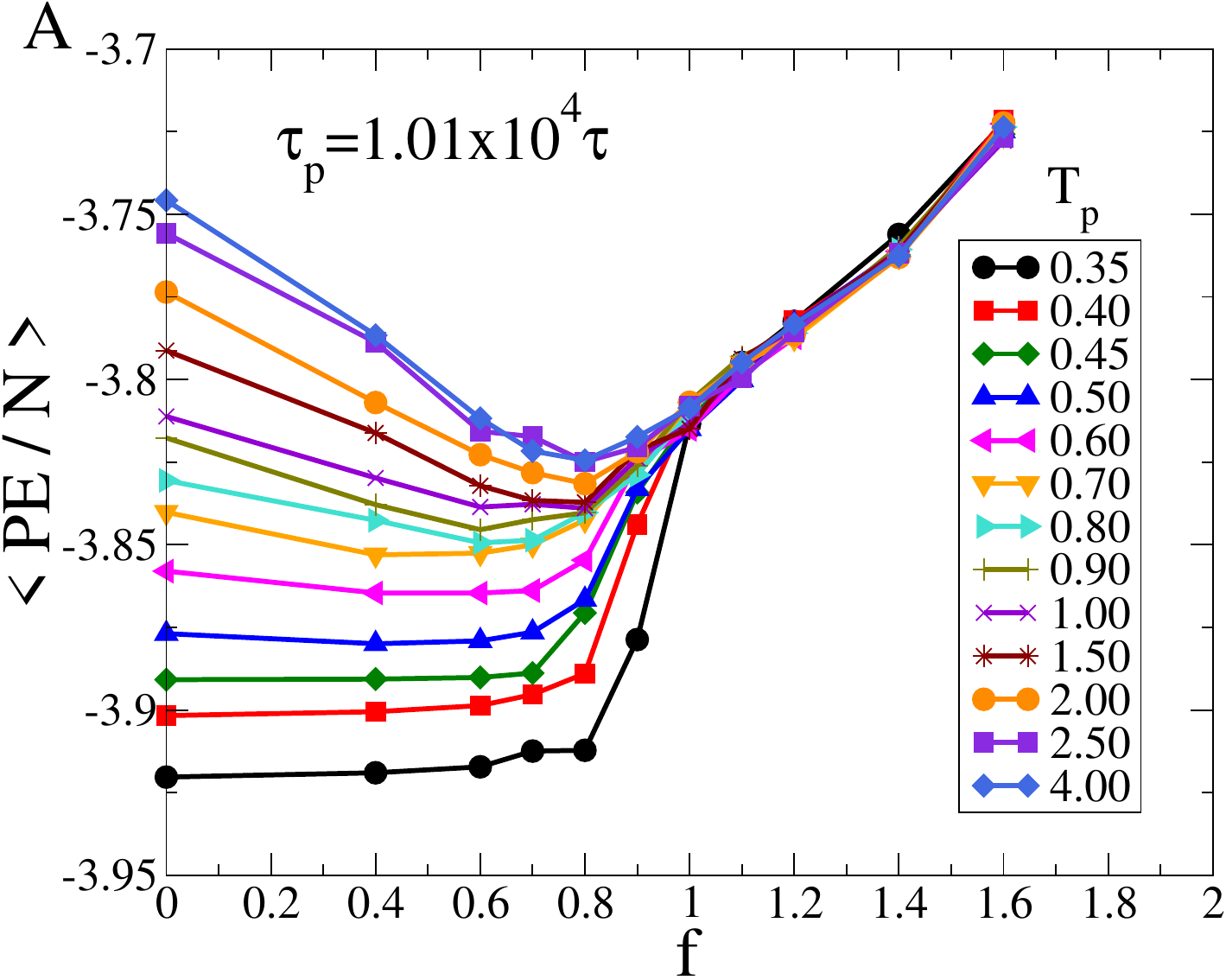}
\includegraphics[scale=0.25]{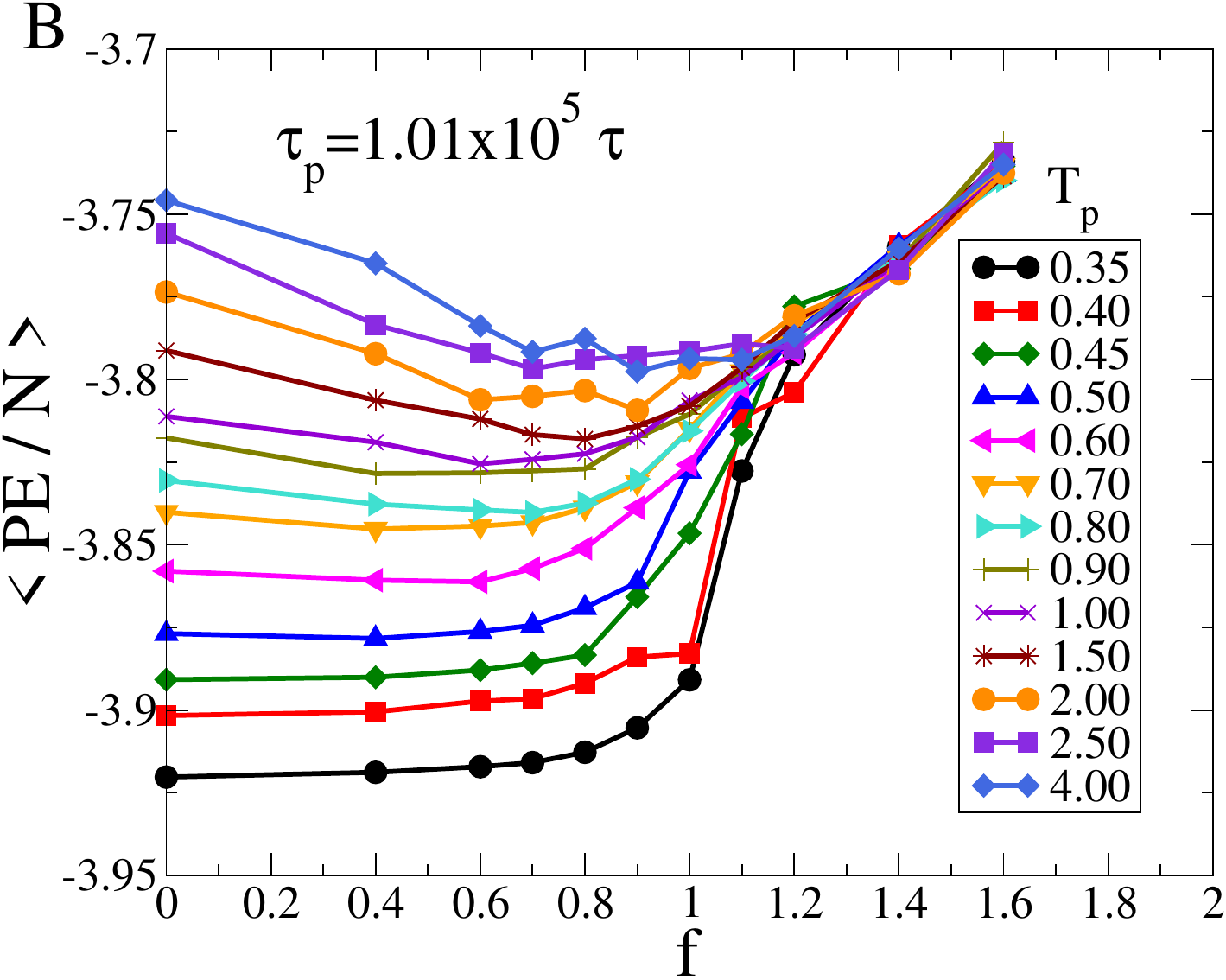}
\caption{Yielding phase diagram for $N=1000$ particles of the $65:35$ binary Lennard-Jones mixture in 2D subjected to active dynamics at $\tau_p= 1.01\times 10^4\tau$ and $\tau_p= 1.01\times10^5 \tau$ (right).}
\label{fig:E_v_f_difftaup}
\end{figure}

In Fig.~\ref{fig:E_v_f_difftaup} we show the full yielding phase diagram with active dynamics at $\tau_p= 1.01\times 10^4\tau$ and $\tau_p= 1.01\times10^5 \tau$. 

\subsection*{Parameteric dependence of change in potential energy, $\Delta E$, on strain step $\Delta \gamma$}
The alignment of instantaneous particle velocities with the respective directions of active forcing yield a measure of the instantaneous strain rate as in Eq.~\ref{eq:active_strain_rate_SI} below~\cite{During2021}. In systems driven by  active forces or using a shear stress, the strain rate response changes drastically across the yield point~\cite{xu2021yielding,During2021}. We define the instantaneous strain rate from the alignment of velocities with the active force direction as 
\begin{align}
    v_{par}^{act} &= \frac{1}{N} \left  \langle \sum\limits_{i=1}^{N}\mathbf{v_i(t)}\cdot\mathbf{n_i(t)} \right \rangle_t \nonumber \\
    \dot \gamma_{act}  &= \frac{12\sqrt{N}}{L} v_{par}^{act}.
    \label{eq:active_strain_rate_SI}
\end{align}
From the instantaneous strain rate, $\dot \gamma_{act}(t)$, we  define an instantaneous step $\Delta \gamma = \dot \gamma_{act} \Delta t$ and consider the change in potential energy with respect to it. We assume a linear parametric dependence of $\Delta E$ on $\Delta \gamma$ of the form $y = \beta x + \xi$, where $\xi$ is a noise term, and identify the slope $\beta$, taken as the Pearson correlation times the ratio of the standard deviations, $\sigma_{\Delta \gamma}/\sigma_{\Delta E}$. 
As mentioned in the main text, this is an approximate procedure that provides estimates of the stress using the response of the change in potential energy to changes in the measured strain step. As shown in Fig.~1 (d) of the main manuscript, and discussed below, such an approximation is found to give estimates in reasonable agreement close to the maximal strain $\gamma_{max}$ when used in the context of cyclic shear simulations at finite rate. 

% \clearpage
\subsection*{Development of non-zero strain rate in active yielding}
\begin{figure}[hb!]
\centering
\includegraphics[scale=0.3]{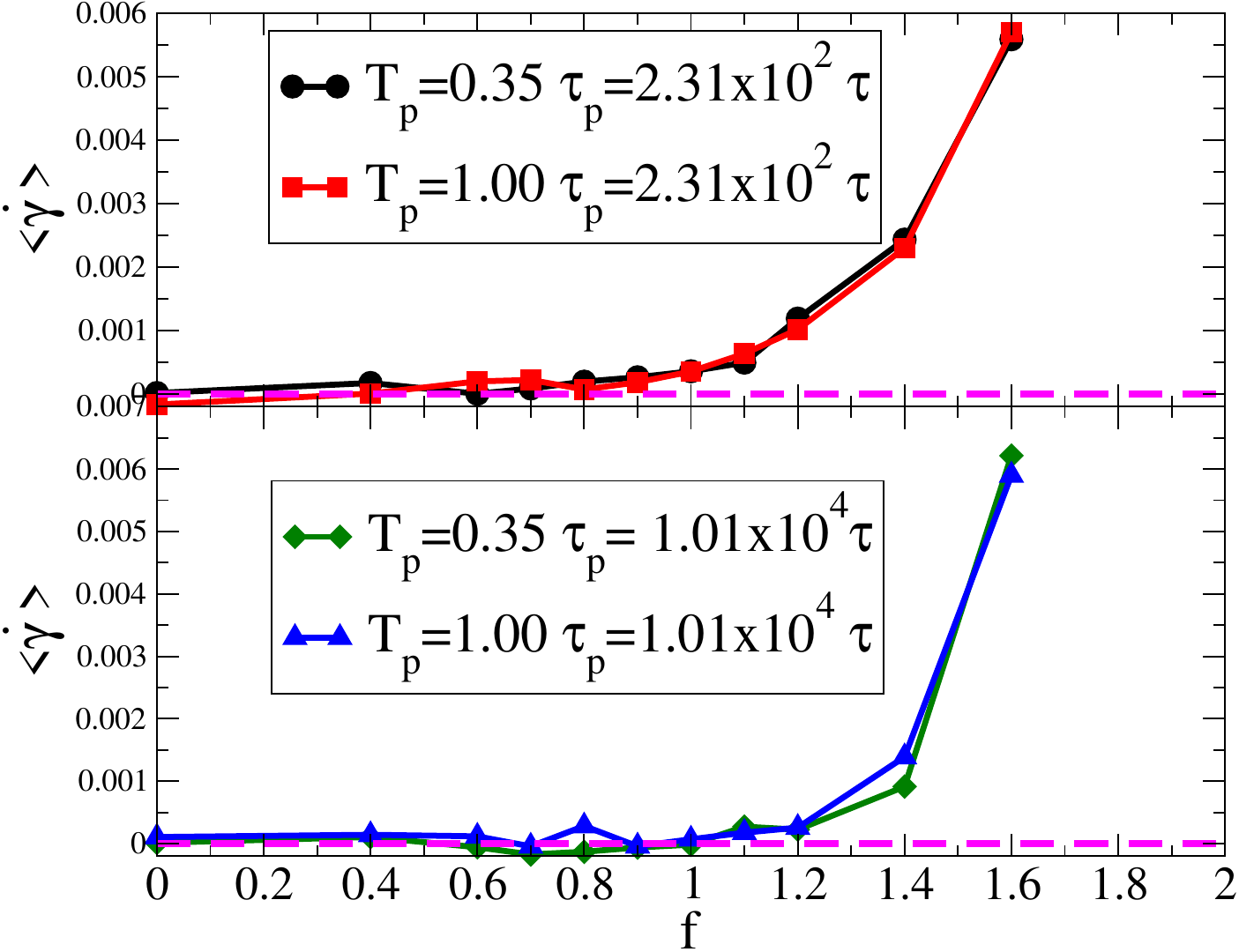}
\caption{We compute the resultant strain rate at different values of the active force, $f$, using the definition in Eq.~\ref{eq:active_strain_rate_SI}. The strain rate fluctuates around $0$ at all pre-yield values of force, subsequently rising non-linearly beyond the yield point ({\it Top}). The onset of non-zero strain rate shifts with the yield point at higher persistence time, $\tau_p$, ({\it Bottom}). }
\label{fig:strain_rate}
\end{figure}

In Fig.~\ref{fig:strain_rate}, we show the average measured strain rate in the steady state condition for differently annealed samples subjected to active driving at two values of the persistence time, $\tau_p$. We find that the strain rate is approximately zero below the yield value of the active force and increases monotonically beyond. The point at which this change in behaviour occurs shifts with the yield point as $\tau_p$ is changed. In both cases, the increase in strain rate occurs for $f$ at the yielding value identified from Fig.~1 in the main manuscript, instead occurring at slightly larger $f$.

\subsection*{Calculating stress for cyclic shear from the parametric dependence of $\Delta E$ on $\Delta \gamma$}
In Fig.~1 (d) of the main text, we show the dependence of the maximum stress on $\gamma_{max}$ in the case of cyclic shear, where the stress is computed by finding the slope of the best linear fit to the scatter of $\Delta E$ vs $\Delta \gamma$ from cyclic shear simulations at constant shear rate. The velocity stress component, $\sum\limits_{i=1}^{N} \delta v_x^i \delta v_y^i$ (the $\delta$ represents change with respect to any global flow), is subsequently added. This stress value is compared to that obtained directly from the virial stress tensor for $\gamma$ near $\gamma_{max}$.
More specifically, we first consider a reference $\gamma$ close to $\gamma_{max}$ ($\gamma_{max}-10^{-4}$ and note the energy $E_{ref}$. We then consider the energy of configurations within a neighbourhood of $\Delta \gamma = 10^{-4}$ of $\gamma_{ref}$ (identified from the strain rate multiplied with time windows $\Delta t$) and store the corresponding $\Delta \gamma$ and $\Delta E = E - E_{ref}$. The parametric dependence of $\Delta E$ and $\Delta \gamma$ is used as in active systems to determine the slope. Such a procedure is limited in this case to conditions where the strain step is very small, such that plastic rearrangements and other non-affine motion does not obscure the dependence of $\Delta E$ on the strain step.
The data of $\Delta E$ vs $\Delta \gamma$ is aggregated over $10$ steady state cycles from $16$ independent simulations.

\subsection*{Diffusivity beyond the yield point}
\begin{figure}[ht!]
\centering
\includegraphics[scale=0.25]{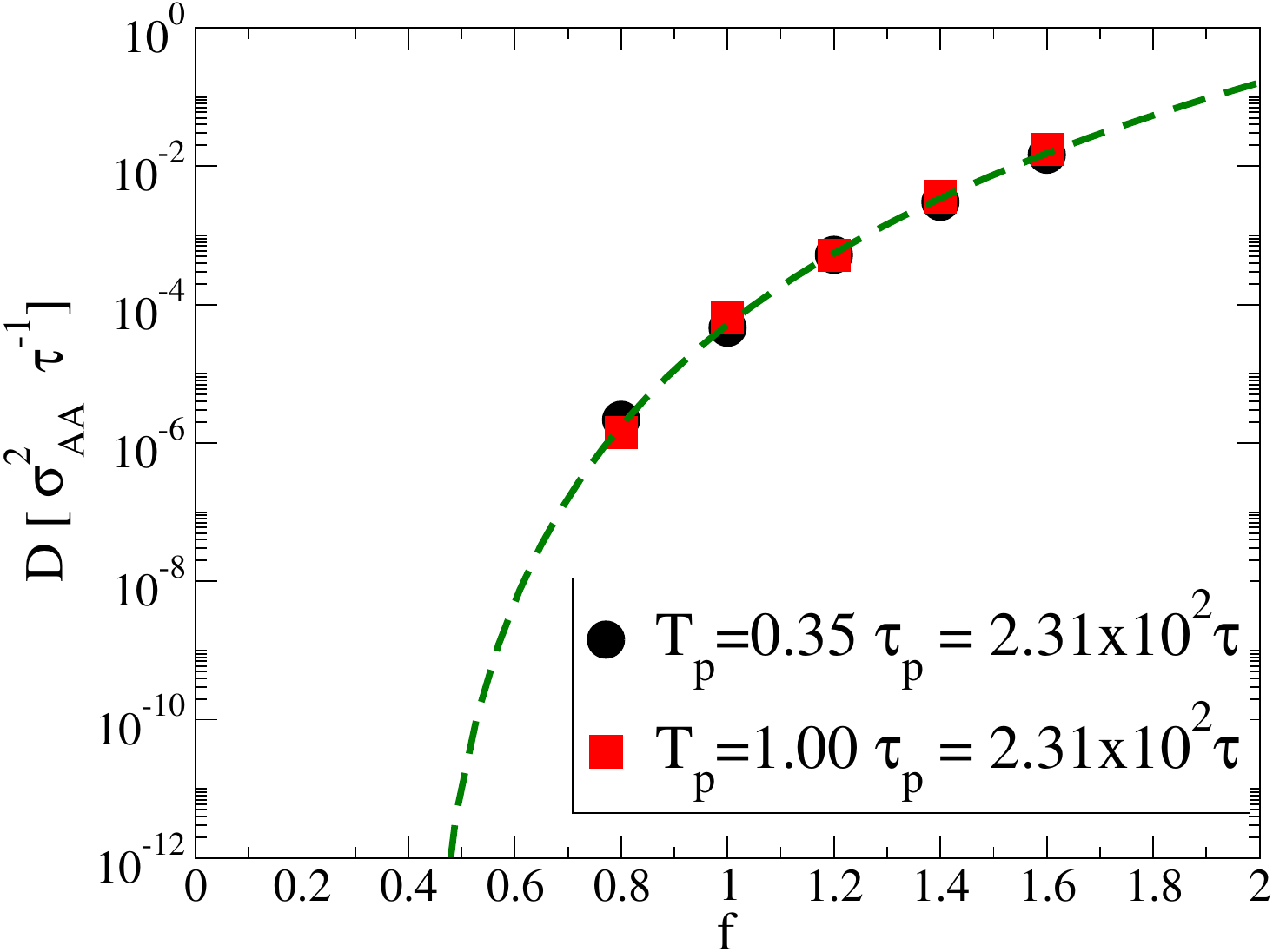}
\caption{Diffusivity obtained from the mean squared displacement data in Fig.~2 of the main manuscript shown for cases where the MSD shows at or near linear scaling with time.  }
\label{fig:diffu}
\end{figure}

In Fig.~\ref{fig:diffu}, we show the diffusivities extracted from the mean squared displacements shown in Fig.~2 of the main manuscript. Data is shown for trajectories simulated with large active forces displaying a diffusive regime where the scaling of the MSD with time is close to, or exactly linear. For both the well-annealed and poorly annealed cases, a free fit of the MSD yields an exponent less than $0.1$ in the pre-yield regime where the MSD is nearly flat. For the case of $T_p=1.00$ and $f=0.7$, the measured exponent is close to $0.8$, which we interpret as reflecting the difficulty of reaching the steady state in this case. For the case of $T_p=1.00$ and $f=0.7$ (not shown), the measured exponent is close to $0.8$, which we interpret as reflecting the difficulty of reaching the steady state in this case.
A best fit through the aggregated data in Fig.~\ref{fig:diffu} reveals that the diffusivity vanishes at $f<0.5$. However, the data in Fig.~1 and Fig.~2 in the main text show that the system remains in an absorbing state up to a higher value of $f=0.6$, suggesting a discontinuity in the range between $f = 0.6$ and $0.8$.

% \clearpage
\FloatBarrier
\subsection*{Energy relaxation and timescale to steady state}
\begin{figure}[htb!]
\centering
\includegraphics[scale=0.22]{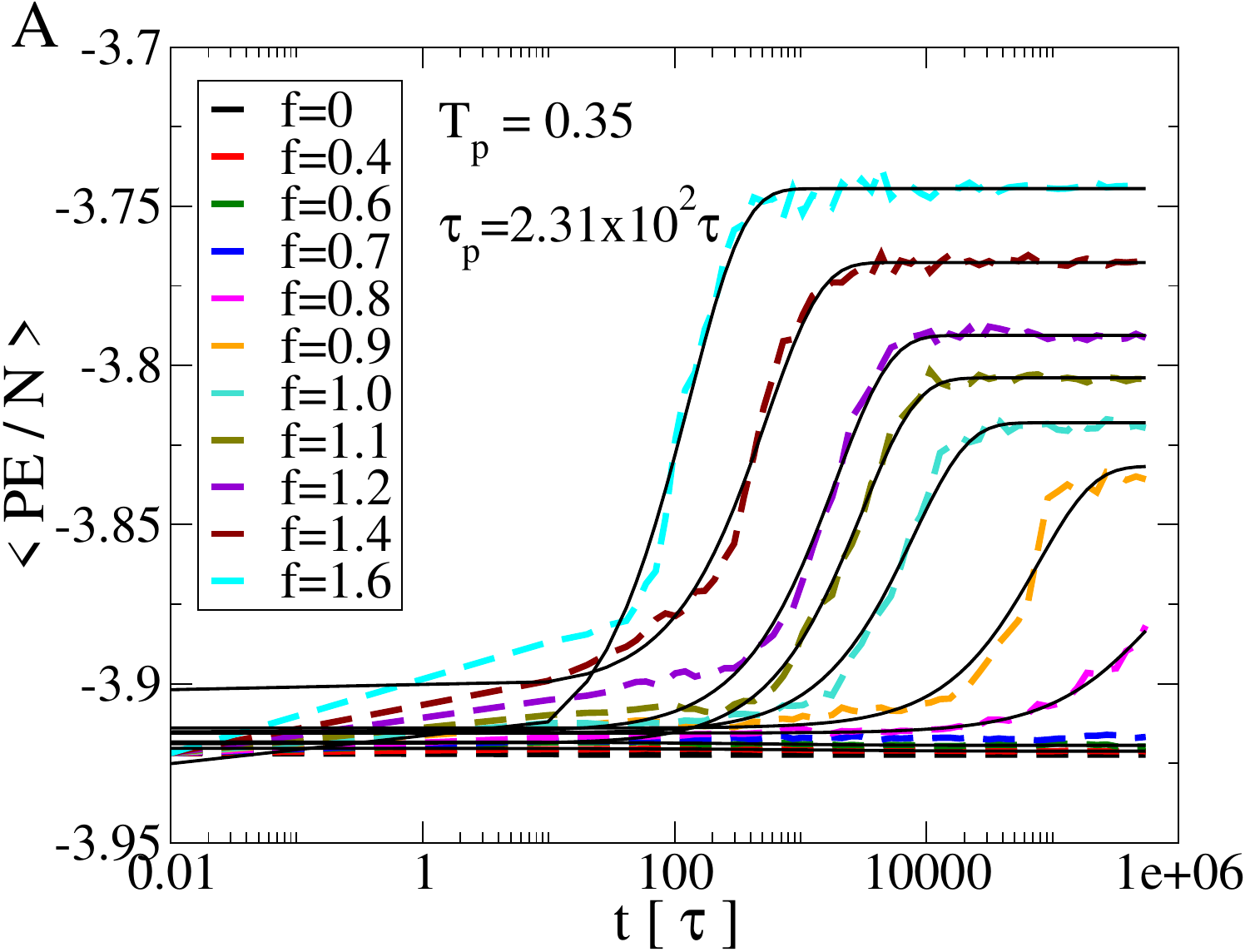}
\includegraphics[scale=0.22]{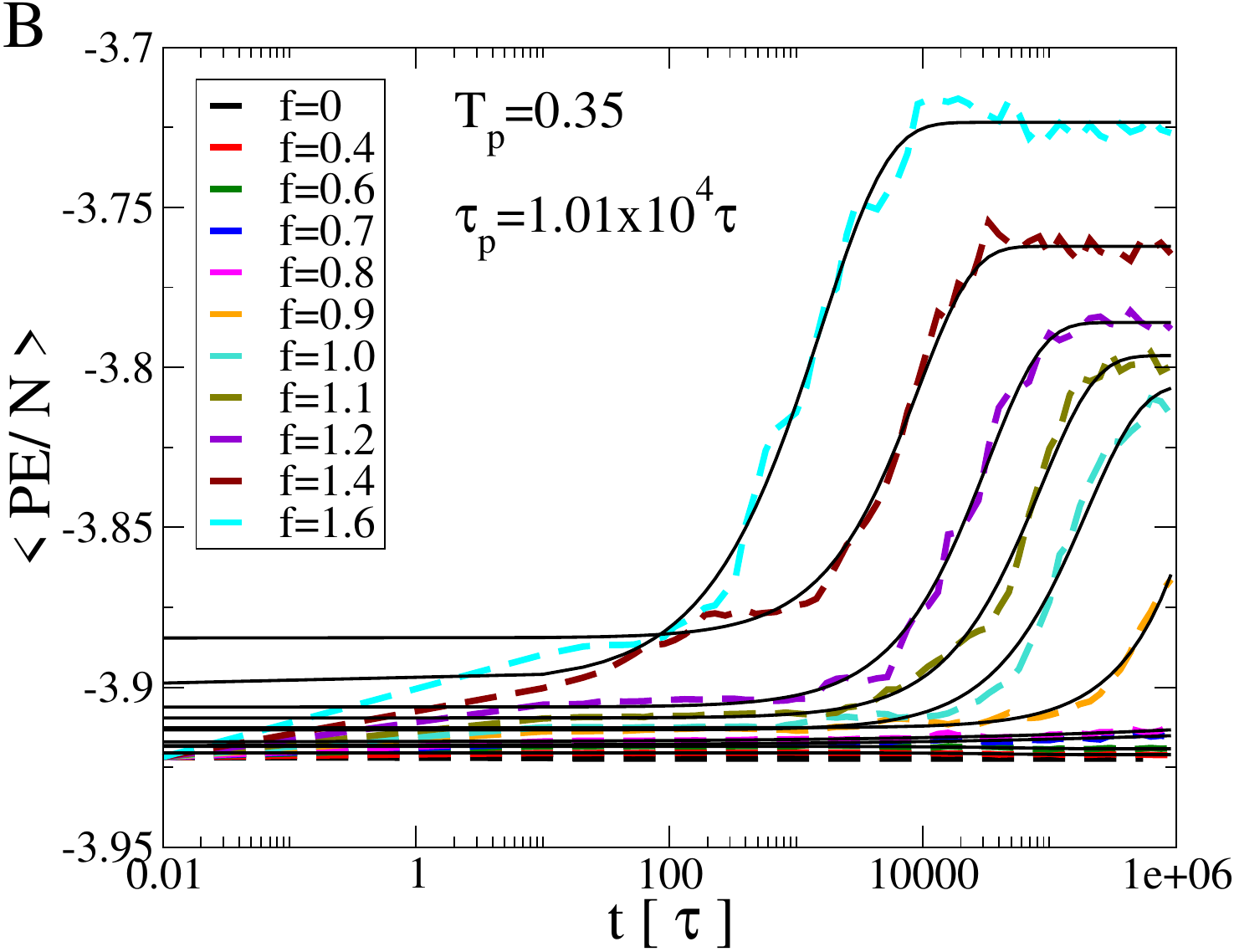}
\includegraphics[scale=0.22]{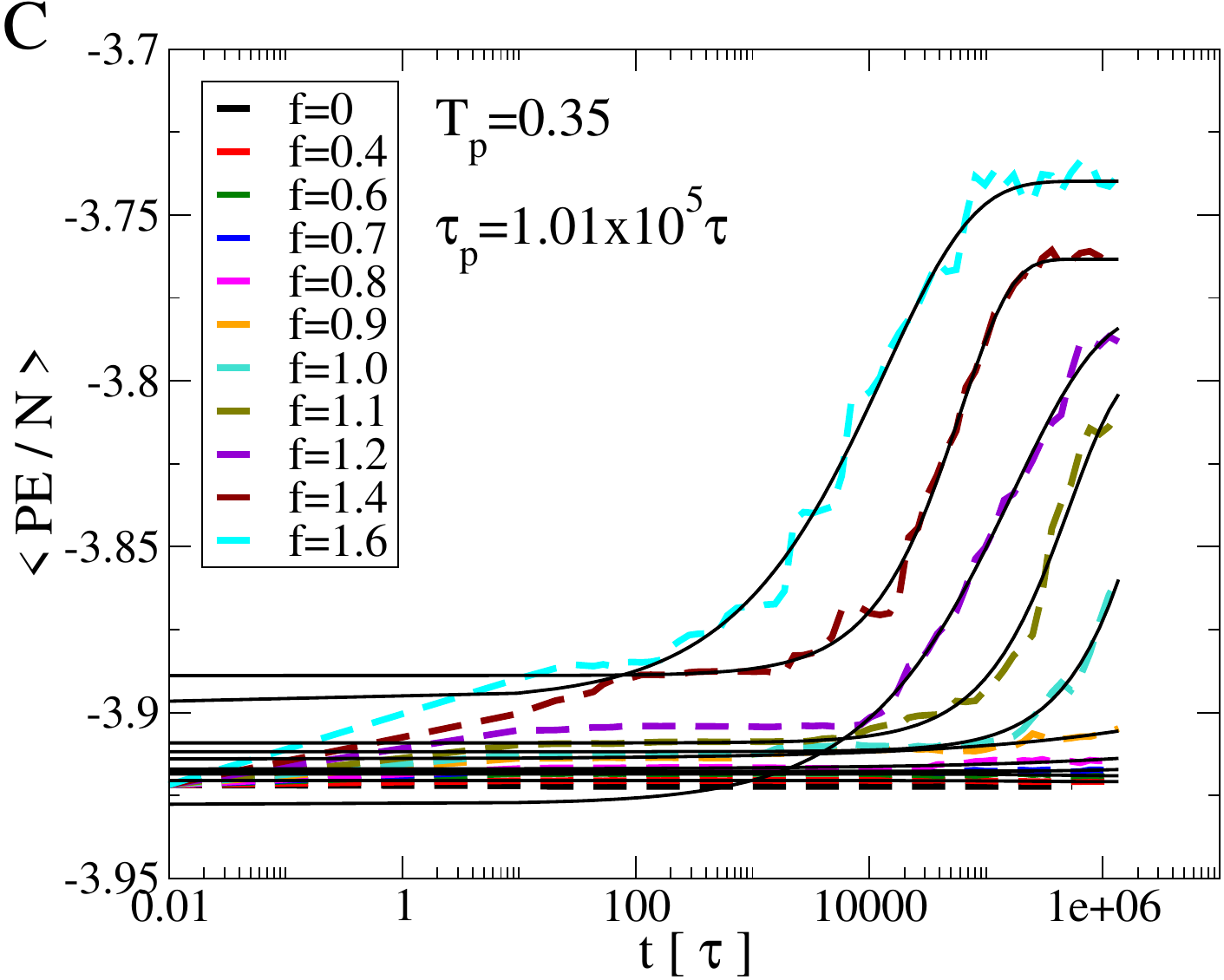}
\caption{Potential energy vs time, averaged over $8$ independent simulations, starting from a well-annealed sample at $T_p=0.35$ at $3$ values of the persistence time, $\tau_p$. Data are obtained from active dynamics simulations performed with a timestep of $dt=0.01$. Dashed lines are stretched exponential fits to the data.}
\label{fig:E_v_timeTp35}
\end{figure}

\begin{figure}[htb!]
\centering
\includegraphics[scale=0.22]{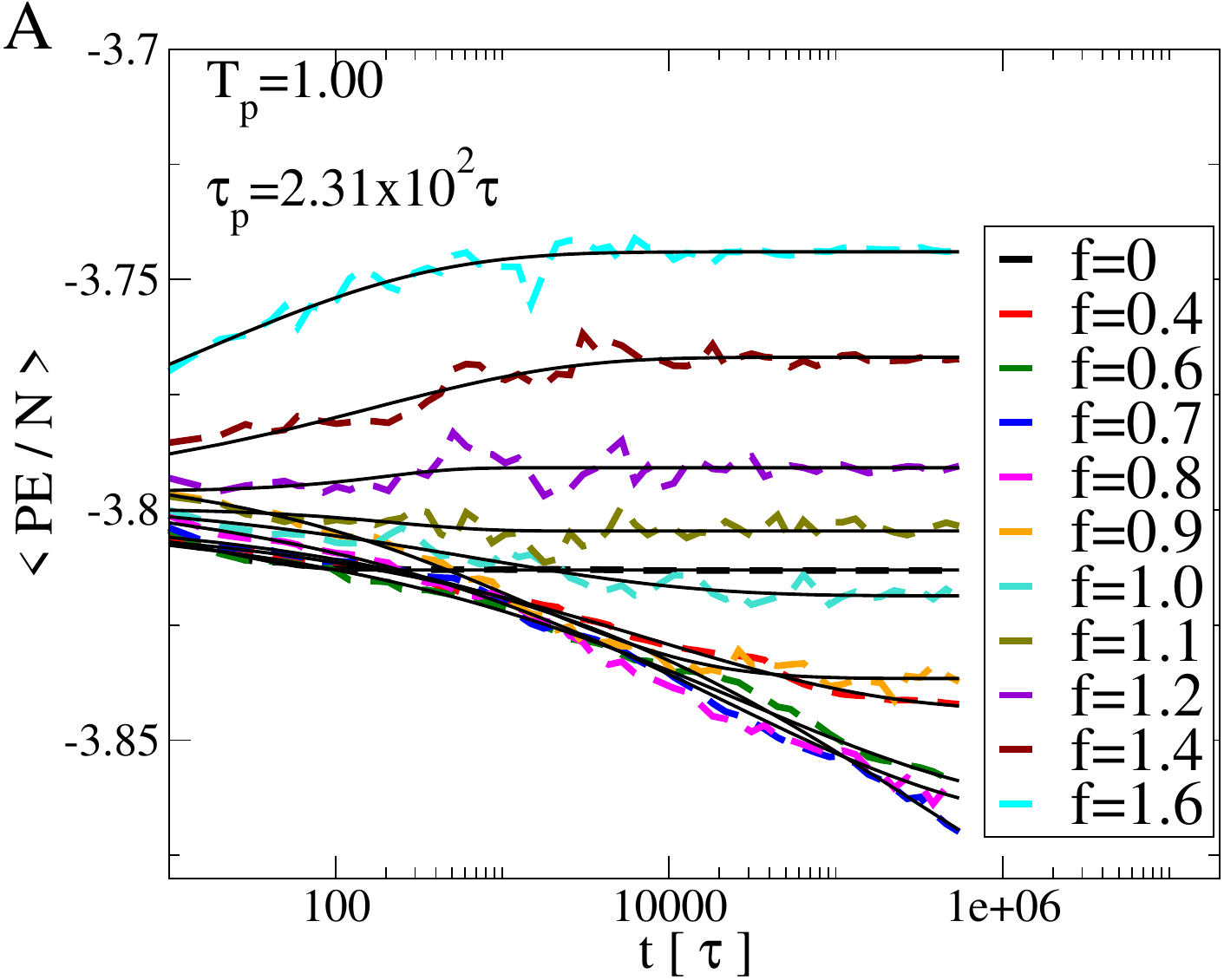}
\includegraphics[scale=0.22]{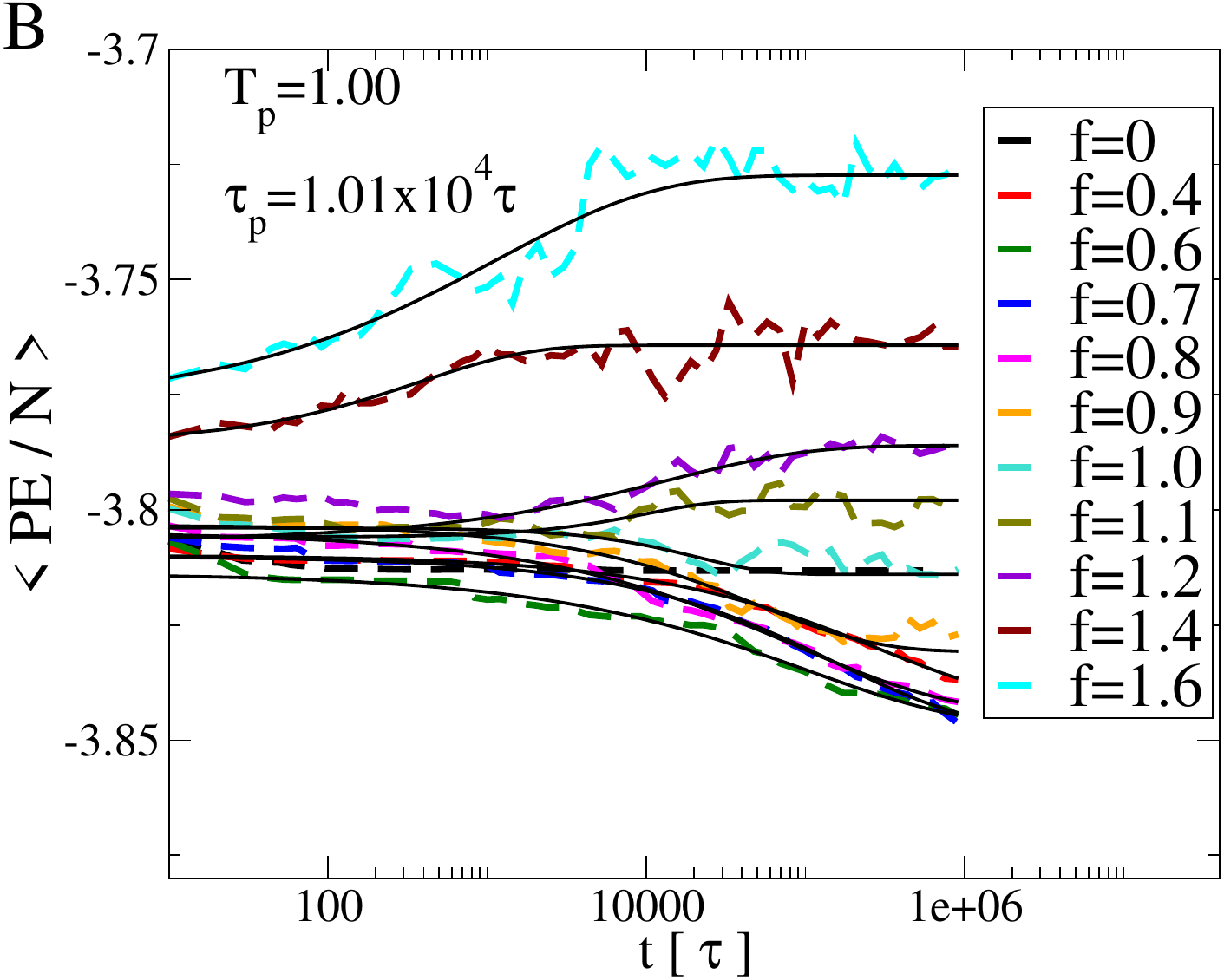}
\includegraphics[scale=0.22]{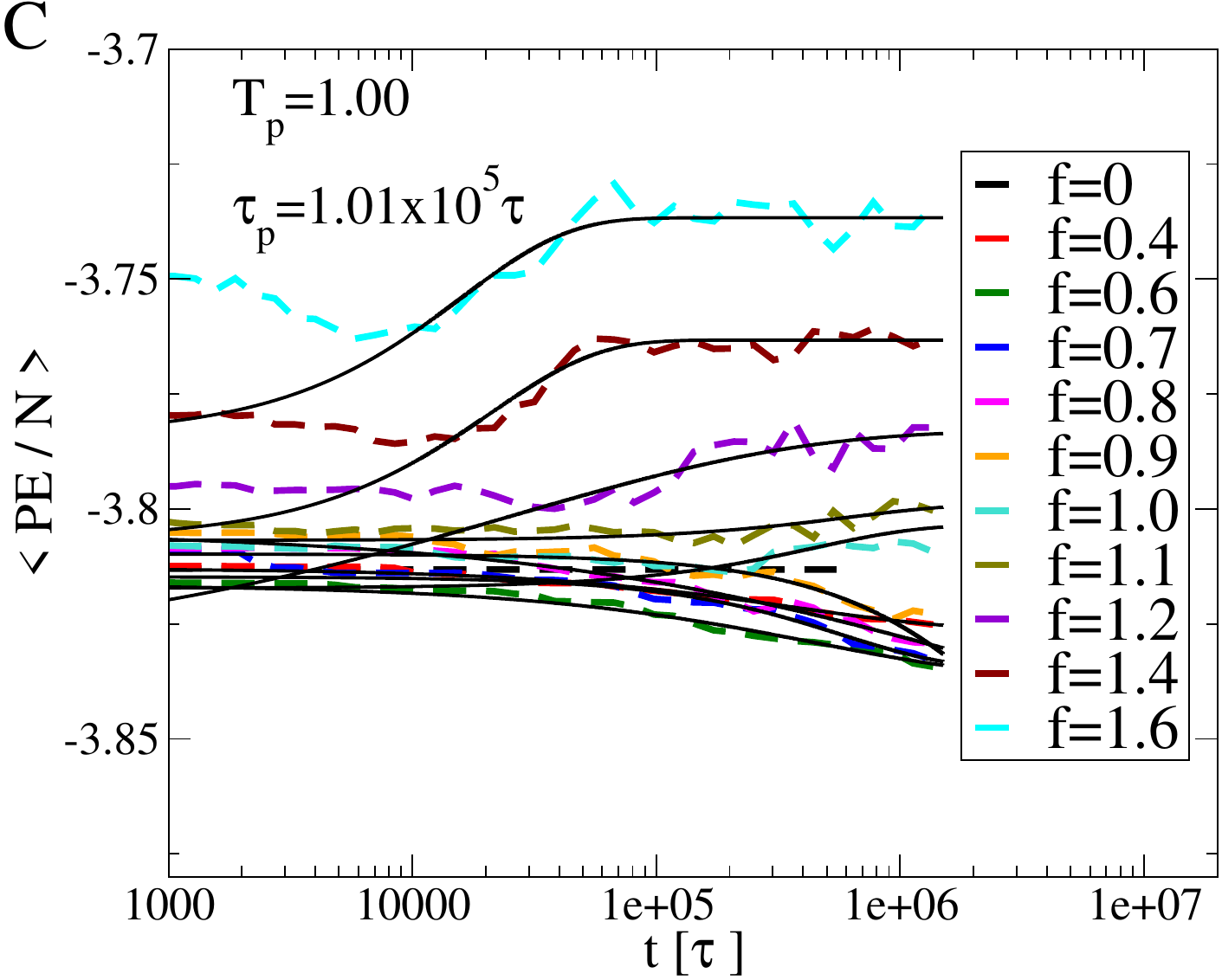}
\caption{Potential energy vs time, averaged over $8$ independent simulations, starting from a poorly-annealed sample at $T_p=1.00$ at $3$ values of the persistence time, $\tau_p$. Data are obtained from active dynamics simulations performed with a timestep of $dt=0.01$. Dashed lines are stretched exponential fits to the data.}
\label{fig:E_v_timeTp1}
\end{figure}

We show the time series of average potential energy per particle over time for $8$ independent trajectories simulated with different values of the active force. The energy vs time curves are fit to stretched exponential fits as shown in Fig.~\ref{fig:E_v_timeTp35} for configurations prepared at $T_p=0.35$ and in Fig.~\ref{fig:E_v_timeTp1} for configurations prepared at $T_p=1.00$.
The fitting function has the following form:
\begin{equation}
    \langle E(t) \rangle = E_{0} + \Delta E \exp(- (t/\tau)^{\beta}),
    \label{eq:stretched_exp}
\end{equation}
controlling for $E_0$ and $\Delta E$ based on the bounds in the trajectory. In each case, the region used for the fit includes the time interval immediately preceding the sharp change that signals the transition and the steady state reached after the transition (see for example the case of $T_p=1.00$, $\tau_p=1.01\times 10^5\tau$ and $f \geq 1.2$, where the system undergoes an initial relaxation before transitioning to the fluidised state).
Simulations performed with $f$ values close to the transition point ($f=0.8$ in Fig.~\ref{fig:E_v_timeTp35} (A), $f=0.9$ in Fig.~\ref{fig:E_v_timeTp35} (B) and $f=1.0$ in Fig.~\ref{fig:E_v_timeTp35} (C)) reach the steady state very slowly, as a result of which the observed final values of $\langle PE / N \rangle$, reported in Fig.~3 of the main manuscript, are apparently intermediate to the values in the pre-yield and the post-yield branches.
% \sscom{Add text about intermediate values.}

\subsection*{Comparison of timescale to steady state from stretched exponential fits and from first passage time identification}
\begin{figure}[htb!]
\centering
\includegraphics[scale=0.3]{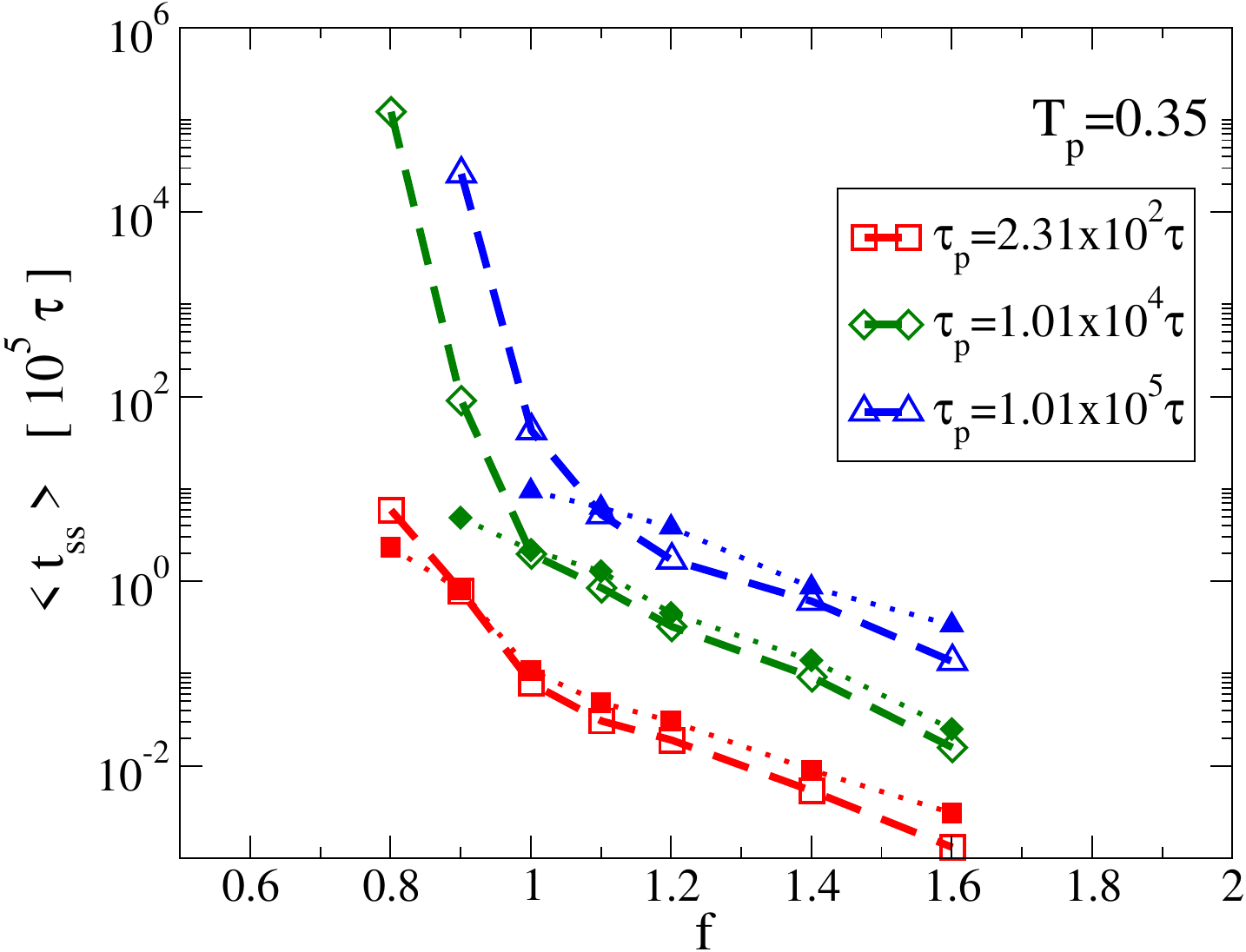}
\caption{Timescale to steady state shown at different values of $\tau_p$ for well-annealed samples at $T_p=0.35$ subjected to active dynamics at forces greater than the yielding force. The hollow symbols with dashed lines are obtained from stretched exponential fits to the aggregate $\langle E \rangle$ vs time curves where the energy at each time point is the average over $8$ independent simulations. The solid symbols with dotted lines are obtained from identifying the average of the first passage time to the steady-state value of the potential energy $E_{ss}$ in each trajectory, where the average is over $8$ independent simulations.}
\label{fig:compare_tss}
\end{figure}

In Fig.~\ref{fig:compare_tss}, we show the average time to reach state obtained from two different procedures. In the first we consider the time series, $\langle E\rangle$ vs time, averaged over $8$ independent simulations and fitted a stretched exponential to the curve, to identify the relaxation timescale $\tau$ from Eq.~\ref{eq:stretched_exp}.
Next, we consider the average first passage time to the steady state energy $ < E_{ss}^i >_{t_{ss}}$ in each trajectory $i$. Here, the average is taken over a window of time $t_{ss}$ where the system is in its final steady state. The average of these first passage times is then taken over $8$ independent simulations. 

We note that for the poorly annealed case, trajectories undergo further slow annealing at small active forces, reminiscent of creep, after they enter into the neighbourhood of its final steady state. Likewise, for trajectories close to the transition point, the approach to steady state is extremely slow. This leads to the appearance of a shorter first passage time to steady state than that obtained from estimates of the relaxation timescale in a stretched exponential process, as seen in Fig.~\ref{fig:compare_tss}. While this may be ameliorated by improved estimates of the steady state energy, $\langle E_{ss}^i \rangle_{t_{ss}}$, reached by each trajectory, we opt for the stretched exponential procedure in our work in order to identify an asymptotic steady state energy. Trajectories subjected to active force significantly larger than the critical force yield similar estimates for $\langle t_{ss} \rangle$ from these two approaches.
\FloatBarrier

\subsection*{Comparison with passive dynamics}
\begin{figure}[ht!]
    \centering
    \includegraphics[scale=0.25]{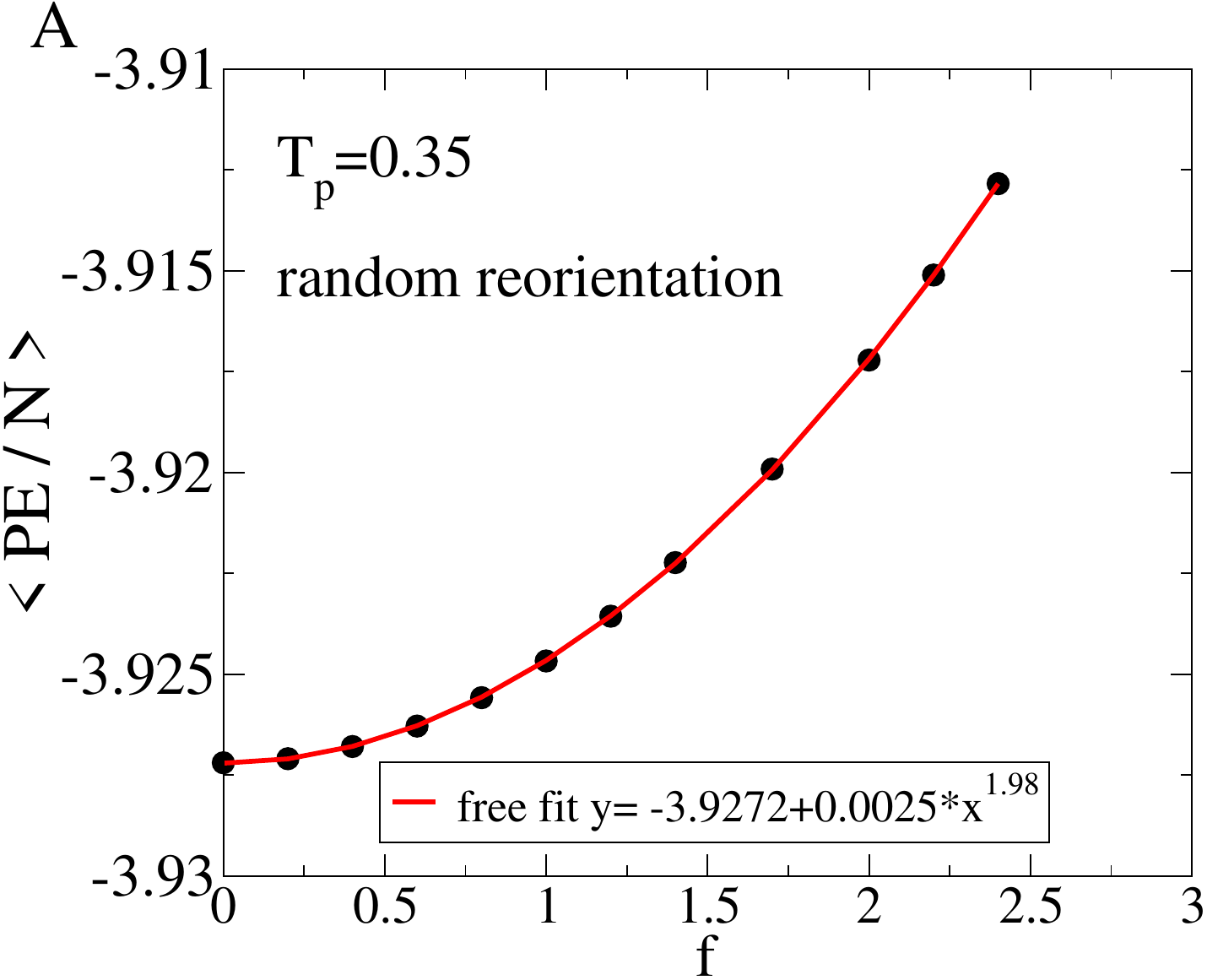}
    \includegraphics[scale=0.25]{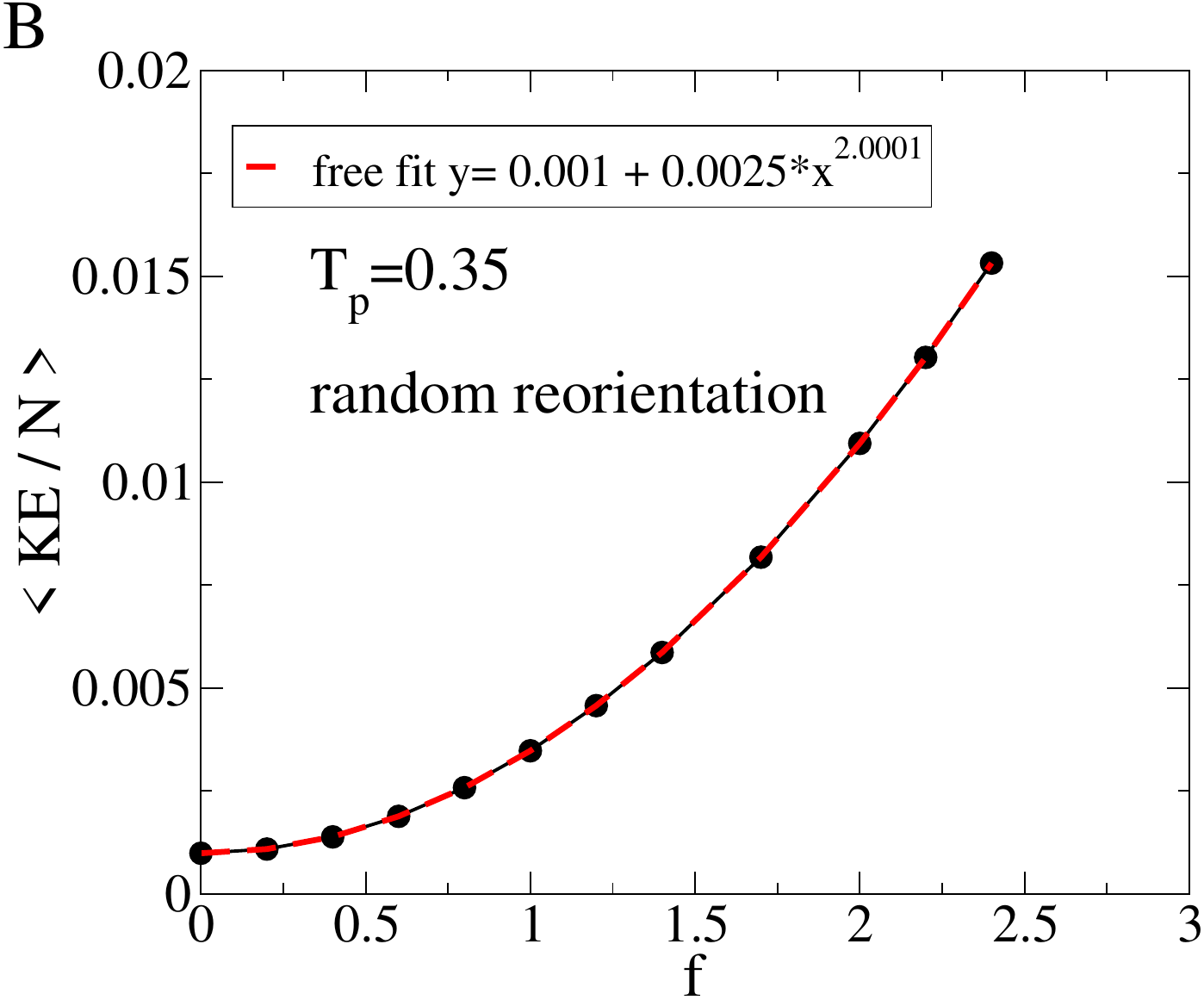}
    \caption{No discontinuity in energy for zero persistence active force with a quadratic dependence of potential energy indicative of increased thermal motion. Quadratic dependence of kinetic energy on active force. }
    \label{fig:zero_persist}
\end{figure}

We address the question of whether purely passive dynamics at comparable temperatures can lead to fluidisation in the same way that active dynamics does. In order to do this, we consider the average kinetic energy per particle across different cases.
In Fig.~\ref{fig:zero_persist}, we show the average potential energy per particle for active dynamics where the active direction re-orients every timestep, $dt$, which we use to represent the limit of zero persistence where the direction of active force is delta-correlated in time. In this limit of small persistence, the active temperature is given by the expression $T_a = f^2 \tau_p/4$ in 2D and the effective temperature is given by $T_{eff} = T_{bath}+T_{a}$. This relation holds exactly in the small persistence time limit, as we see in Fig.~\ref{fig:zero_persist}. We reiterate here that the kinetic energies at finite persistence are considerably lower than that for the small persistence time case. One does not, however, expect the same expression for $T_a$ to hold at large $\tau_p$. It has been argued in the literature that $T_a$ goes as $A f^2 \tau_p/(1+ B \tau_p)$\cite{nandi2018random,mandal2022random}, but results reported in \cite{mandal2020extreme,mandal2022random} suggest that for the large persistence times we consider in the present manuscript, the above expression may not be valid, with the arrest line exhibiting non-monotonic dependence on $f$.

\begin{figure}[htb!]
\centering
\includegraphics[scale=0.3]{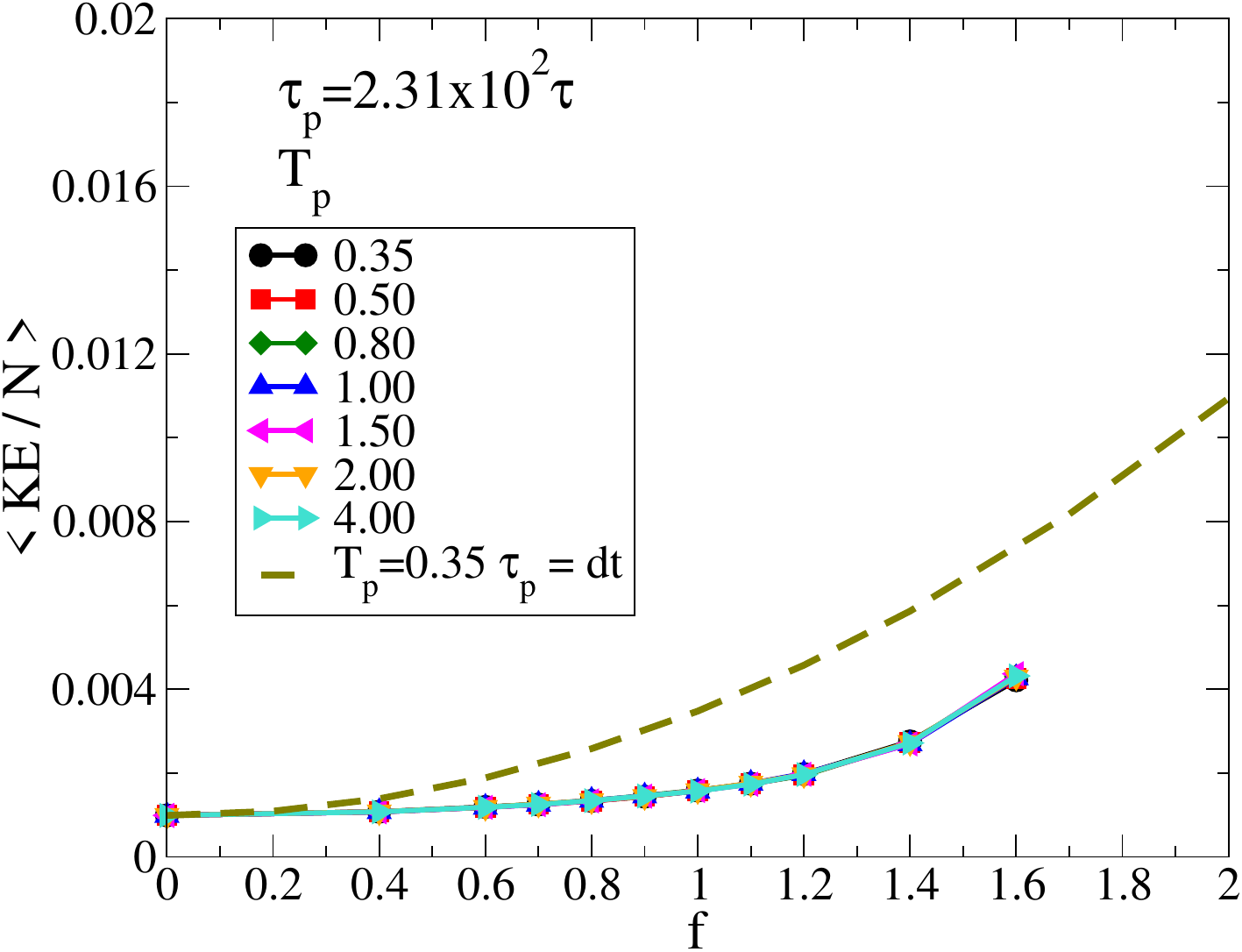}
\caption{Effective temperature from the kinetic energy per particle for differently annealed samples subjected to active forces at either finite persistence time, or at zero persistence for $T_p=0.35$.}
\label{fig:KE_v_f}
\end{figure}

\begin{figure}[htpb!]
\centering
\includegraphics[height=4.5cm,width=5.5cm]{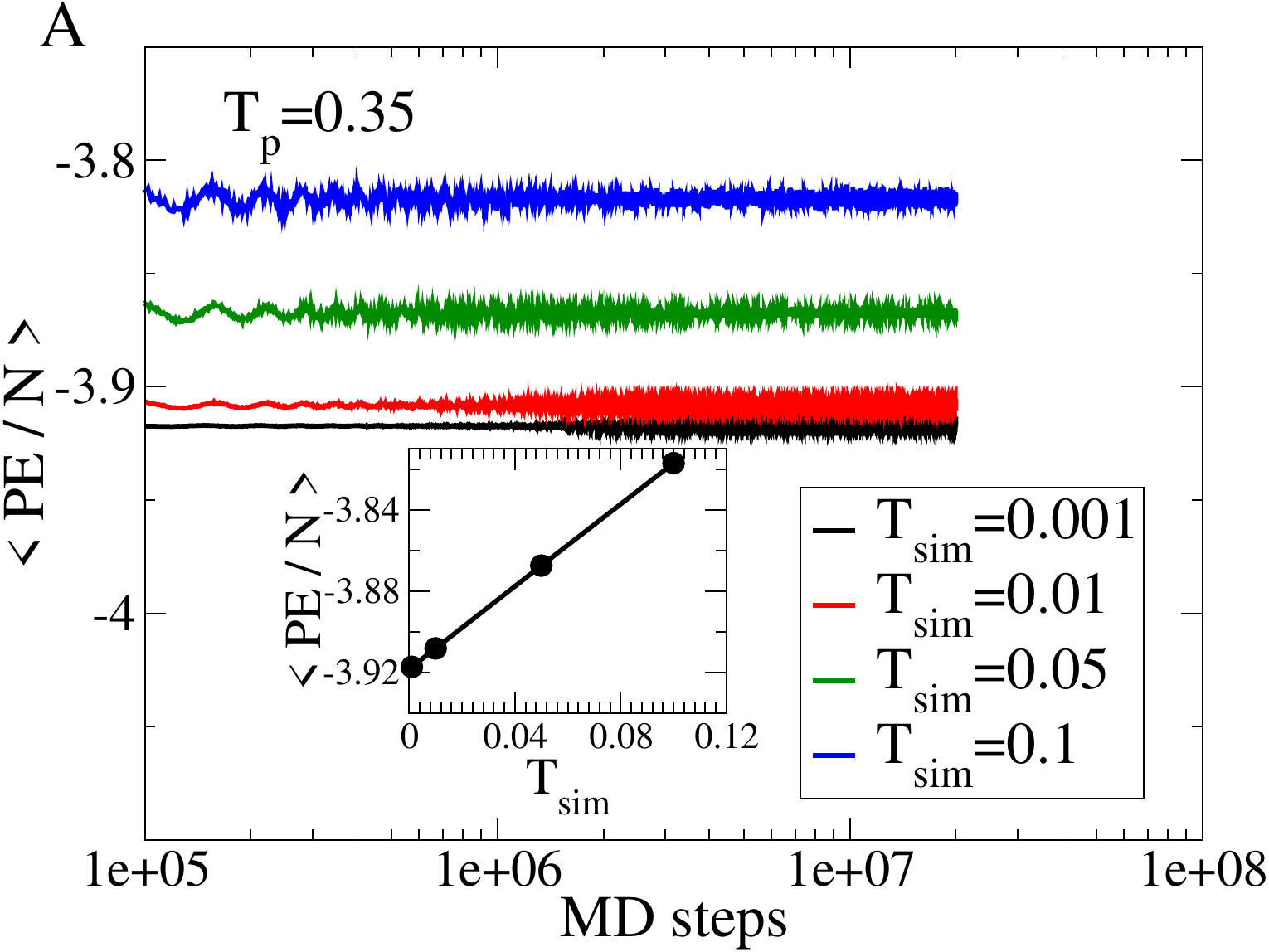}
\includegraphics[height=4.5cm,width=5.5cm]{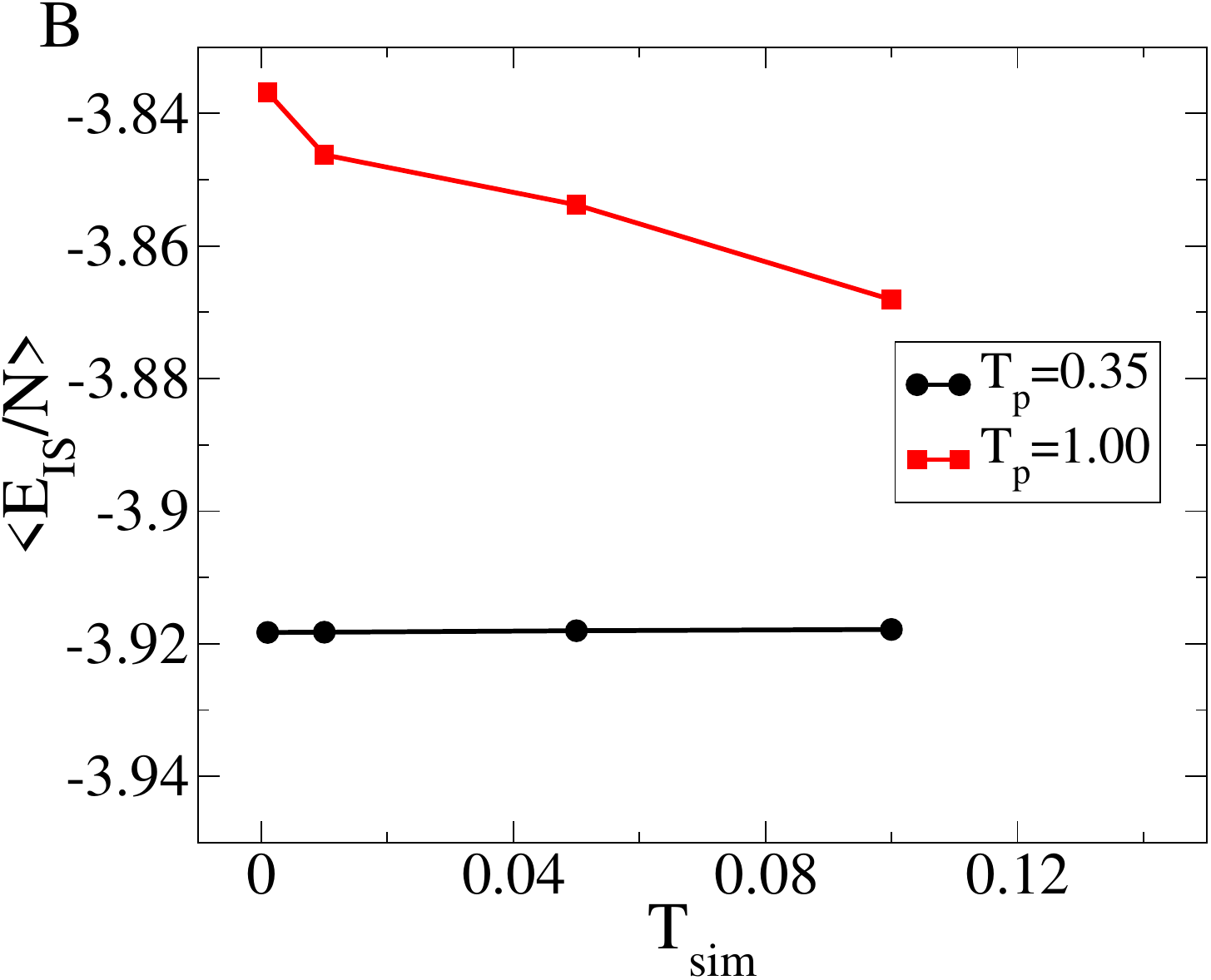}
\includegraphics[height=4.5cm,width=5.5cm]{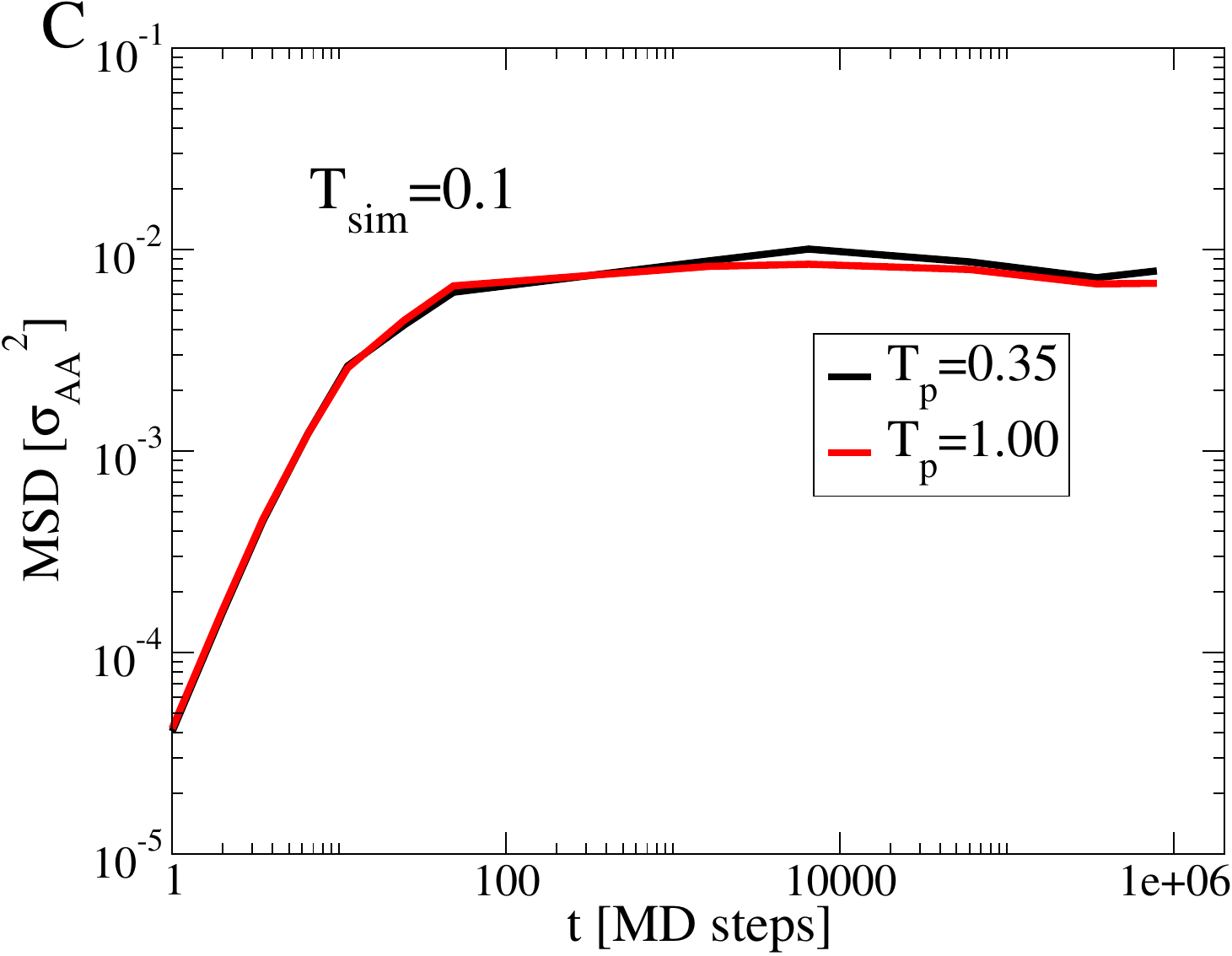}
\caption{Bi-disperse assemblies heated through the temperatures corresponding to the average kinetic energy of the actively driven systems. We perform NVT MD of the bidisperse assembly (see methods) and heat the differently annealed initial samples directly to target temperatures, $T_{sim}$, in the desired range. We find a linear dependence between potential energy and temperature (left). A steady state is reached in each case with the inherent structure energies being either constant, or showing annealing depending on whether we start from a high initial energy or a low initial energy (middle). The MSD obtained after ignoring a transient of $2\times10^7$ MD steps for the raw trajectories shows no diffusive behaviour at even the highest target temperature (right). }
\label{fig:heating}
\end{figure}

In Fig.~\ref{fig:KE_v_f} we show the average kinetic energy for simulations at different active force at finite and small $\tau_p$. The average kinetic energy per particle remains low in these cases, and this is underscored by the fact that it remains below the kinetic energy for the small persistence case, where the kinetic energy can be compared with an exact expression of the effective temperature.
The average kinetic energy per particle is higher in this low persistence case compared to that at finite persistence time.
In order to determine whether these, or comparable, temperatures are sufficient to induce fluidisation with passive dynamics, we subject differently annealed samples, i.e., different parent temperature $T_p$, to $4\times10^7$ MD steps of Langevin dynamics at constant volume at a range of simulation temperatures, $T_{sim}$, the highest of which ($T_{sim}=0.1$), exceeds the highest observed average kinetic energy ($\approx~0.01$) by almost an order of magnitude.
In Fig.~\ref{fig:heating} A, we show the potential energy per particle averaged over $4$ independent samples at different thermostat temperatures, $T_{sim}$. We find that a steady state is reached in each case with a linear dependence of the potential energy per particle on simulation temperature (inset of Fig.~\ref{fig:heating} A).
% Similarly, in Fig.~\ref{fig:heating} B, we show the corresponding data for $T_p = 1.00$. 
Examining the inherent structure energies per particle in each case shows a constant energy at even $T_{sim}=0.1$ for $T_p=0.35$, and thermal annealing for $T_p=1.00$, as shown in Fig.~\ref{fig:heating} B. Note that the highest simulation temperature, $T_{sim}$, exceeds the highest $KE~/~N$ in active simulations by a factor of $\sim 25$. Finally, we consider trajectories at the highest simulation temperature, $T_{sim}=0.1$, and observe the mean squared displacements for the two samples, finding that the trajectories are in an absorbing state, as shown in Fig.~\ref{fig:heating} C. Based on this, we conclude that the fluidisation observed in the actively driven simulations at finite persistence is not correlated with the temperature of the system reflected in the average kinetic energy per particle. The nontrivial scaling of the kinetic energy with active force $f$ and $\tau_p$ has been discussed in \cite{mandal2020extreme} and more generally, the departure from equilibrium-like behaviour for large persistence times in \cite{fodorPRL2016}.

\FloatBarrier
% \clearpage
\subsection*{Confinement boundary implementation}
\begin{figure}[htb!]
\centering
\includegraphics[trim=10 10 10 10,clip,width=5.5cm,height=4.5cm]{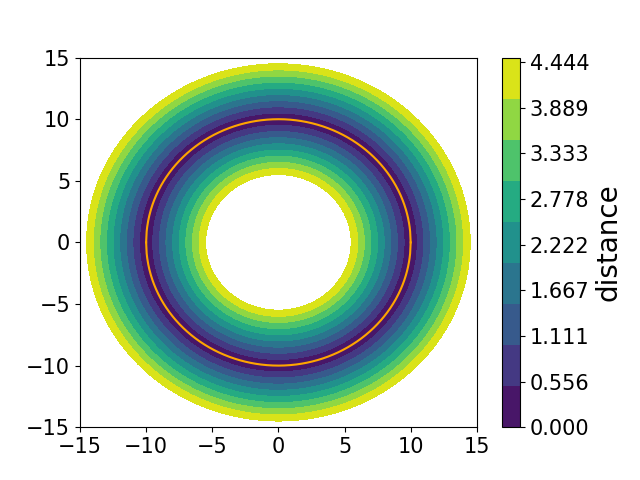}
\includegraphics[trim=10 10 10 10,clip,width=5.5cm,height=4.5cm]{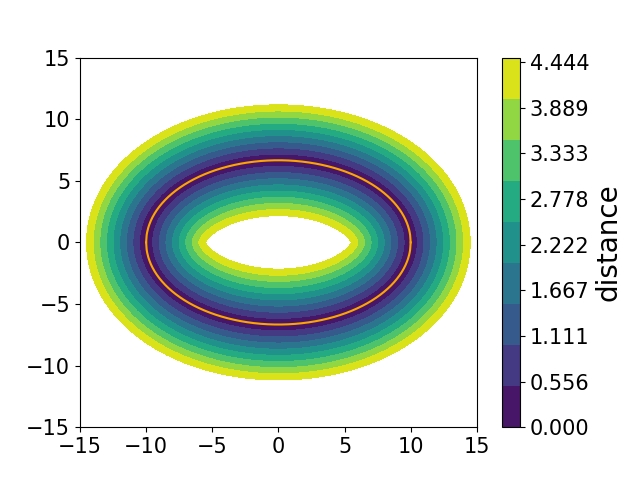}
\includegraphics[trim=10 10 10 10,clip,width=5.5cm,height=4.5cm]{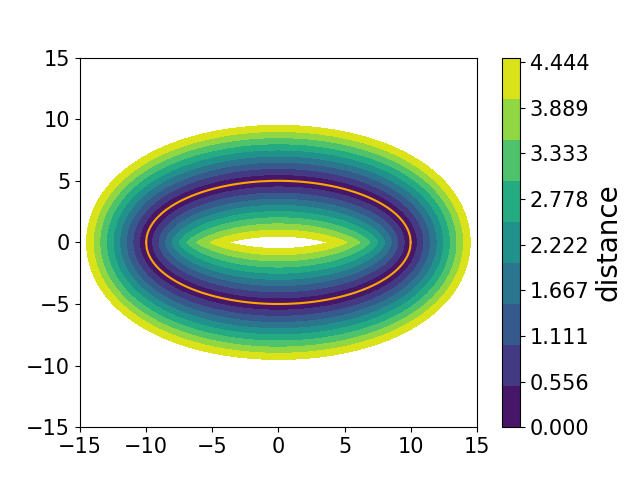}
\caption{Heat map of distance from the boundary of the ellipse, as computed by the optimisation procedure to identify the closest point on a given ellipse from a chosen query point.}
\label{fig:ellipse_distance}
\end{figure}

In order to compute the wall interaction, we need to identify the vector from the centre of any particle, $p$, to the closest point, $q$, on the elliptical confinement boundary. We follow an iterative procedure with the following steps~\cite{maisonobe2006quick,Chatfield2017}:
\begin{enumerate}
    \item Pick a random initial $a$ on the ellipse.
    \item Compute the distance $r_{pa}$ between $p$ and $a$.
    \item The circle of radius $r_{pa}$ centred at $p$ will intersect the ellipse at another point $b$.
    \item Identify the midpoint $a'$ between $a$ and $b$.
    \item Calculate $r_{pa'}$ and go to step 3.
\end{enumerate}
In order to follow this procedure, we describe the relevant details from~\cite{Chatfield2017} for completeness:
\begin{align}
    x &= a\cos(t) \nonumber \\
    y &= b\sin(t)
\end{align}
We can approximate the curvature of the ellipse at $a$ to a circle centred at $ev_x,ev_y$ with radius $|\omega_a|$. The centres of curvature of the ellipse are obtained using
\begin{align}
   ev_x &= \frac{(a^2 - b^2)}{a}\cos^3(t) \nonumber \\
   ev_y &= \frac{(b^2 - a^2)}{b}\sin^3(t)
\end{align}
Vectors from the centre of curvature to $a$ and to $a'$, $\omega_a$ and $\omega_{a'}$ are separated by an arclength $\Delta c$. We can write:
\begin{equation}
    sin \left (\frac{\Delta c}{|\omega_a|} \right )= \frac{\omega_a \times \omega_{a'}}{|\omega_a||\omega_{a'}|}
\end{equation}
One can then write
\begin{align}
    \frac{dc}{dt} &= \sqrt{\left ( \frac{dx}{dt} \right)^2 - \left ( \frac{dy}{dt} \right)^2} \nonumber \\
    \frac{\Delta c}{\Delta t} &= \sqrt{a^2\sin^2(t) + b^2\cos^2(t)}
\end{align}
We can then identify $\Delta t$ as
\begin{equation}
    \Delta t = \frac{\Delta c}{\sqrt{a^2 + b^2 - x^2 - y^2}}
\end{equation}
Finally, one can find the coordinates of $a'$ as 
\begin{align}
x' &= a\cos(t+\Delta t) \nonumber \\
y' &= b\sin(t+\Delta t)
\end{align}
Once $a'$ has been identified, we can iterate the procedure till convergence is achieved. In Fig.~\ref{fig:ellipse_distance}, we show heatmaps of the distance computed from ellipses of different aspect ratio, including the symmetric circular case.

% \clearpage

\subsection*{Sample preparation in confinement}
\begin{figure}[htb!] 
\centering
\includegraphics[scale=0.28]{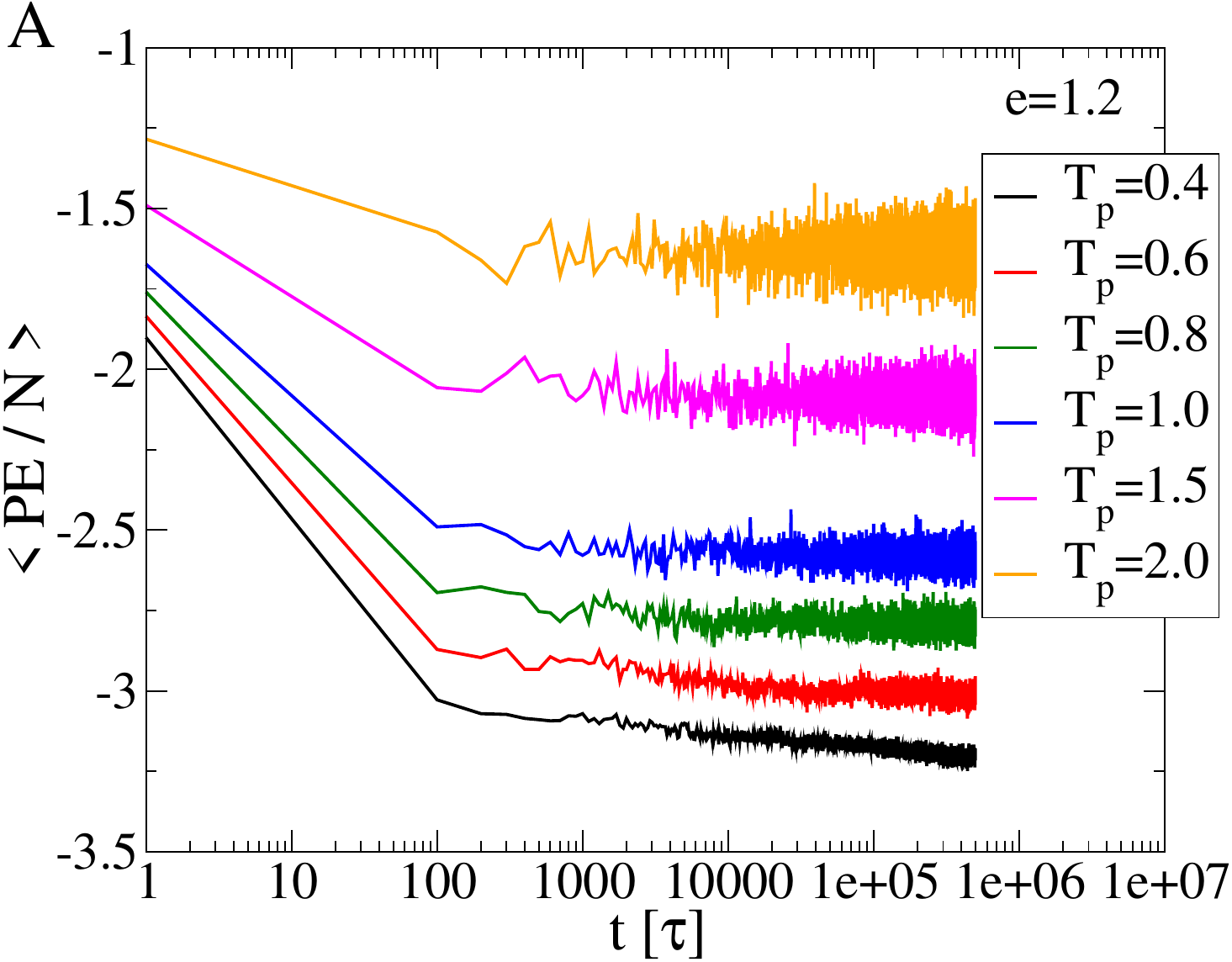}
\includegraphics[scale=0.28]{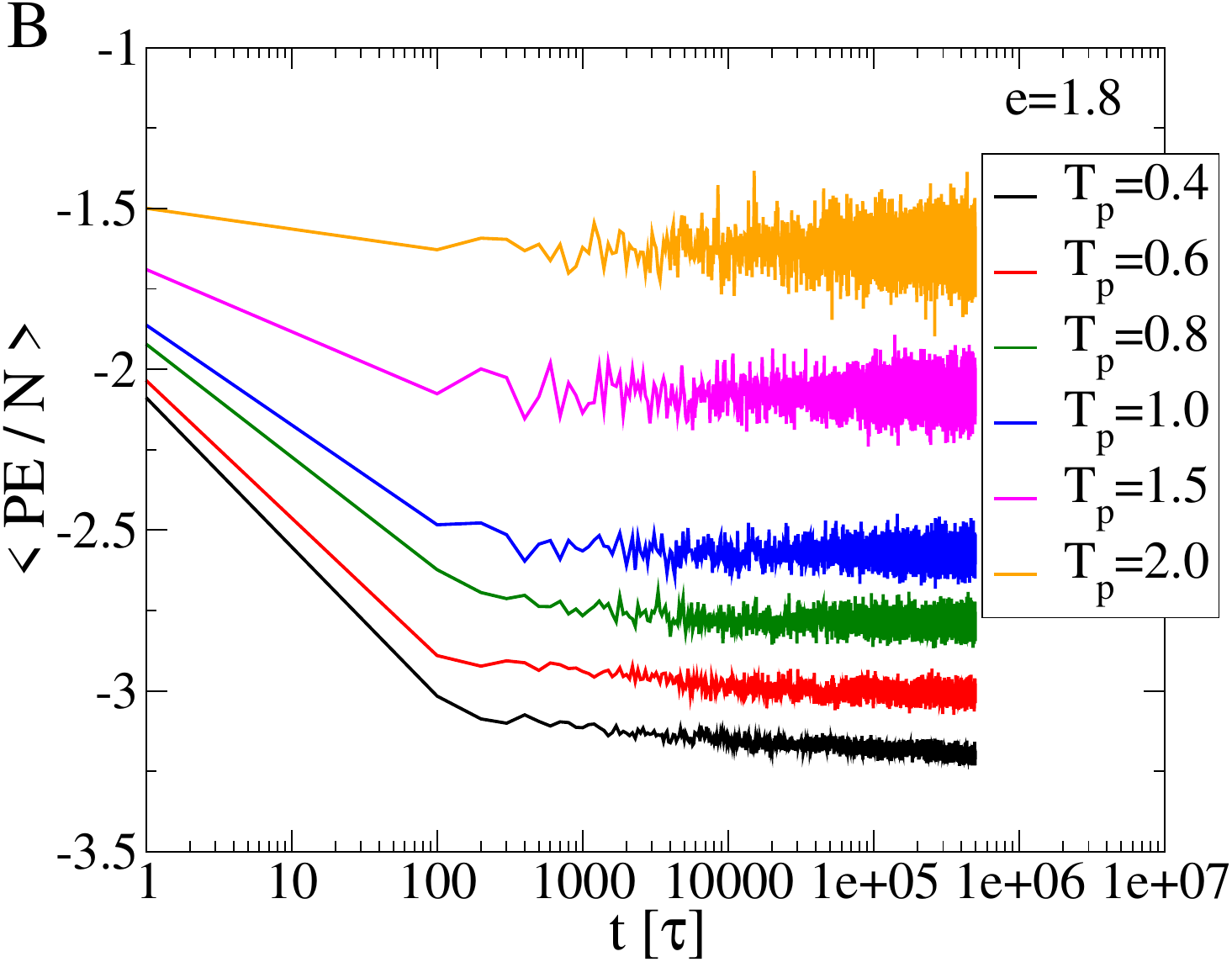}
\caption{Evolution of potential energy per particle over time for a system in two confinement geometries, $e=1.2$ (left), and $e=1.8$ (right) at different preparation temperatures, $T_p$.}
\label{fig:E_age_confine}
\end{figure}

In order to prepare initial samples in confinement, we first thermally anneal the binary mixture at different simulation temperatures, $T$, shown in Fig.~\ref{fig:E_age_confine}. 
Steady state configurations from the annealed trajectories are then subjected to an instantaneous thermal quench to $T_q=0.0001$ and simulated for $5\times10^4$ MD steps to approximate a mechanically stable initial configuration.

\subsection*{Time evolution of energy in confinement}
\begin{figure}[htb!]
\centering
\subfloat{\includegraphics[scale=0.28]{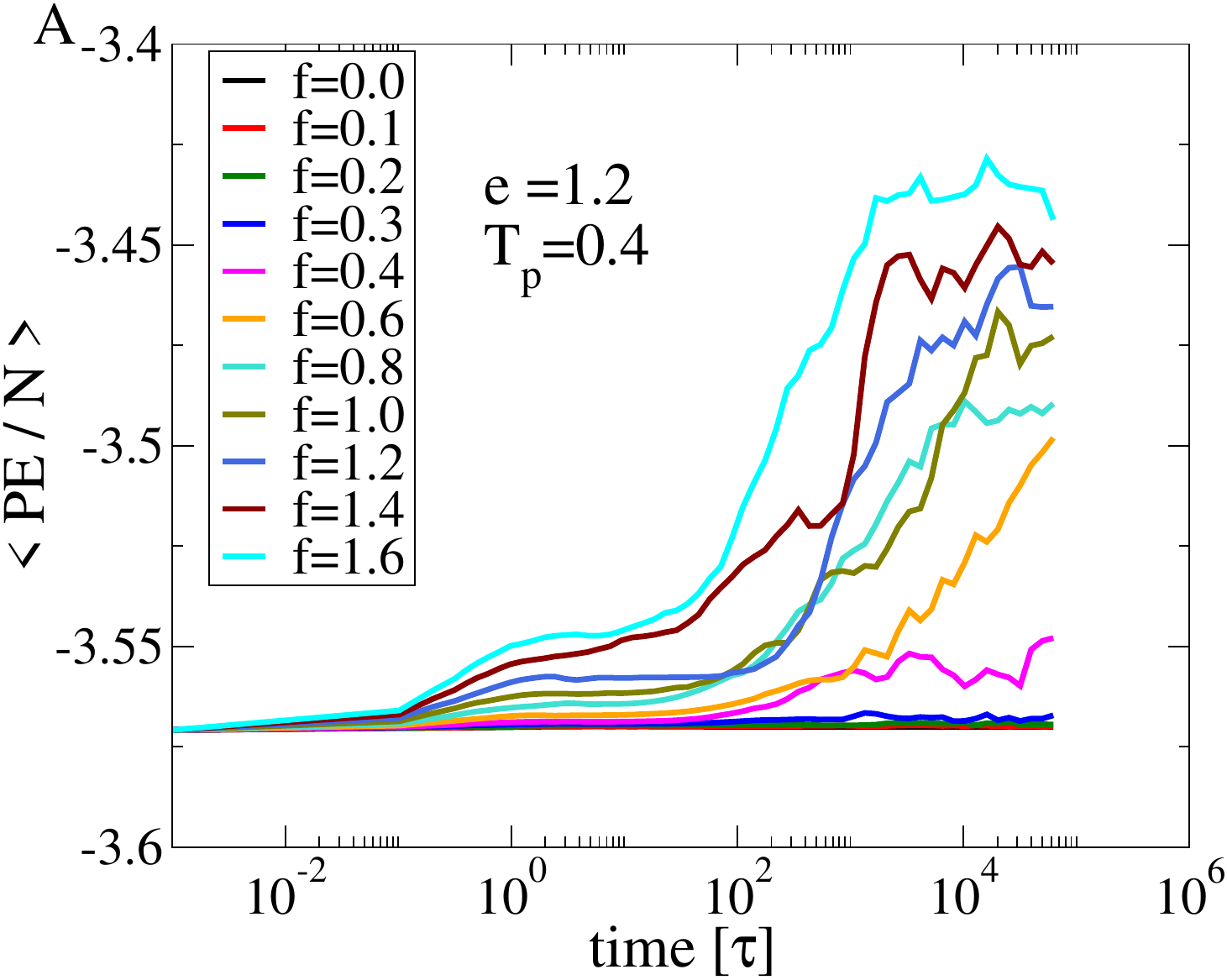}}
\subfloat{\includegraphics[scale=0.28]{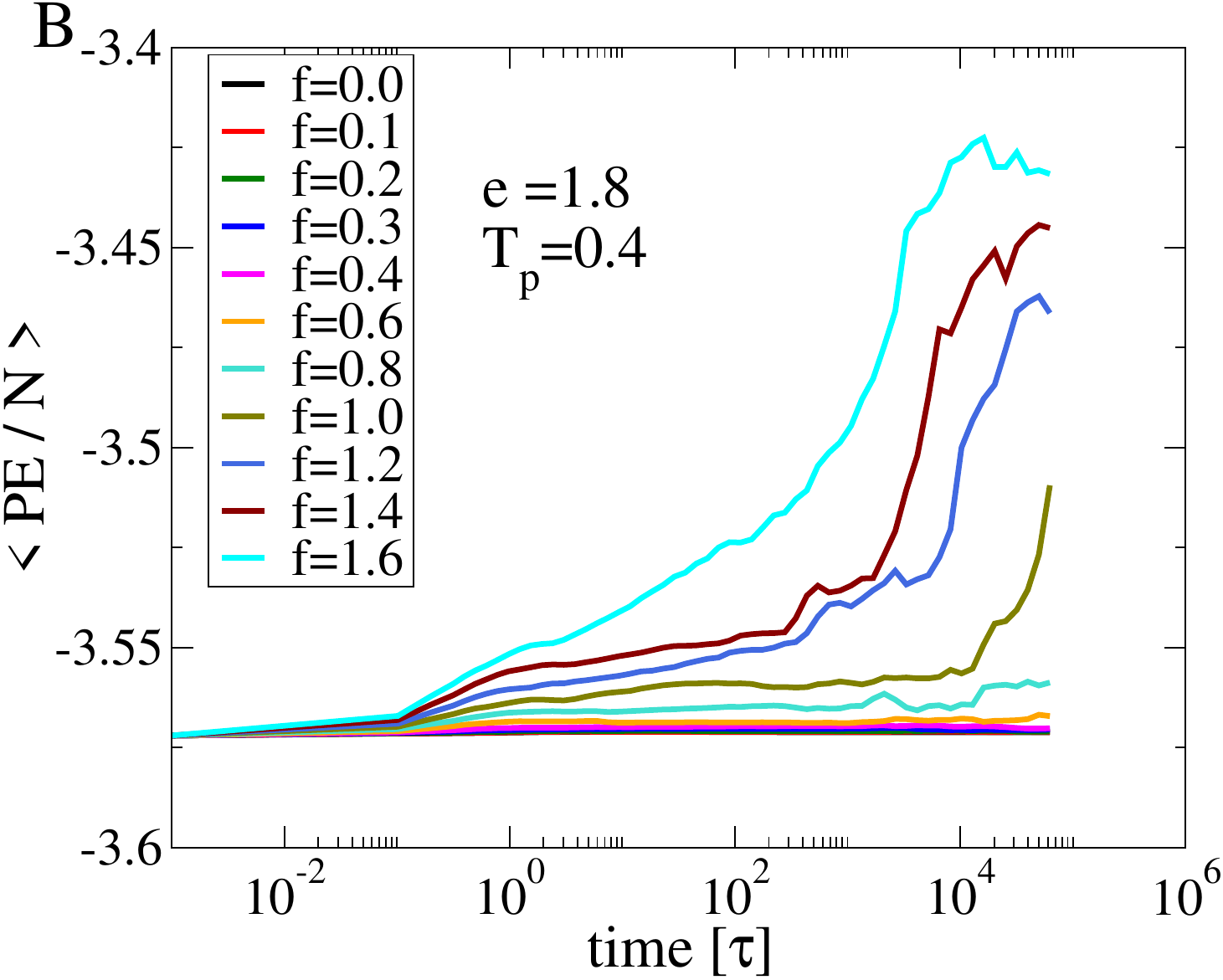}}

\subfloat{\includegraphics[scale=0.28]{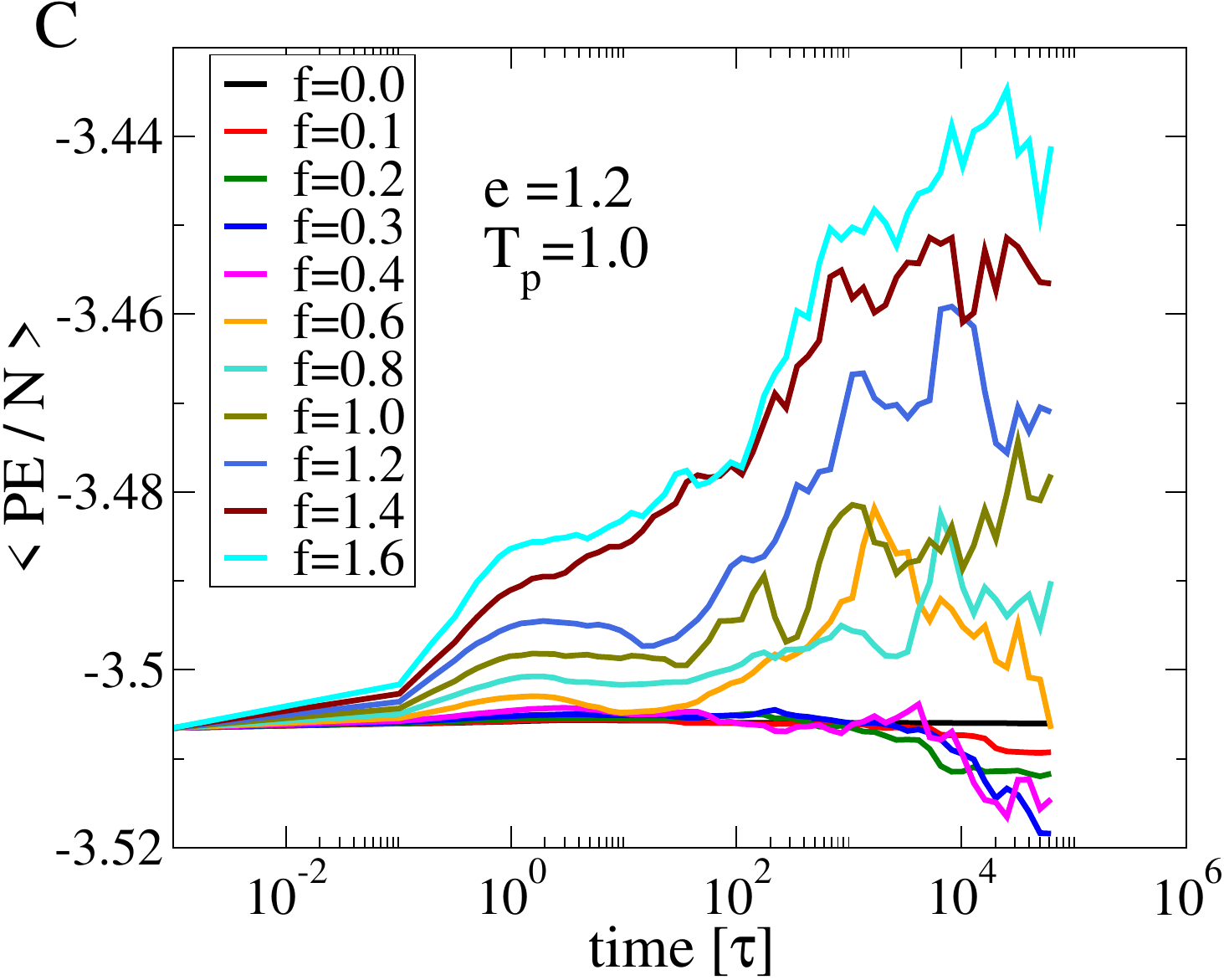}}
\subfloat{\includegraphics[scale=0.28]{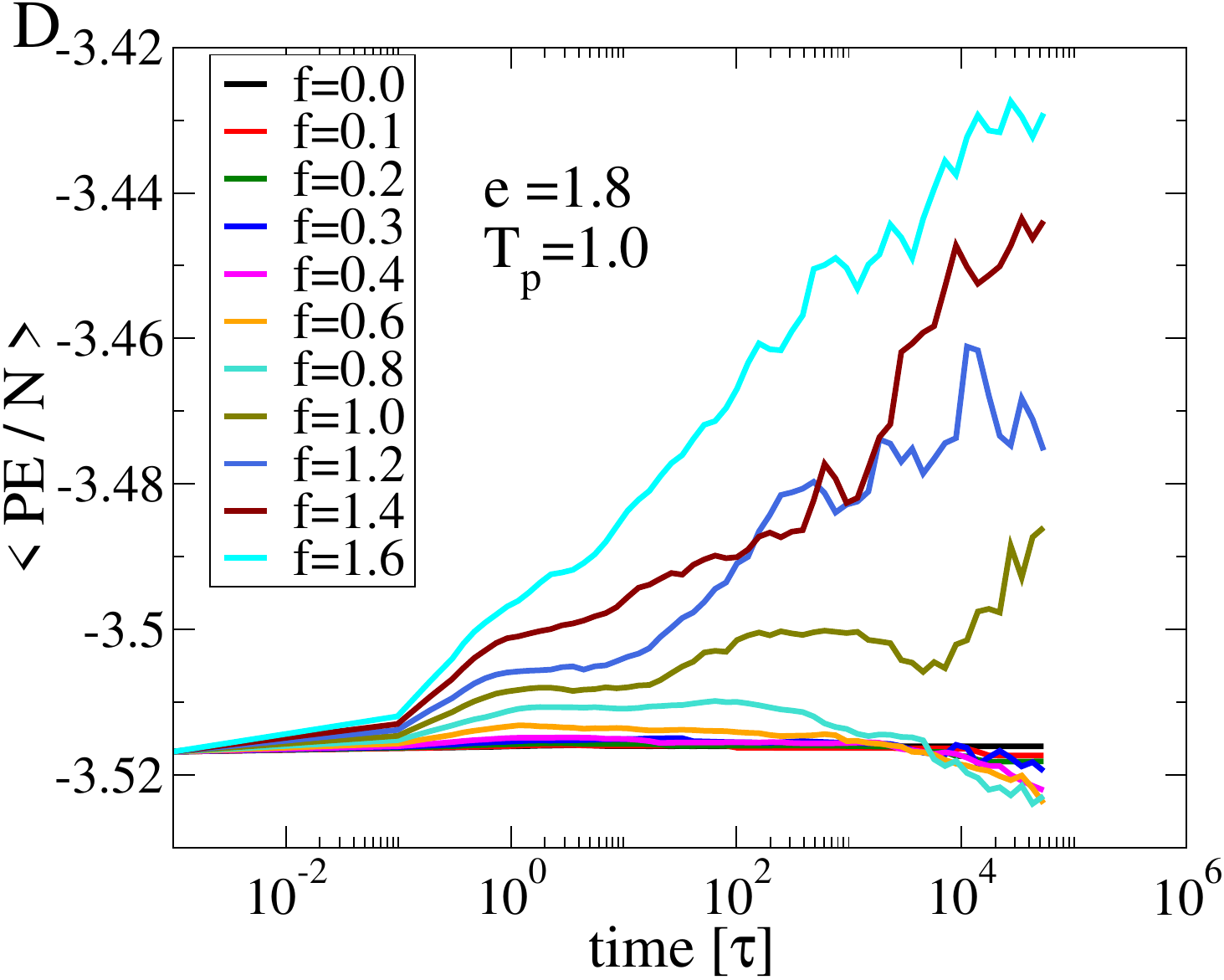}}
\caption{Time evolution of average energy per particle at the values of $e$ and $T_p$ shown in the plots, averaged in logarithmically spaced bins. Curves reach a steady state for all values of active driving force, $f$, except those close to the respective transition values.}
\label{fig:E_v_time_confine}
\end{figure}

In Fig.~\ref{fig:E_v_time_confine} we show the time evolution of the average energy per particle for the two confinement geometries for both the well-annealed sample ($T_p=0.4$) and the poorly annealed one ($T_p=1.0$).
Well-annealed samples show a constant $\langle PE / N \rangle$ with increasing $f$ for small $f$, followed by a steep rise as $f$ is increased beyond the yield point. For the poorly annealed sample, the system undergoes further annealing with increasing $f$ upto the yield point, with a corresponding monotonic decrease in $\langle PE / N \rangle$.
The time taken to reach the steady state is longer close to the yield point, which we interpret will introduce difficulty in obtaining reliable estimates of the final steady state value of the potential energy, as can be seen for the case of $f=0.6$ in Fig.~\ref{fig:E_v_time_confine} (a) and (c) and for the case of $f=0.8$ in Fig.~\ref{fig:E_v_time_confine} (b) and (d).

\FloatBarrier

\end{document}